\documentclass{aa}  
\usepackage{lineno}
\usepackage{float}
\usepackage{printlen}
\usepackage{color}
\usepackage[breaklinks=true]{hyperref}
\usepackage{caption}
\usepackage{subcaption}
\usepackage{multirow}
\usepackage[export]{adjustbox}
\usepackage{multirow}
\usepackage{natbib,twoopt}
\usepackage{tabu}
\usepackage{xcolor}
\usepackage{graphicx,colortbl}

\usepackage{txfonts}
\usepackage{amsmath} 
\usepackage{indentfirst}
\usepackage{mathtools}
\usepackage{ulem} 
\bibpunct{(}{)}{;}{a}{}{,} 
\makeatletter
  \newcommandtwoopt{\citeads}[3][][]{\href{http://adsabs.harvard.edu/abs/#3}%
    {\def\hyper@linkstart##1##2{}%
     \let\hyper@linkend\@empty\citealp[#1][#2]{#3}}}
  \newcommandtwoopt{\citepads}[3][][]{\href{http://adsabs.harvard.edu/abs/#3}%
    {\def\hyper@linkstart##1##2{}%
     \let\hyper@linkend\@empty\citep[#1][#2]{#3}}}
  \newcommandtwoopt{\citetads}[3][][]{\href{http://adsabs.harvard.edu/abs/#3}%
    {\def\hyper@linkstart##1##2{}%
     \let\hyper@linkend\@empty\citet[#1][#2]{#3}}}
  \newcommandtwoopt{\citeyearads}[3][][]%
    {\href{http://adsabs.harvard.edu/abs/#3}
    {\def\hyper@linkstart##1##2{}%
     \let\hyper@linkend\@empty\citeyear[#1][#2]{#3}}}
\makeatother

\titlerunning{}
\authorrunning{Siu-Tapia et al.}
\begin{document}

   \title{Temporal evolution of short-lived penumbral microjets}

   \subtitle{}

   \author{A. Siu-Tapia
          \inst{1}
           \and
          L. R. Bellot Rubio\inst{1} \and
          D. Orozco Su\'arez \inst{1}
          \and
          R. Gafeira \inst{1, 2}
          }

   \institute{
Instituto de Astrof\'isica de Andaluc\'ia (IAA-CSIC), Apdo. 3004, E-18080 Granada, Spain\\
              \email{siu@iaa.es}
               \and
Univ Coimbra, CITEUC - Center for Earth and Space Research of the University of Coimbra, Geophysical and Astronomical Observatory, 3040-004 Coimbra, Portugal
\\
             }


 
  \abstract
   {Penumbral microjets (PMJs) is the name given to elongated jet-like brightenings observed in the chromosphere above sunspot penumbrae. They are transient events that last from a few seconds to several minutes, and their origin is presumed to be related to magnetic reconnection processes. Previous studies have mainly focused on their morphological and spectral characteristics, and more recently on their spectropolarimetric signals during the maximum brightness stage. Studies addressing the  temporal evolution of PMJs have also been carried out, but they are based on  spatial and spectral time variations only.}
   {Here we investigate, for the first time, the temporal evolution of the polarization signals produced by short-lived PMJs (lifetimes $<2$ minutes)  to infer how the magnetic field vector evolves in the upper photosphere and mid-chromosphere. }
   {We use fast-cadence spectropolarimetric observations of the Ca II 854.2 nm line taken with the CRisp Imaging Spectropolarimeter  at the Swedish 1-m Solar Telescope.
   The weak-field approximation (WFA) is used to estimate the  strength and inclination of the magnetic field vector. By separating the Ca II 854.2 nm line into two different wavelength domains to account for the chromospheric origin of the line core and the photospheric contribution to the wings, we infer the height variation of the magnetic field vector. 
   }
   {The WFA reveals larger magnetic field changes in the upper photosphere than in the chromosphere  during the PMJ maximum brightness stage. 
   In the photosphere, the magnetic field inclination and strength undergo a transient increase  for most PMJs, but in $25\%$ of the cases the field strength decreases during the brightening.
   In the chromosphere, the magnetic field  tends to be slightly stronger during the PMJs. 
   }
   {
    The propagation of compressive perturbation fronts followed by a rarefaction phase in the aftershock region may explain the observed behavior of the magnetic field vector. The fact that such behavior varies among the analyzed PMJs could be a consequence of the limited temporal resolution of the observations and the fast-evolving nature of the PMJs.  
}

   \maketitle
%

\newpage

\section{Introduction}

Penumbral microjets (PMJs) are among the first discoveries made with the Solar Optical Telescope \citep[SOT;  ][]{Tsuneta2008} aboard the Hinode spacecraft \citep[][]{Kosugi2007}. \citet{Katsukawa2007} observed the ubiquitous occurrence of small-scale elongated jet-like brightenings above a sunspot penumbra in Ca II H images taken with the Broadband Filter Imager. 
PMJ brightenings were characterized as short transients about $10-20\%$ brighter than the surrounding environment, with lifetimes of $\sim$1 minute. They appear suddenly in image sequences, with apparent horizontal speeds of the order of 100 km s$^{-1}$, and then fade gradually. 

\citet{Katsukawa2007} also found that PMJs have typical lengths between 1000 and 4000 km (but up to 10000 km), widths around 400 km, and that they appear to be well aligned with the superpenumbral fibrils. The brightenings are more easily identified when the sunspot is located closer to the limb, given that they make a larger angle with  the bright photospheric penumbral filaments. 
Near the disk center, the PMJ brightenings are closely aligned with 
the bright penumbral filaments and so running-difference images are advantageous to detect them there. The center-to-limb variation of the PMJ orientation with respect to the penumbral filaments is due to the expansion of the magnetic field with height, as pointed out by \citet{Jurcak2008}. This interpretation can also explain the dependence of the PMJ inclination on position within the penumbra \citep{Jurcak2008}.

Magnetic reconnection between differently inclined field lines has been suggested as a possible driver of PMJs \citep[e.g., ][]{Katsukawa2007, Katsukawa2010, Jurcak2010}, given that the complex configuration of the magnetic field in sunspot penumbrae  involves strong variations of the field strength and inclination over small spatial scales \citep[e. g., ][]{Borrero2011, Rempel2011b}. This scenario has been supported by magnetohydrodynamic simulations  \citep{Magara2010}.
In particular, \citet{Katsukawa2010} and \citet{ Jurcak2010} studied the photospheric response associated with PMJs and found some  downflows in the lower photosphere  to be related to chromospheric brightenings. They claimed that these could be a possible signature of magnetic reconnection in the middle or low photosphere.
\citet{Tiwari2016} also reported small-scale photospheric downflows and opposite polarity patches related to PMJs, while \citet{Vissers2015} found a progressive heating to transition region temperatures along the PMJs. These features might be a byproduct of magnetic reconnection occurring in the low atmosphere. However, \citet{Samanta2017} suggested that reconnection might occur higher in the atmosphere, in the low corona or transition region, based on observations of inward-moving elongated bright dots that appear before and superposed to some PMJs.

 \begin{figure*}
   \centering
   \includegraphics[width=\hsize]{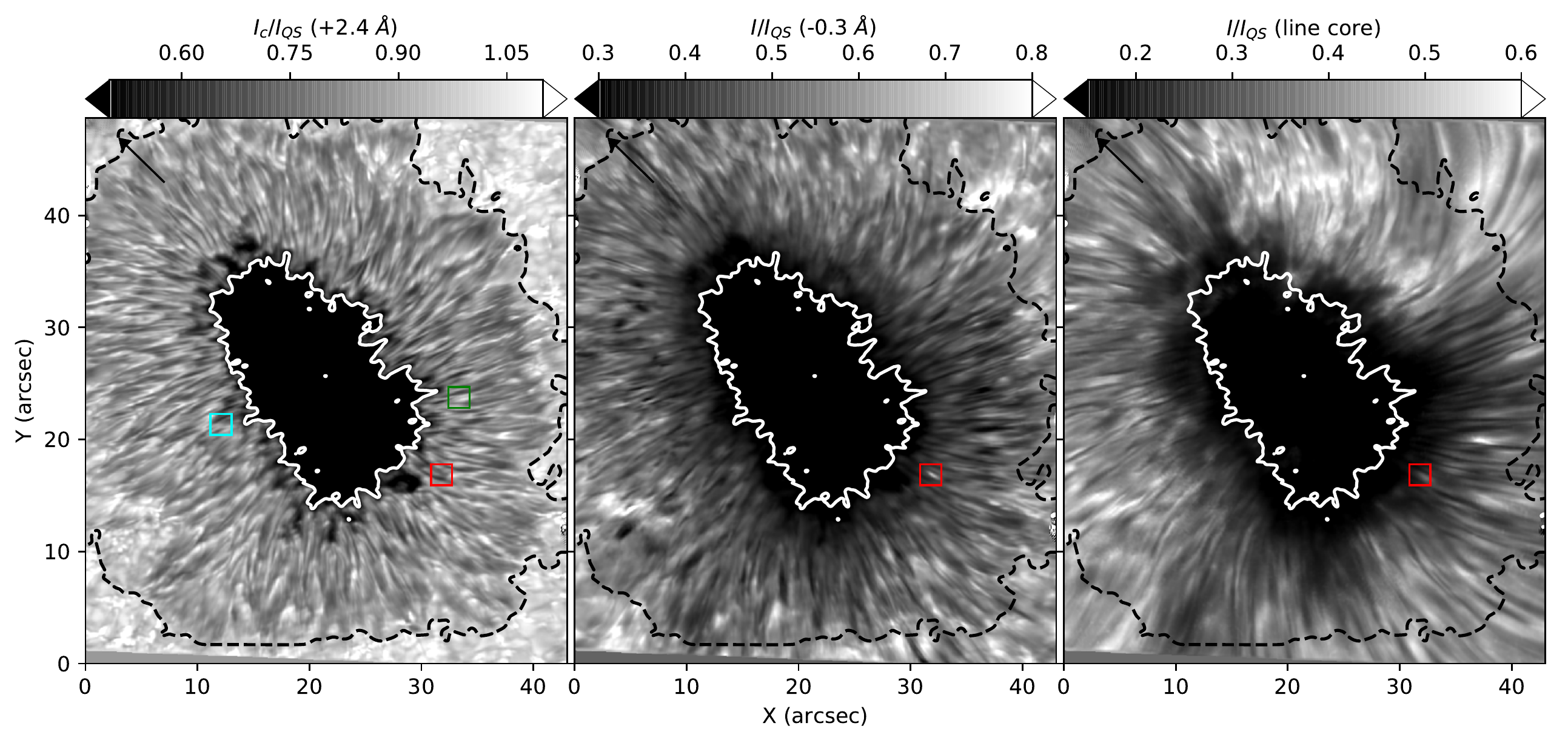}
      \caption{Images of the main sunspot in active region 12553 on 2016 June 16 recorded with the CRISP instrument at different wavelengths within the Ca II 8542 $\r{A}$ line. Left panel: Continuum intensity  $I_c$ image, at $+2.4$ $\r{A}$ from the line core. Central panel: Blue-wing image, at $-0.3$ $\r{A}$ from the line core. Right panel: Line core image. All the images are normalized to the mean continuum value in the quiet sun, $I_{QS}$. The umbra-penumbra boundary (white  contours) is at $I_c/I_{QS}$=0.45 and the outer penumbral boundary (black dashed contours) at $I_c/I_{QS}$=0.98. Black arrows point towards the disk center. Small colored squares on the continuum image highlight the location of three PMJs referred to as PMJ 1 (red), PMJ 2 (green), and PMJ 3 (cyan) in the main text. The images correspond to the frame in which PMJ 1 displays the maximum blue-wing brightness. 
              }
         \label{Fig:1}
   \end{figure*}

 \citet{Reardon2013}  performed the first spectroscopic analysis of PMJs using Ca II 8542 $\r{A}$  intensity profiles from the Dunn Solar Telescope and co-temporal Hinode Ca II H imaging data. They found a wide variety of temporal, spatial, and spectral behaviors among penumbral brightenings observed in the chromosphere, 
 and  argued that  different physical processes could drive these transients.  However, their  temporal light curves in integrated intensity are very similar. They also found several events whose characteristics  matched well the lifetimes and orientations of PMJs described in works based on SOT Ca II H filtergrams only. Such events were characterized by clear Ca II 8542 line wing emissions and almost unchanged line cores, similar to the spectral signatures of Ellerman bombs \citep[e.g., ][]{Vissers2013}.

 \citet{Reardon2013} also reported the presence of ``precursors''  around 1 minute before the rapid impulsive brightening of some PMJs, which could be the signature of a shock, as proposed by \citet{Ryutova2008} in their bow-shock model. 
 
 The first statistical analysis of PMJ properties was performed by \citet{Drews2017} by means of an automatic detection method using Ca II 8542 $\r{A}$   spectroscopic data and Ca II H filtergrams taken at the SST. Their algorithm was set to select those PMJs whose spectral characteristics resemble Ellerman bombs in H$\alpha$, i. e. enhanced line wings and an almost undisturbed line core, finding that the wing enhancements could extend to about  0.6 $\r{A}$ from the line core. Most PMJs displayed stronger 
  enhancements in the inner blue wing compared to the red wings. 
 The average lifetime of the 453 identified PMJs was 90 s, for an upper cutoff value of 8 minutes, but the distribution peaked at shorter durations. 
 
 Recently, \citet{Esteban2019}  investigated the polarization signals of 37 PMJs with lifetimes between 1 and 6.5 minutes, and found that they tend to appear in regions where the inclination of the photospheric penumbral magnetic field undergoes significant horizontal variations, such as at the interface between spines and intra-spines \citep[e.g., ][]{Borrero2011} or in the outer penumbra where the field lines bend over at the end of the filaments \citep[see also][]{Tiwari2018}. However, no strong Doppler shifts were observed and the estimated line-of-sight (LOS) motions had speeds lower than 4 km s$^{-1}$, value that is well below the apparent speeds of the order of 100 km s$^{-1}$ previously inferred.
 Nonetheless, velocity gradients along the LOS were deduced from the shapes of the emission peaks of the Ca II 8542 $\r{A}$  and Ca II K intensity profiles.  The authors also found an increase of the temperature at chromospheric heights in their inversion results. Therefore, they suggested that PMJs could be a consequence of magnetic reconnection in the low photosphere, a process that might trigger an upwardly propagating perturbation responsible for heating the chromosphere.
 In addition,  \citet{Esteban2019} found that PMJs exhibit enhanced polarization signals, particularly the Ca II 8542 $\r{A}$ circular polarization profiles which in some cases displayed extra lobes. The later is explained by the authors to be a consequence of the typical moustache shapes of the intensity profiles, given that they can be reproduced  using the weak field approximation (WFA).
 
 In this work we investigate for the first time the temporal evolution of the polarization profiles of the most common PMJs, which, according to the statistical study by \citet{Drews2017}, are those with lifetimes under 2 minutes. To our knowledge, the magnetic properties and evolution of these very short events have not been analyzed before. The reason is that such a study requires challenging observations (ultra-fast spectropolarimetry) due to their highly dynamic nature. 
 
 Studying the temporal changes in the magnetic field configuration and in the plasma properties occurring in and around PMJs can give us insight on their physical nature  and on where they lie on the solar atmosphere.
However, such an analysis faces various observational limitations.
Because of the extremely short lifetimes of the most frequent PMJs, observations with the fastest temporal cadence possible are needed. This requirement puts restrictions on other observational aspects, such as the number of spectral lines that can be scanned, the width
 of the observed spectral range and the spectral sampling. Moreover, it implies short exposure times which can substantially increase the noise. 
 
 Here, we approach this problem by making trade offs between the different observational parameters as described in Sect. 2. 
 In Sect. 3 we explain our identification criteria and the diagnostic tools we use. In Sect. 4  we present individual examples and statistical results, which are summarized in Sect. 5. Finally, in Sect. 6 we discuss our findings and draw conclusions.

 \begin{figure}
   \centering
    
    \includegraphics[width=0.9\hsize]{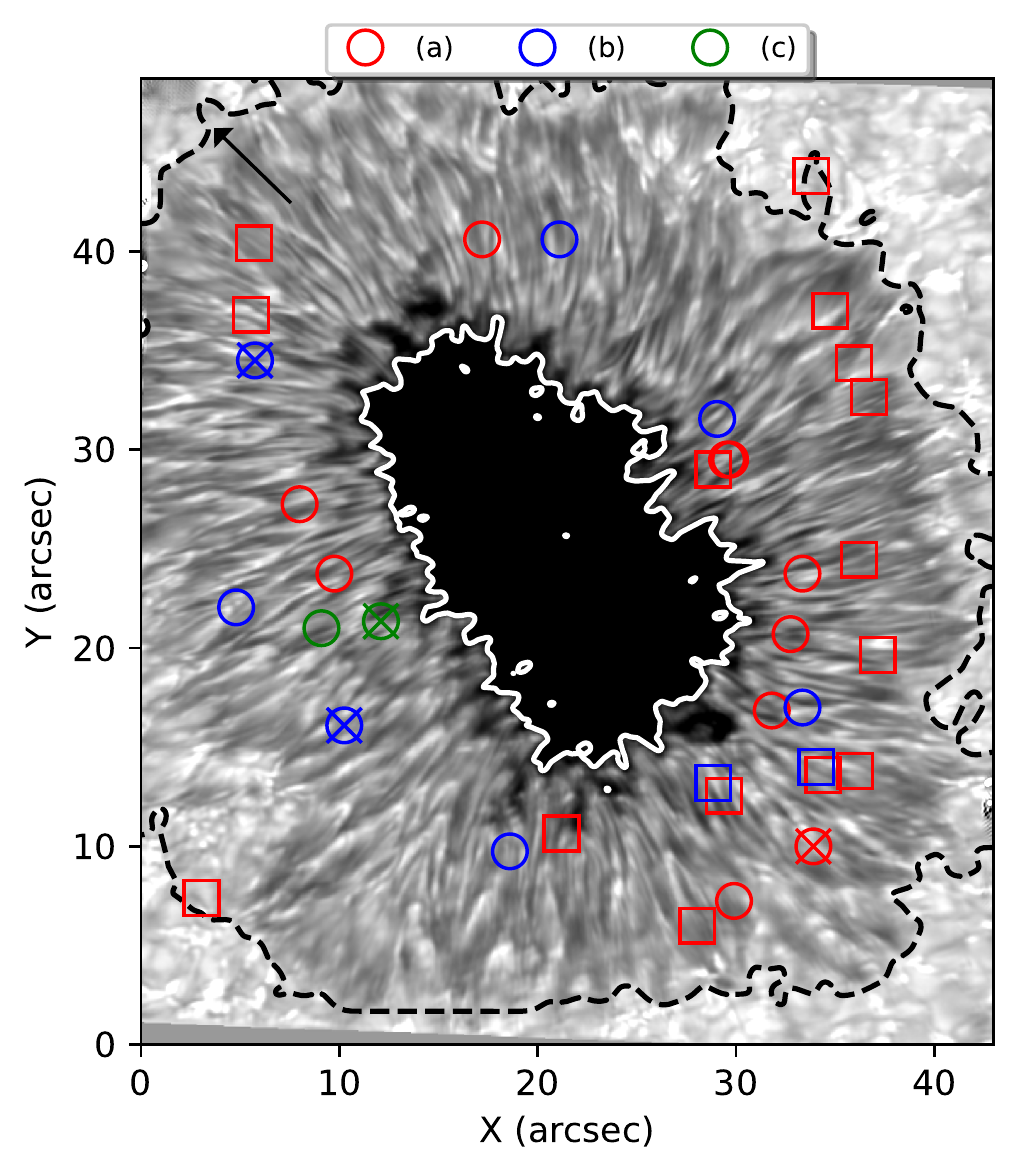}
    \includegraphics[width=0.9\hsize]{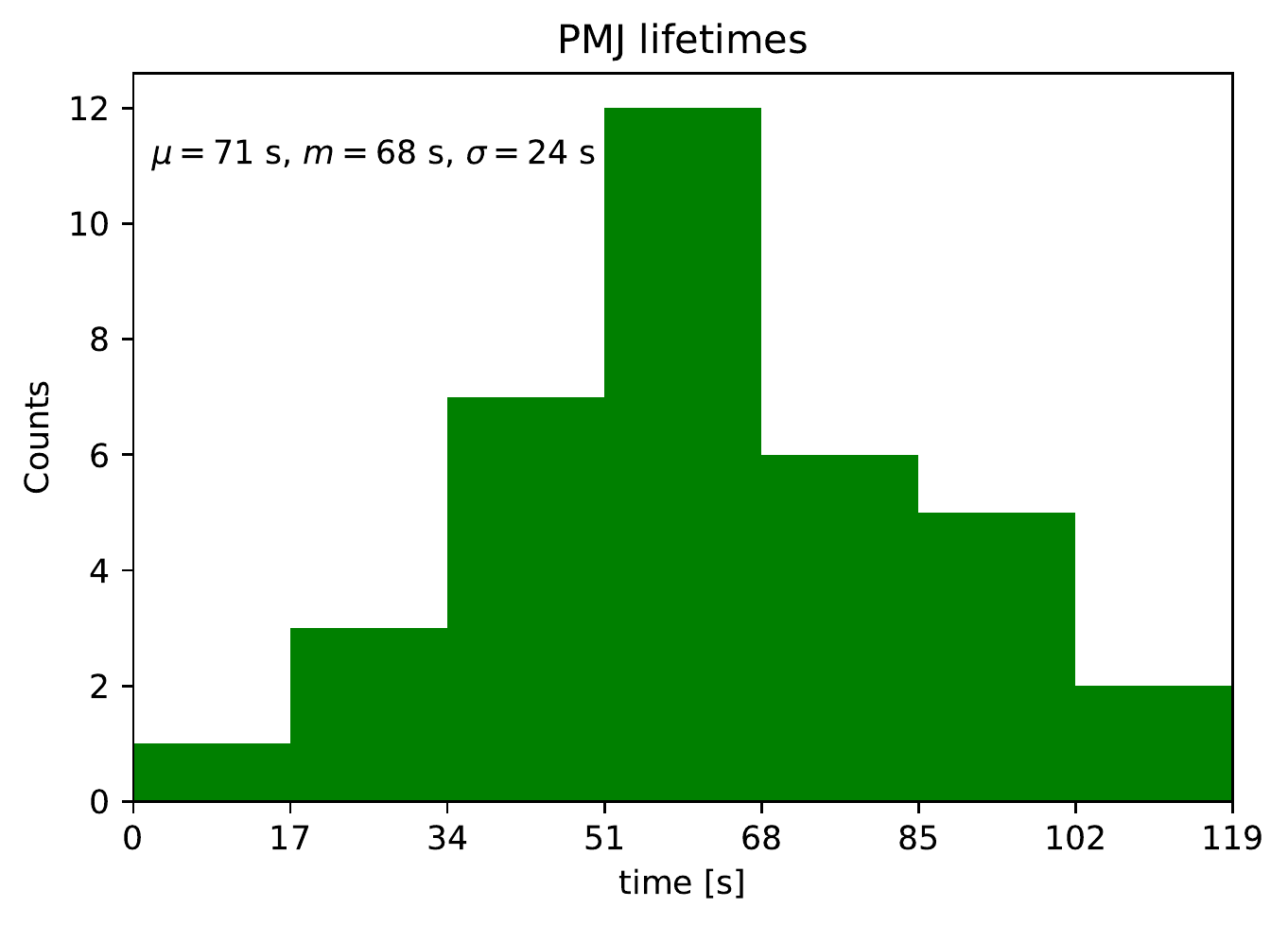}
      \caption{Top: Continuum intensity image of the sunspot in the same format as Fig. \ref{Fig:1}, displaying the location of the 36 short-lived PMJs that were identified in the complete temporal sequence,  with lifetimes shorter than 2 minutes. The different colors of the markers indicate different types of evolution of the photospheric magnetic field as described in the main text. 
      PMJs displaying line core brightness enhancements larger than $10\%$ are shown with circles; otherwise, they are indicated with squares. The crossed symbols stand for PMJs displaying clear changes of the chromospheric magnetic field during their brightening.
      Bottom: Lifetime distribution for the selected PMJs. 
      The mean $\mu$, median $m$, and standard deviation $\sigma$ of the distribution are indicated in the upper left corner.
              }
         \label{Fig:2}
   \end{figure}

\section{Observations }\label{Sec:obs}

Spectropolarimetric observations of the main sunspot in active region 12553  were carried out on 2016 June 16 at the Swedish 1-m Solar Telescope \citep[SST; ][]{Scharmer2003} between 08:06:00  and 09:27:53 UT  using the CRisp Imaging Spectrometer \citep[CRISP;][]{Scharmer2008b}. The sunspot displayed positive magnetic polarity and  was observed very close to disk center, at a heliocentric angle of $\sim8^{\circ}$ (19$^{\prime\prime}$ W, 125$^{\prime\prime}$ S).

Our CRISP data consist of time series of full Stokes measurements in the chromospheric Ca II 8542 $\r{A}$ line and in the photospheric Fe I 6173 $\r{A}$ line. 
In order to perform observations with a temporal cadence fast enough to study the evolution of the shortest-duration PMJs, the Ca II line was sampled at 9 wavelength positions in steps of 150 m$\r{A}$, from -600 to +600 m$\r{A}$ around the line core, plus a 
point at +2.4 $\r{A}$. Similarly, the photospheric Fe I  line was sampled at 5 wavelength positions in steps of 40 m$\r{A}$, from -80 to +80 m$\r{A}$ around the line core, plus one position in the continuum at +400 m$\r{A}$.
Sampling the two lines required 17 s (11 s for the Ca II line and 6 s for the Fe I line), using 9 accumulations for each line. The later was set as the best possible compromise between the total integration time and the photon noise level. The duration of the  time series is $\sim82$ minutes, i.e., 290 time steps. The field of view (FOV) is $\sim50^{\prime\prime}\times50^{\prime\prime}$ with a pixel size of 0.05$^{\prime\prime}$.

The data were reduced using the CRISPRED pipeline \citep{delaCruz2015} and reconstructed with the Multi-Object-Multi-Frame-Blind-Deconvolution  technique \citep[][]{Lofdhal1994,Vannoort2005},  separately for each line.
Unfortunately, the reconstruction of the Fe I 6173 $\r{A}$ line resulted in poor quality data and is therefore not used in this study.
However, the Ca II 8542 $\r{A}$ line has contributions from different layers, i.e., while its line core 
 originates from layers at heights between $\sim800-1000$ km, the wings are formed at considerably deeper layers, mainly below  500 km \citep{Cauzzi2008}. Thus, some information about the photospheric properties of the sunspot is  available from  the  Ca II 8542 $\r{A}$ line and the 
  wavelength point scanned at +2.4 $\r{A}$. 

In order to reduce the noise level of the observations, we have convolved the complete dataset with a 3 by 3  low-pass filter kernel so that each pixel in the resulting images has a value equal to the average value of its neighboring pixels in the original image. The noise levels in the resulting images are 
$\sigma_{Q,U}\sim3\times10^{-3}$, and $\sigma_V\sim4\times10^{-3}$, as measured on the continuum wavelength maps. 
 Finally, the  data have been corrected for residual crosstalk by using the average value of the $\alpha_{s}$ maps at each time step,  where $\alpha_{s}=s(\lambda_c)/I(\lambda_c)$  with $s=Q,U,V$ and $\lambda_c$ corresponding to the continuum wavelength.

Figure \ref{Fig:1} displays the general appearance of the observed sunspot at different wavelengths within the Ca II 8542 $\r{A}$ line: continuum (left), blue wing (central), and line core (right panel). The image on the right shows that there is a predominantly brighter region in the northern part of the sunspot where the line core is in emission. This region maintains line core emissions during our complete time series and makes the detection of PMJs 
very difficult in that part of the penumbra.

\section{Analysis}

\subsection{PMJ identification}

We are interested in studying the most common PMJs which, according to \citet{Drews2017}, are the events with lifetimes shorter than 2 minutes.
In particular, the polarimetric properties of the shortest duration PMJs ($<1$ minute) have never been analyzed, inasmuch as the only available study focused on PMJs that live longer than 1 minute \citep[][]{Esteban2019}.

Given that PMJs show a preference for larger brightness enhancements in the blue wing of the Ca II 8542 $\r{A}$ intensity profile, 
with the help of the temporal sequence of images at -0.3 $\r{A}$ from the line core and of the running difference images (intensity differences with respect to the frame recorded 17 seconds before), we identified 36 PMJs with lifetimes shorter than 2 minutes through visual inspection. This means that the associated brightening is observed during a maximum of 7 frames (119 s) and a minimum of 1 frame (17 s) on the blue wing images and/or on their running differences.
The brightenings must display elongated shapes and intensity changes in the blue wing (at -0.3 $\r{A}$) of at least 10$\%$ of their intensity profiles averaged over the full time sequence.

Figure \ref{Fig:2} shows the location and lifetimes of the 36 short-lived PMJs identified in our complete temporal sequence. 
The spatial distribution reveals a larger number of PMJs occurring in the limb-side  penumbra. However, this trend could be the result of intrinsic asymmetries of this particular sunspot, such as the persistent line core emissions detected over the center-side penumbra which hinder the identification of possible wing emissions produced by PMJs in that region.

The PMJs were observed at different radial positions within the penumbra, mainly above the interface between dark and bright filaments (13 cases), but some PMJs occurred at the end of bright penumbral filaments (6 cases) or above the filaments (6 cases), and between tails and heads of different filaments (9 cases).   
Two PMJs emerged just at the outer penumbral boundary.
The lifetime distribution of the 36 PMJs has a mean value $\mu=71$ s and a median $m=68$ s (i.e., 4 frames).

A total of 19 PMJs (circle symbols in Fig. \ref{Fig:2}) displayed clear brightness enhancements in the line core within regions that are slightly shifted in the radial direction from the PMJ regions, as shown in the following subsection. The observed displacements are interpreted as a projection effect of the chromospheric counterparts (CCs) of magnetic field lines that connect with the PMJ regions at photospheric layers. 

 \begin{figure}
   \centering
   
   \includegraphics[width=1\hsize]{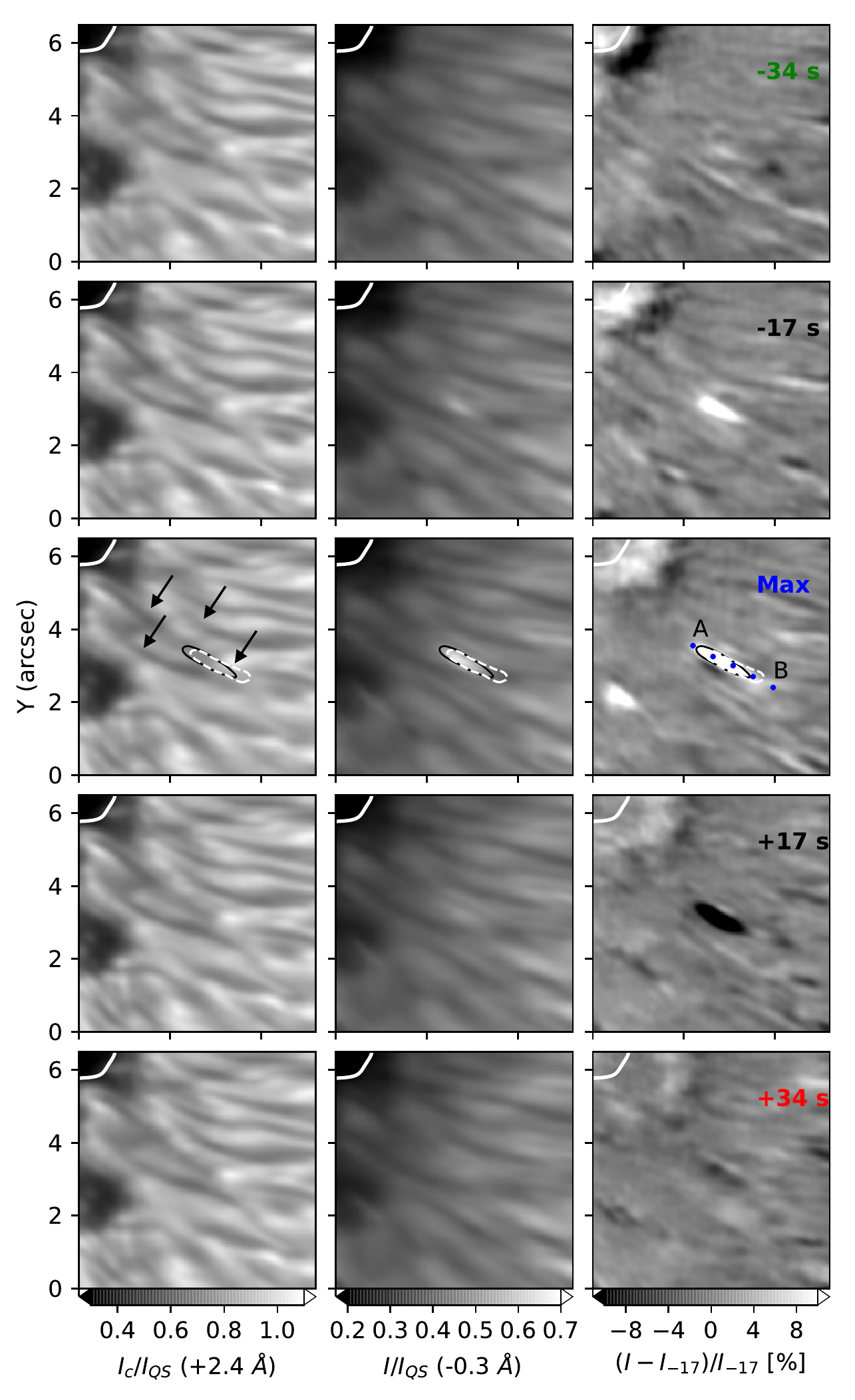}

      \caption{Example of the evolution of a PMJ whose brightening is observed during 3 frames (lifetime of 51 s) in the inner limb-side penumbra (labeled as PMJ 1 in Fig. \ref{Fig:1}). Left: Continuum intensity images at $+2.4$ $\r{A}$ from the line core. Center: Blue-wing images, at $-0.3$ $\r{A}$ from the core. Right: Blue-wing intensity difference images  with respect to the frame recorded 17 s before. Time increases from top to bottom, see labels on the right panels. The middle row displays the maximum brightness stage. 
      Black contours delimit the PMJ area, i.e., the region displaying blue-wing brightness enhancements larger than 10$\%$.  White dashed contours enclose the CC, defined as the region displaying a line-core brightness enhancement larger than $10\%$.
      White solid lines delimit the inner penumbral boundary. Black arrows indicate three penumbral filaments and a bright grain observed in the continuum images near the brightening region.
              }
         \label{Fig:3}
   \end{figure}

 \begin{figure}
   \centering
   
     \includegraphics[width=0.9\hsize]{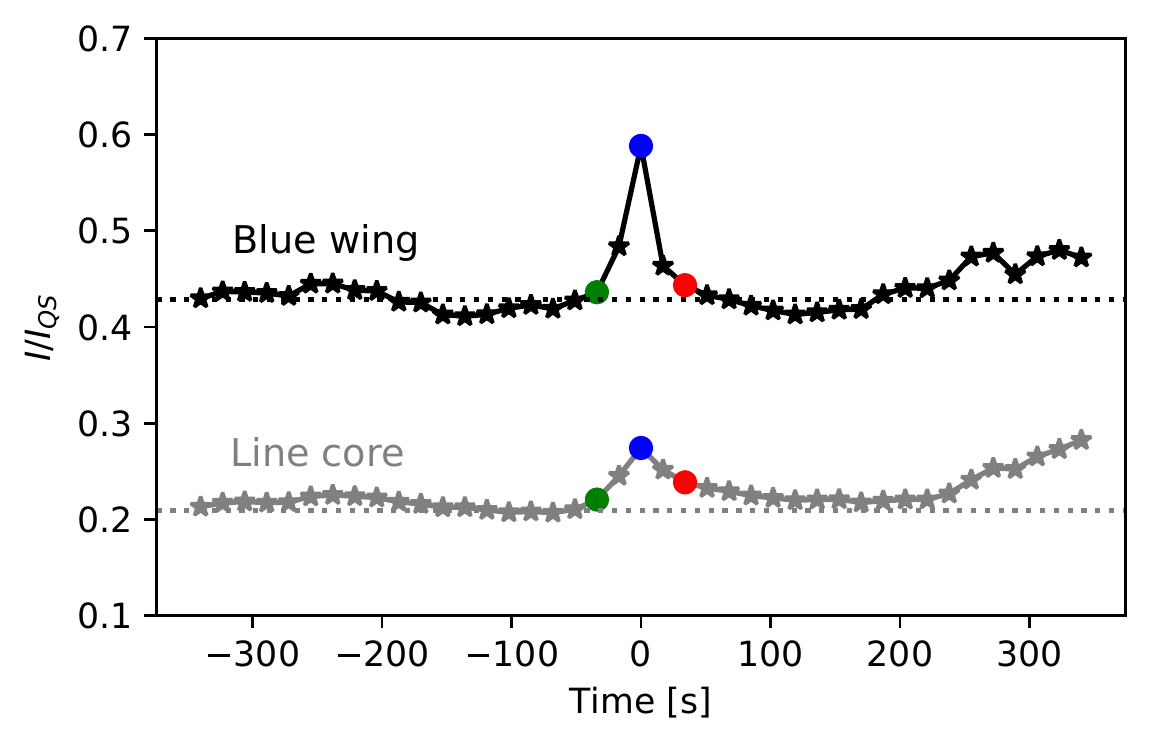}
   \includegraphics[width=0.58\hsize]{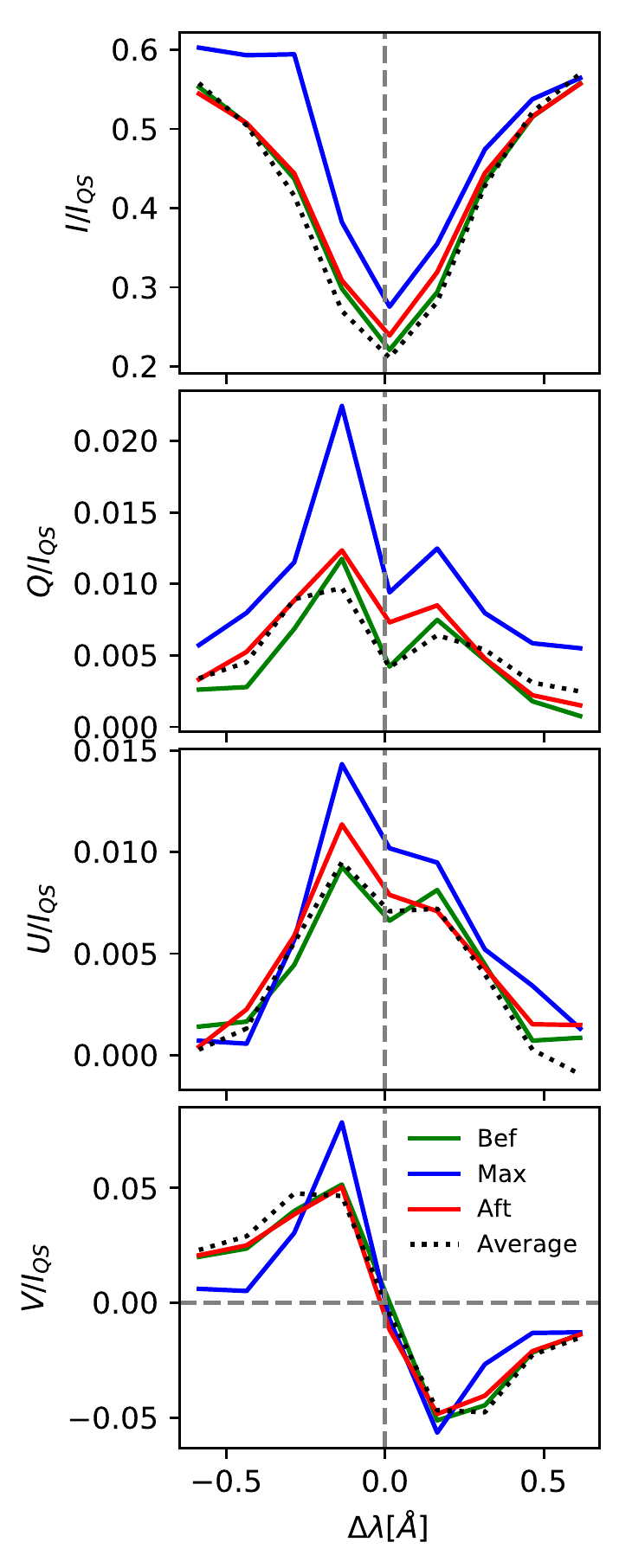}
   \includegraphics[width=0.408\hsize]{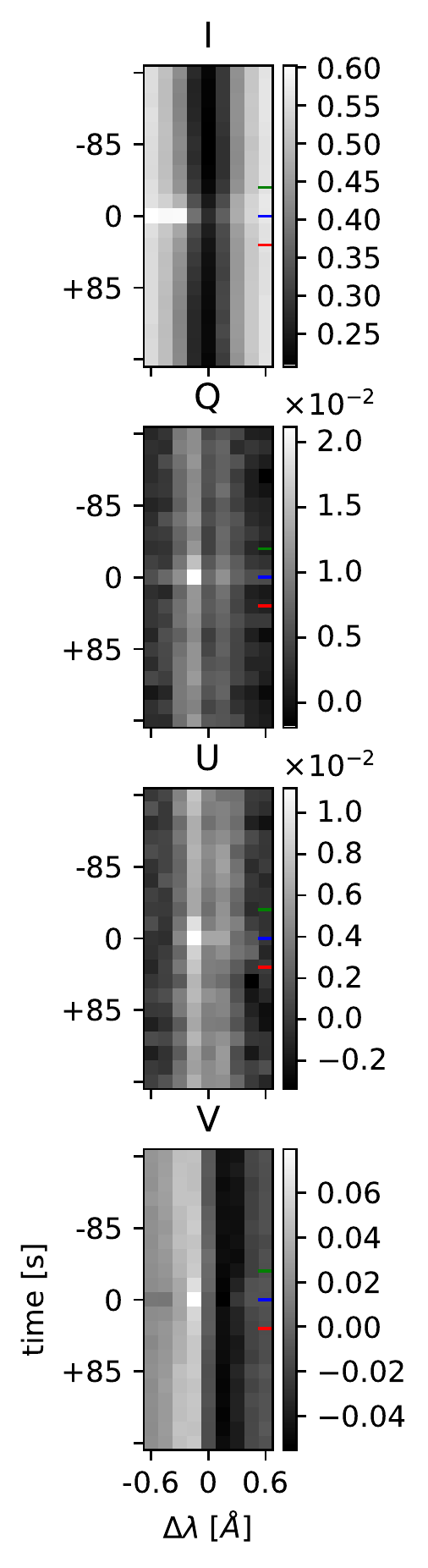}
      \caption{Polarization signals of PMJ 1. Top: Temporal evolution of the blue-wing intensity 
       (black) and line-core intensity (gray)
       at the maximum brightness pixel (MBP), averaged over its 8 closest neighboring pixels. The green dots correspond to the time step before the PMJ becomes visible  (first row in Fig. \ref{Fig:3}), the blue dots to the maximum brightness stage (third row  in Fig. \ref{Fig:3}), and the red dots to the frame after the PMJ disappears (last row  in Fig. \ref{Fig:3}). 
      The black and gray dotted lines indicate the  blue-wing and line-core intensities, respectively, averaged over the full time sequence. 
      Bottom left: Stokes profiles emerging from the MBP  before (green), during (blue), and after (red) the PMJ is visible. The black dotted  profiles correspond to the average Stokes profiles over the complete time series at the MBP. Bottom right: Temporal evolution of the Stokes profiles at the MBP. The colored dashes on the right indicate the 3 time steps of interest, with the same color code. 
              }
         \label{Fig:4}
   \end{figure}

    \begin{figure}
   \centering
   
   \includegraphics[width=\hsize]{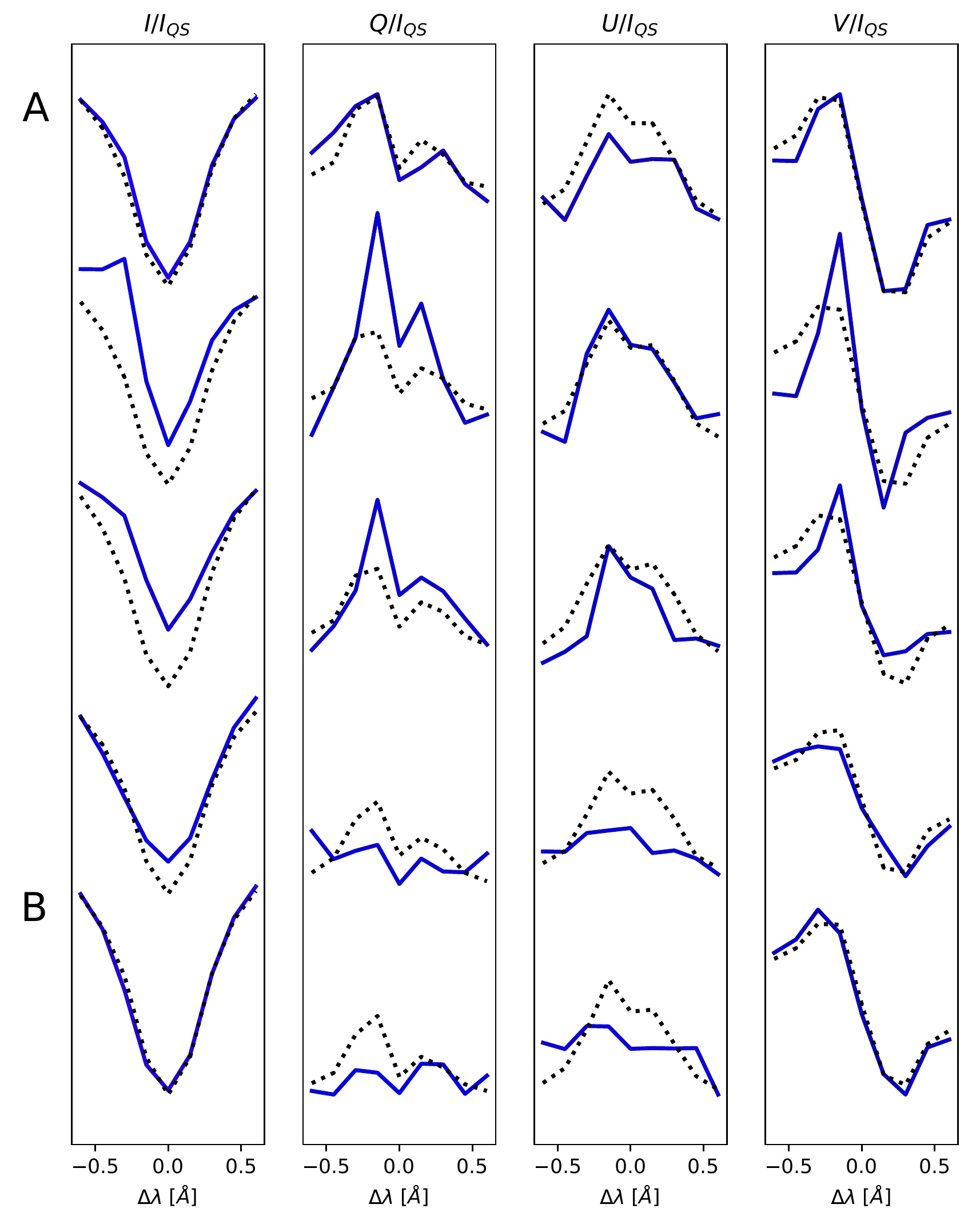}
  
      \caption{Spatial variation of the Stokes profiles along PMJ 1. The profiles emerged from the locations marked with blue symbols in Fig. \ref{Fig:3}, from point A (top) to point B (bottom). The solid profiles correspond to the maximum brightness stage and the dotted profiles to the temporal averages over the complete time series at the same locations. }
         \label{Fig:4_new}
   \end{figure}

\subsection{Spectral and polarimetric characteristics}

An example of a short-lived PMJ which is visible during less than 1 minute, hereafter referred to as PMJ 1, is displayed in Figure \ref{Fig:3}.
The figure shows images of the continuum intensity (left column), the blue-wing intensity at -0.3 $\r{A}$  (central column), and the normalized running differences of the blue-wing images during 5 consecutive frames (right column). The PMJ brightening is visible in 3 frames (from row 2 to 4) in the blue-wing images, so that its estimated lifetime is  51 s. The intensity change in the blue wing is larger than $10\%$ between the first and the second frame, and it is nearly $36\%$ between the first and the maximum brightness frame (row 3). The intensity change is negative after the maximum brightness stage (row 4). 

We refer to the frame preceding the first appearance of the PMJ brightening as  ``before'' (frame in row 1 of Fig. \ref{Fig:3}), and to the frame after the PMJ brightening vanishes as ``after'' (frame in row 5 of Fig. \ref{Fig:3}).

The PMJ brightening is not visible in the continuum  images. 
The black elongated contours in the plots enclose the PMJ during the maximum brightness stage. 
 Three penumbral filaments with dark lanes  meet in such region.
The PMJ brightening occurs near the end (or tail) of the central filament, which protrudes from the inner penumbral boundary into the umbra. In addition, a bright penumbral grain (which could be either the head of a fourth filament or a supersonic downflow at the end of one of the filaments, see \citealt{Esteban2016}) is also visible  in the radial direction further away from the tail of the central filament.

\begin{figure*}[hbt!]
   \centering
   
   \includegraphics[width=0.67\hsize]{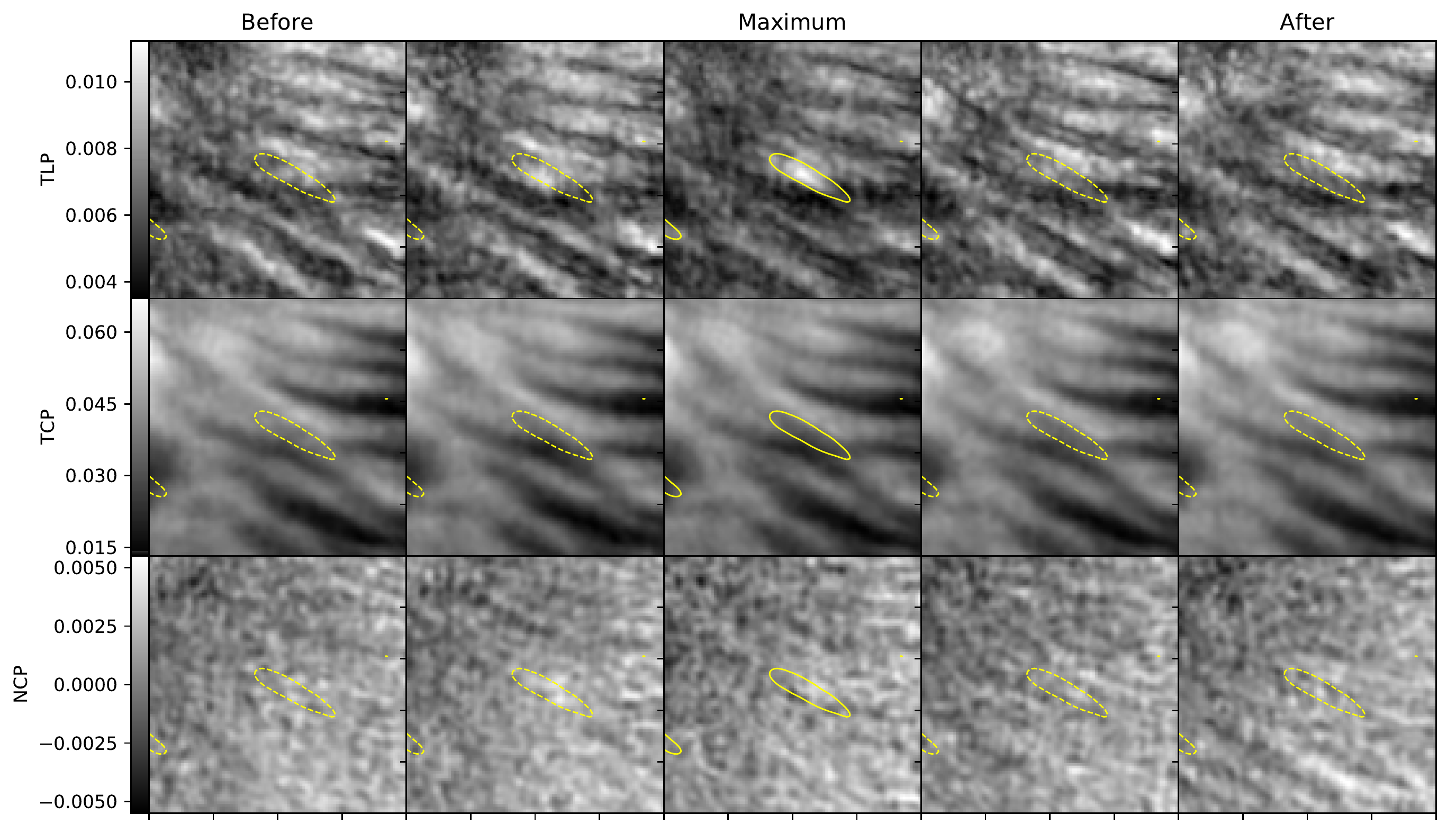}
  \includegraphics[height=0.345\hsize]{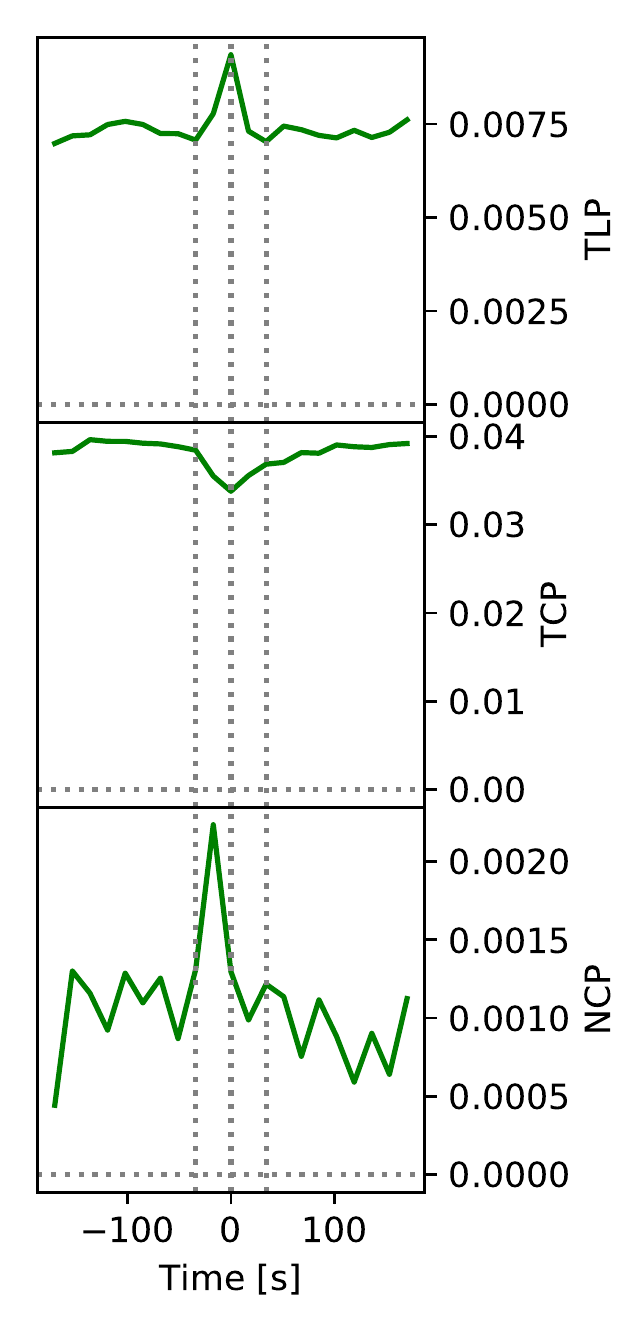}
~
   \includegraphics[width=0.67\hsize]{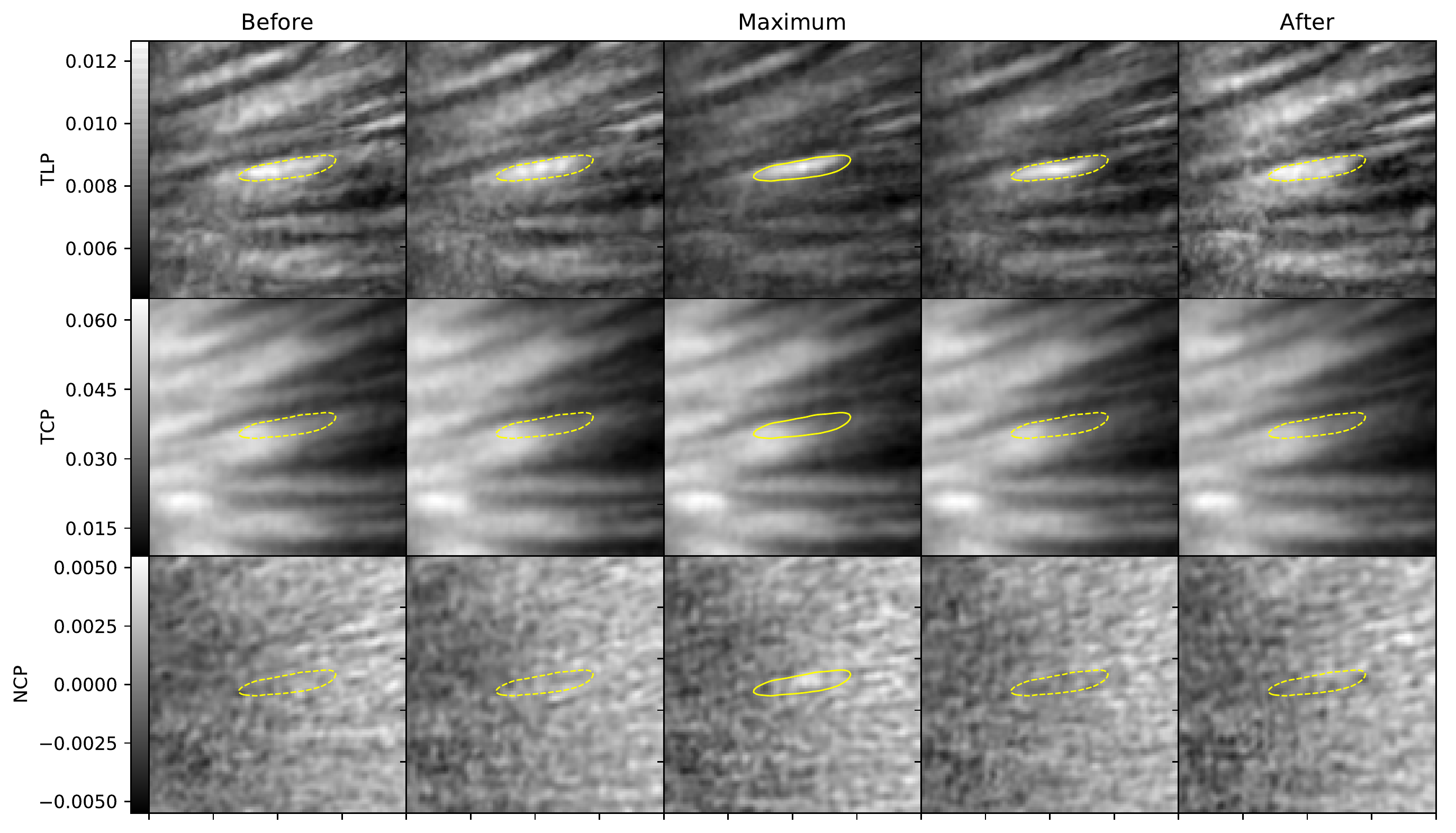}
   \includegraphics[height=0.345\hsize]{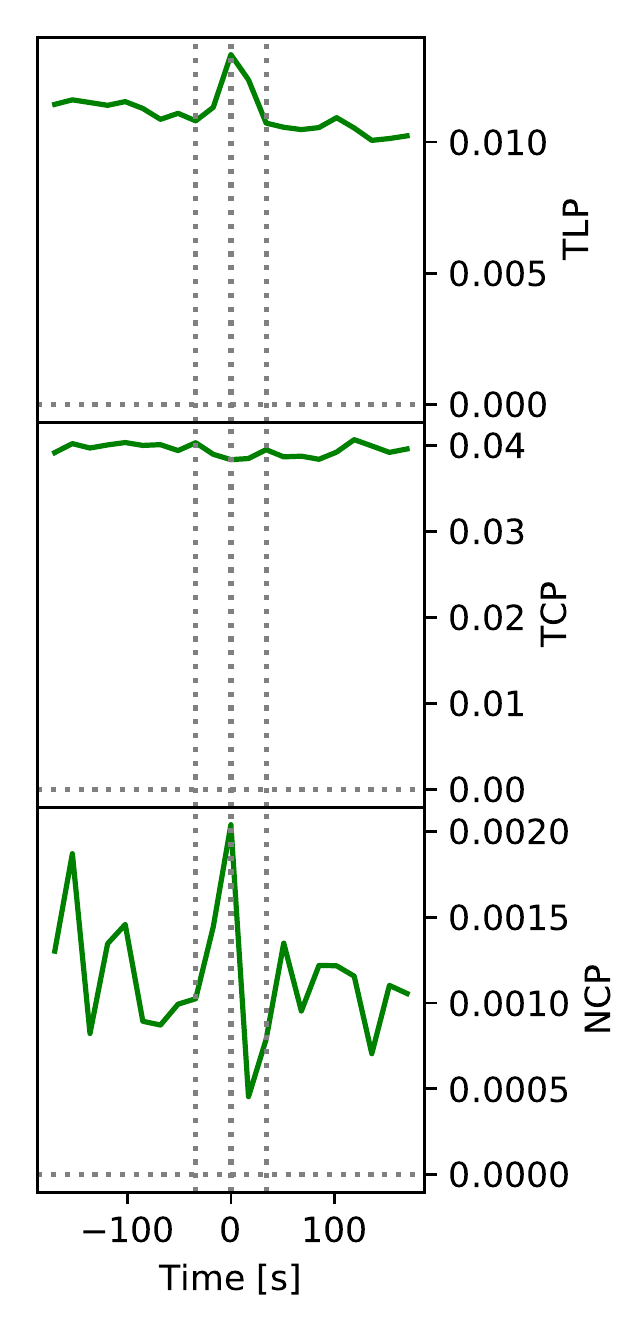}
~
   \includegraphics[width=0.7\hsize]{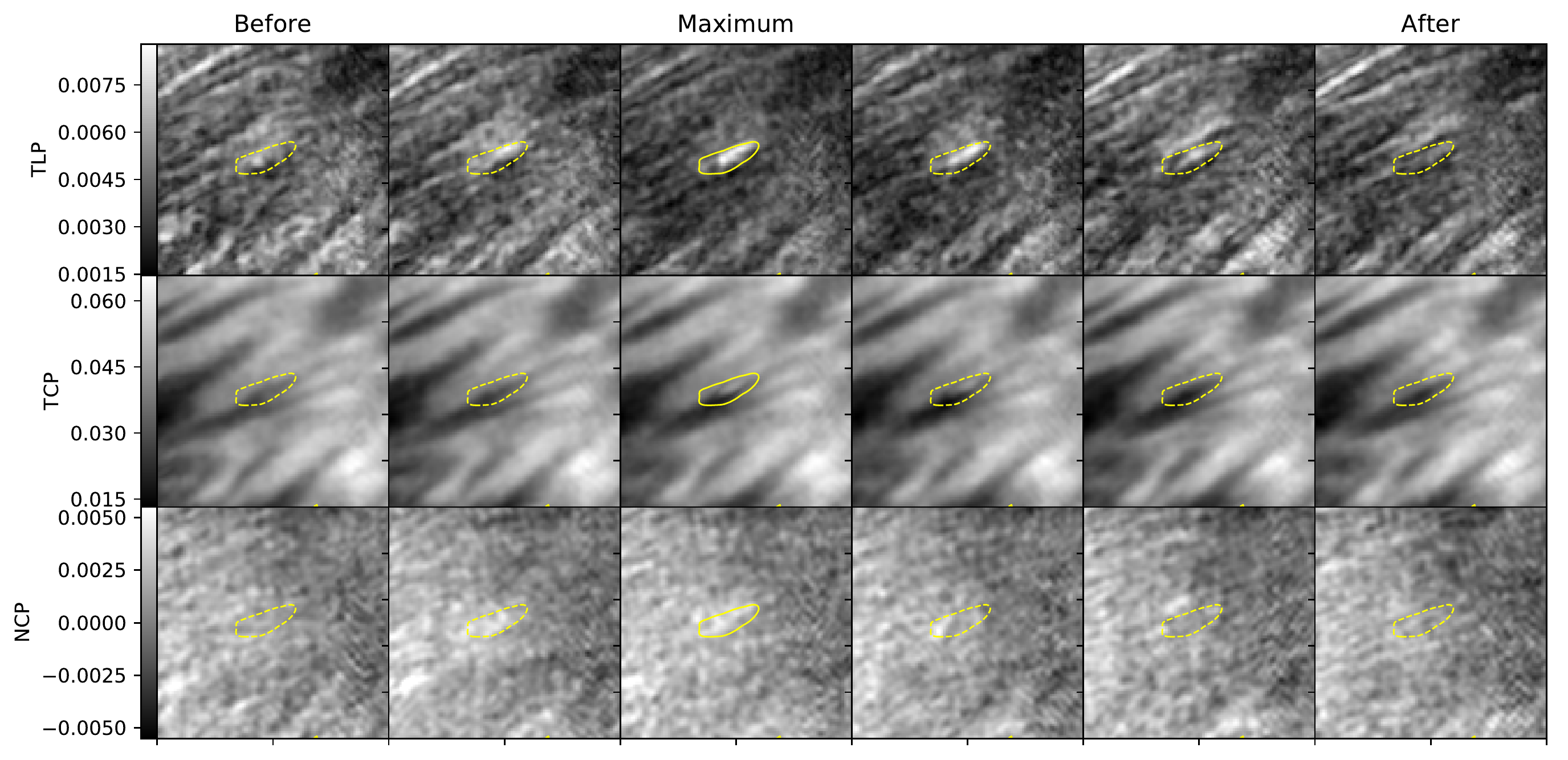}  
   \includegraphics[height=0.325\hsize]{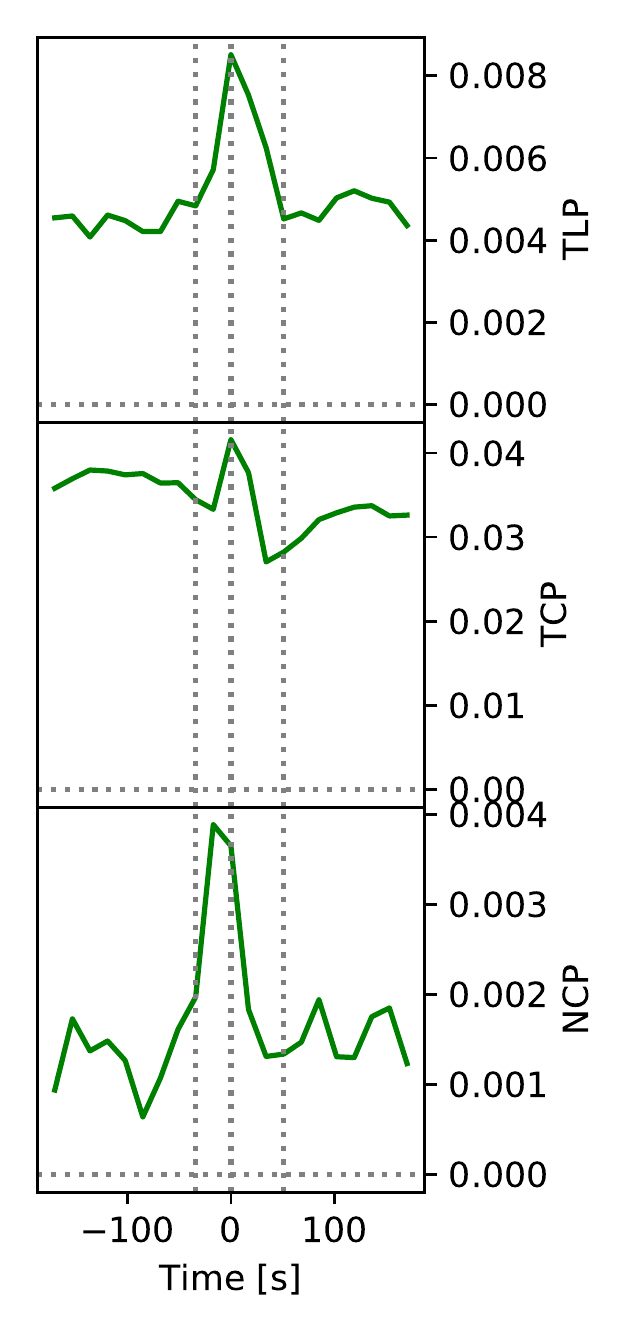}  
      \caption{Temporal evolution of three PMJs whose positions within the penumbra are indicated in Fig \ref{Fig:1}:  PMJ 1 (top), PMJ 2 (middle), and PMJ 3 (bottom). Right: for each case the panels show, from top to bottom, maps of total linear polarization (TLP), total circular polarization (TCP), and net circular polarization (NCP). Contours enclose the PMJ area. 
       Time runs from left to right with 17 s cadence. Left: Evolution of the average TLP, TCP, and NCP in the microjet area. The vertical dotted lines correspond to the frames `before', `maximum', and `after'.
              }
         \label{Fig:4tlp}
   \end{figure*}

The top panel of Figure \ref{Fig:4}  shows the temporal evolution of the average blue-wing intensity (black) and the average line-core intensity (gray) in the vicinity of  the maximum brightness pixel (MBP), i.e.  averaged over its eight closest neighbors. These pixels display brightness fluctuations before and after the PMJ, with amplitudes that are up to 4$\%$ of their mean 
intensity over the full time sequence (horizontal dotted lines). Moreover, about 200 s after the maximum brightness stage, these pixels present a generally enhanced brightness of around 8$\%$ compared with their average over the full time sequence.
However, the PMJ brightening stands well above such enhanced intensity,  the noise level and  other brightness fluctuations that might be caused by, for example,  Doppler shifts of the line, waves, bad seeing conditions, etc. Also, the PMJ occurs abruptly while other brightness fluctuations vary more smoothly, which allows for a precise identification of PMJs. Notably, the line-core intensity remains slightly enhanced after the PMJ disappears in the blue wing wavelengths.

Figure \ref{Fig:4} also shows the polarization signals of the Ca II 8542  $\r{A}$ line as observed in the MBP  during the three stages of interest (colored solid lines in the left panels) as well as the Stokes profiles at the MBP averaged over the complete time sequence (black dotted lines).  
The Stokes profiles observed during the stages `before' (green) and `after' (red) are remarkably similar to the temporally averaged profiles, but display slightly narrower intensity profiles as well as marginally brighter line cores.

At the maximum brightness stage (blue), all four Stokes profiles display enhanced signals. In particular, the intensity profile shows an emission in the blue wing, which peaks near $\Delta\lambda=-0.3 \r{A}$. The line core and red wing intensities are also enhanced, but their increase is considerably smaller.
The circular  and linear polarization profiles display larger signals at the maximum brightness stage, particularly in the blue wing. However, Stokes $V$ decreases toward the outer wings.
 The wavelength with maximum polarization signal in Stokes $Q$, $U$, and $V$ occurs around $\Delta\lambda=-0.15 \r{A}$.  Such enhanced polarization signals produced by the PMJ are also clearly distinguishable from the spectro-temporal plots (right panels). In particular, both Stokes $I$ and $V$  stand  well above the noise level, whilst enhanced Stokes $Q$ and $U$ signals appear on a noisier background but with amplitudes that are still above $3\sigma$ at the maximum brightness stage.

 In Figure \ref{Fig:4_new}, we show the Stokes profiles emerging from selected pixels along the PMJ, indicated by the blue symbols (from point A to B) in Fig. \ref{Fig:3}. 
 The first set of profiles, which emerged from a pixel located outside the PMJ contour (point A),  display regular shapes and are very similar to their temporally averaged profiles. The next set of profiles emerged from the PMJ region and show enhanced intensity as well as larger polarization signals in the blue wing wavelengths.  
 As we keep moving radially outwards, the blue-wing intensity enhancements gradually decrease and those in the line core wavelengths become more notorious, 
 reaching a point where the profiles only show an enhancement in the core but not  in the blue wing.  
 The profiles show more regular shapes outside the CC region (near point B).

  The radial shift 
between  
the PMJ and CC regions  is roughly 12 pixels or $\sim 500$ km, since our pixel size corresponds to $\sim 41$ km. It
 can be explained as the result of the three-dimensional structure of the magnetic field. 
 Due to the expansion with height of the penumbral magnetic field, a given field line will be observed slightly displaced radially outwards in the chromosphere with respect to the photosphere, which is what we detect in this case. The more inclined the field line, the larger the radial displacement with height. This suggests that the brightness enhancements detected in the line wing and the line core correspond to a perturbation or process that occurs on the same field lines at different heights in the atmosphere.
The dependence of the intensity enhancements on wavelength and radial position during the microjet might be the result of
  plasma heating occurring along a penumbral structure, from the upper photosphere to the low chromosphere.

   \begin{figure*}
   \centering
   
\includegraphics[width=\hsize]{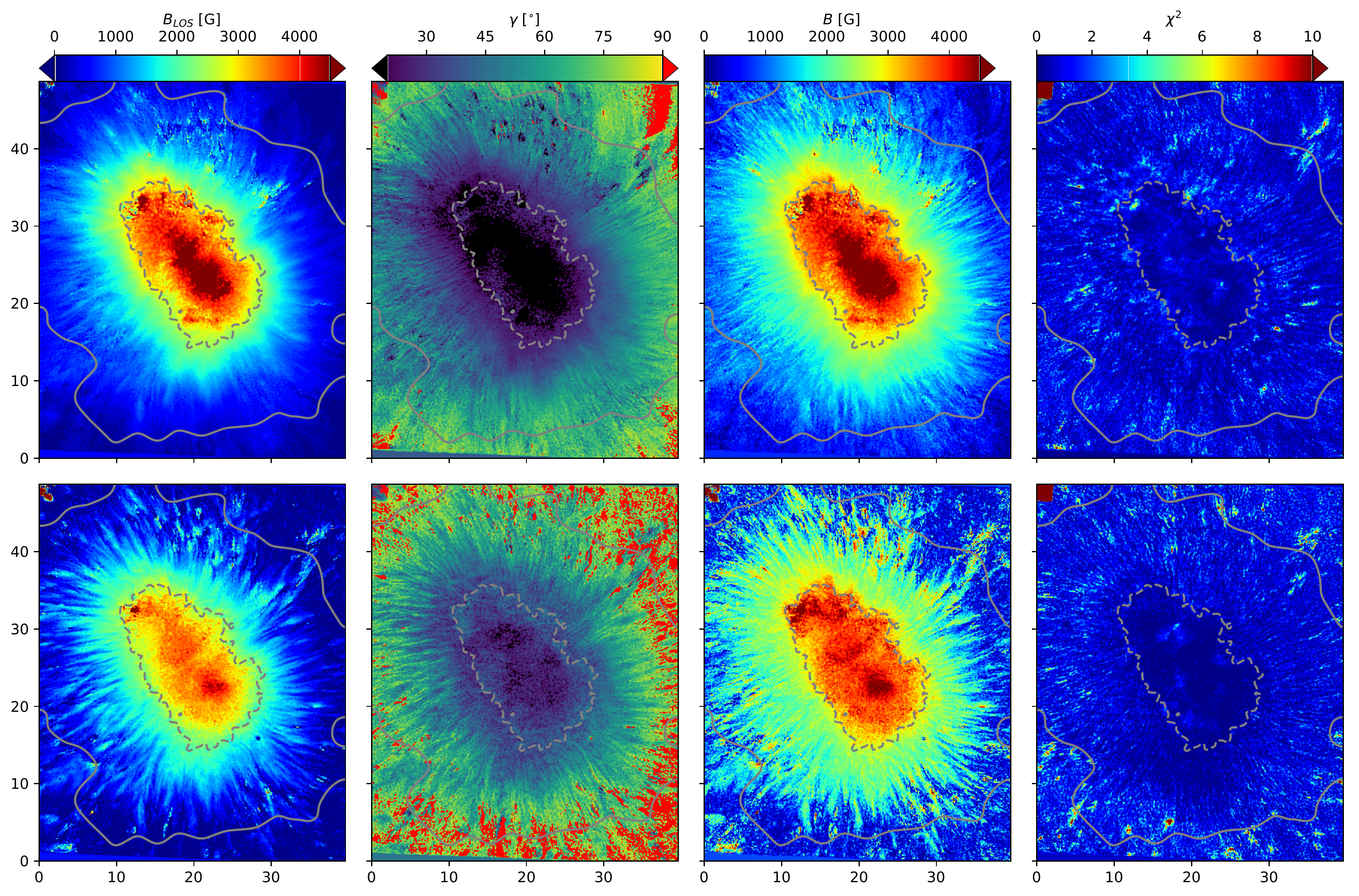}

      \caption{From left to right: maps of the longitudinal component of the magnetic field $B_{LOS}$, the field inclination $\gamma$ with respect to the LOS, the magnetic field strength $B$, and the merit function $\chi^2$ that result from the WFA applied to the line core wavelengths (top panels) and to the line wing wavelengths (bottom panels) for the entire sunspot as recorded at 08:13:22 UT. Red patches on the inclination maps are pixels where $\gamma>90^{\circ}$. 
      Smoothed gray contours delimit the inner (dashed) and outer (solid) penumbral boundaries.
              }
         \label{Fig:5}
   \end{figure*}

We also inspected the total linear polarization $TLP$, total circular polarization $TCP$, and net circular polarization $NCP$ in the PMJ environment, defined as:

\begin{equation} 
TLP=\int_{-0.6\r{A}}^{+0.6\r{A}} \sqrt{\frac{Q^2(\lambda)+U^2(\lambda)}{I^2_{QS}}} \,d\lambda,
\end{equation}

\begin{equation}
TCP=\int_{-0.6\r{A}}^{+0.6\r{A}} \frac{|V(\lambda)|}{I_{QS}} \,d\lambda,
\end{equation} 

\begin{equation}
NCP=\int_{-0.6\r{A}}^{+0.6\r{A}} \frac{V(\lambda)}{I_{QS}} \,d\lambda,
\end{equation}

 Figure \ref{Fig:4tlp} shows  the temporal evolution of these quantities for three PMJs. 
As in all 36 PMJs in our sample, the TLP maps (transverse magnetograms) show enhanced signals inside the PMJ area (yellow contours). These enhanced signals generally reach a maximum at the maximum brightness stage. Moreover, in many cases, such as in PMJ 2 (middle panels) and PMJ 3 (bottom panels), some patches of positive TLP signal can be detected since the stage `before' within the yellow contour lines. Such preexisting  features of enhanced linear polarization  are observed in more than $60\%$ of the cases, mainly inside the PMJ area but also in  regions adjacent to it, which might indicate the presence of an enhanced horizontal component of the magnetic field in the pre-PMJ phase.

The TCP signals display different behaviors among the 36 PMJs. For PMJ 1, as in most of the cases ($69\%$ of the total sample), 
the TCP undergoes a slight decrease at the maximum brightness stage, despite the fact that the Stokes $V$ profiles display enhanced signals in their inner wings. However, the outer wings display a notable signal reduction at `maximum' (see Fig. \ref{Fig:4} for PMJ 1) which explains why the TCP  decreases in these cases. 
In contrast, the TCP in PMJ 2 displays no clear changes. This is the behavior observed in 4 PMJs from our sample. 
Finally, as in PMJ 3, a total of 7 PMJs 
 displayed an increase inside the microjet region at `maximum' with respect to the stage `before'. 
This indicates that there are different types of evolution in the longitudinal component of the magnetic field for each case, which will be further investigated in the following sections. 

The NCP is a measure of how asymmetric the Stokes V profiles are. It can be used as a proxy for gradients 
along the LOS of the magnetic field and of the LOS velocities. We found an increase of the NCP inside the PMJ regions in all 36 cases.  
The NCP values  are generally positive inside the PMJs and reach a maximum value at the maximum brightness stage, such as in PMJ 2. However, there are also a few cases in which the maximum NCP occurs one frame before or after the maximum brightness phase, such as in PMJs 1 and 3.

\subsection{Magnetic field configuration}
To investigate how the magnetic field configuration evolves during PMJs, we consider the weak-field approximation (WFA). 

The WFA can be applied when the thermal broadening $\Delta\lambda_{D}$ of the line is much larger than the Zeeman splitting $\Delta\lambda_{B}$ 
\citep[e. g., ][]{Deglinnocenti2004}.
Under the assumption of a constant and sufficiently weak magnetic field in the region of the solar atmosphere where the spectral line is formed, 
the radiative transfer equation (RTE) can be solved analytically by considering the magnetic field to be a perturbation of the zero-field case. In such case, the analytical solution is the WFA 
and the first order perturbation (in which Stokes $Q$ and $U$ are zero) yields the following expression:

\begin{equation}
V=-\Delta\lambda_{B}fg\cos(\gamma)\frac{\partial I}{\partial \lambda}=\alpha f B_{LOS}\frac{\partial I}{\partial \lambda},
\label{eq:1}
\end{equation}

\noindent where $g$ is the effective Land\'e factor of the line, $f$ is the filling factor which represents the fraction of the resolution element covered by the magnetic field, and $\gamma$ is the inclination of the magnetic field vector with respect to the LOS.

 Expression \ref{eq:1} implies that Stokes $V$ is proportional to the partial derivative of Stokes $I$ with respect to wavelength, and that the factor of proportionality depends on the longitudinal component of the magnetic field, i. e., on $B_{LOS}=B\cos(\gamma)$.
 Therefore, $B_{LOS}$ can be estimated by  applying a linear least-squares fit to Eq. \ref{eq:1}, which yields:
 
 \begin{equation}
B_{LOS}=\frac{\sum_i \frac{\partial I (\lambda_i)}{\partial \lambda_i}V(\lambda_i)}{\alpha \sum_i (\frac{\partial I (\lambda_i)}{\partial \lambda_i})^2},
\label{eq:2}
\end{equation}

\noindent with $\alpha=-4.67\times10^{-13}g\lambda_{0}^2$ \citep[e. g., ][]{Martinez2009}. Note that in Eq. \ref{eq:2} we have set $f=1$ since we are in a sunspot.

Similarly, the expressions for the linear polarization (Stokes $Q$ and $U$) can be derived by perturbing the magnetic field to second order in the RTE \citep[e. g., ][]{Deglinnocenti2004}. Such expressions involve the first and second derivatives of the intensity profile with respect to wavelength. Their combination with Eq. \ref{eq:1}, along with the application of a least-squares minimization procedure, lead to the following relation for the inclination angle $\gamma$:

 \begin{equation}
\tan^2(\gamma)=\frac{4}{3}\frac{g^2}{G}\frac{\sum_i |\lambda_i|  |L_i|  |V_i|^2  |\frac{\partial I}{\partial\lambda}|_i}{\sum_i |V_i|^4},
\label{eq:3}
\end{equation}

\noindent where $L_i=\sqrt(Q_i^2+U_i^2)$  and $G$ is the Land\'e factor for the transverse magnetic field \citep[see e. g., ][]{Hammar2014}.  Eq.  \ref{eq:3} gives only the modulus of the inclination. The polarity of the field is determined by Eq. \ref{eq:2}. Thus,  the value of $\gamma$ (in degrees) can be defined as follows:

 \begin{equation}
\gamma=\begin{cases}
               |\gamma| & \text{if $B_{LOS}>0$}\\
               180^{\circ}-|\gamma| & \text{if $B_{LOS}<0$}
            \end{cases}
\label{eq:3*}
\end{equation}

The validity of the WFA for Ca II 8542 $\r{A}$ ($g\approx1.10$ and $G\approx1.18$) has been studied by, among others,  \citet{Hammar2014} and \citet{Centeno2018}, who concluded that the WFA allows for an efficient and rapid inference of the chromospheric magnetic field vector from Ca II 8542 observations. 
The modest Land\'e factor and broad shape of this line guarantee the validity of the
WFA even when the magnetic field is relatively strong.

\begin{figure*}
   \centering
   \includegraphics[width=0.55\hsize]{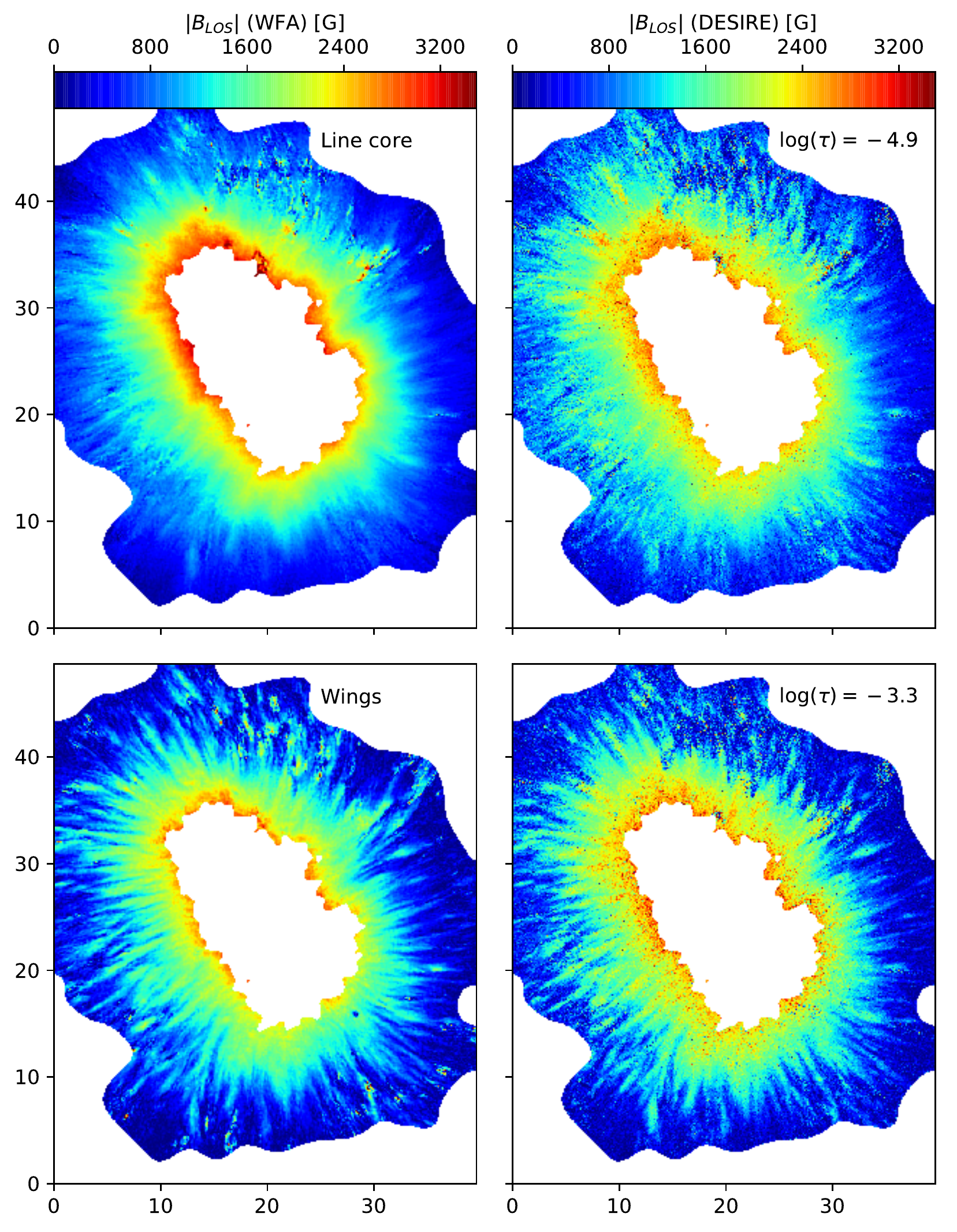}
   \includegraphics[width=0.42\hsize]{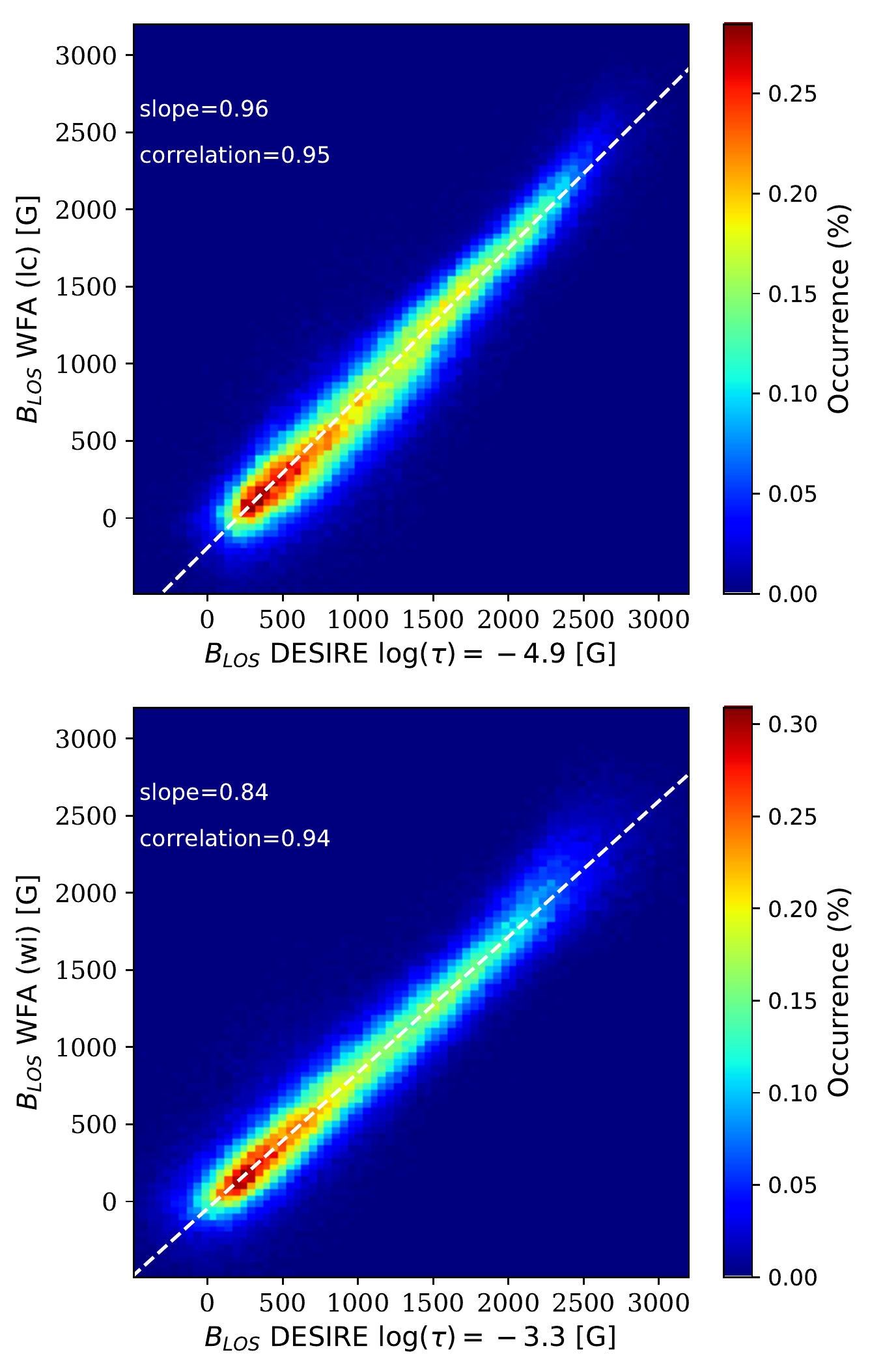}

      \caption{Comparison between the $B_{LOS}$ values obtained in the sunspot penumbra (same timestep as in Fig. \ref{Fig:5}) using the WFA and DeSIRe inversions. The left column shows the results from the WFA applied to the line core (top) and to the wings (bottom). The central column displays the inversion result at $\log(\tau)=-4.9$ (top) and  $\log(\tau)=-3.3$ (bottom). The right column shows density plots comparing the values from the two methods. We consider only penumbral pixels where $\chi^2<3$ is obtained with both methods. White dashed lines are regression lines.
}
         \label{Fig:5*}
   \end{figure*}

\citet{Hammar2014} and \citet{delaCruz2013} estimated that, for a chromospheric model in which the temperature is 4500 K and the micro-turbulent velocity is $3$ km s$^{-1}$, the Ca II 8542 $\r{A}$ line is in the weak-field regime for chromospheric fields weaker than 2500 G.
However, the Doppler width and consequently the upper limit of the magnetic field both increase at higher temperatures. 
 Therefore, the wings of the line can also be treated under the WFA for the stronger magnetic fields of the photosphere. 

Specifically, \citet{Centeno2018} modeled this  line considering scenarios in which the field strength decreased with height and found that the WFA applied to the wings can be used to probe magnetic fields around $\log(\tau)=-1.4$ with remarkable accuracy for a range of field strengths of up to 4000 G.
She also found that when evaluated  in the line core ($\lambda_0 \pm250$ m$\r{A}$, probing an average optical depth of $\log(\tau)=-5.3$)  the WFA errors are only about $10\%$ of the real value for field strengths up to 1200 G.

In Figure \ref{Fig:5}, we display the result of the WFA applied to the entire sunspot as observed at 08:13:22 UT. The figure shows  maps of the longitudinal component of the magnetic field $B_{LOS}$, the inclination angle  $\gamma$, the magnetic field strength $B=B_{LOS}/\cos(\gamma)$, and the merit function $\chi^2$ which represents the sum of the squared differences between the observed Stokes $V$ profiles and the  values resulting from the WFA, normalized to the noise level and number of wavelength points.

The WFA has been applied separately to the core of the 
 line  using the wavelength interval between $\pm0.3$ $\r{A}$ (upper panels in Fig. \ref{Fig:5}) and to the wings using the wavelength points at $\Delta\lambda=[-0.6,-0.45,0.45,0.6]  $ $\r{A}$ (lower panels). In general, Eq. \ref{eq:1} can reproduce the shape of the observed Stokes $V$ profiles, with most fits inside the spot having  $\chi^2<3$ in both  the wing and the line core regimes.
However, the WFA returns magnetic fields larger than 2500 G in the umbral areas. 
These values are above  the  field strength limit for the validity of the WFA in the chromosphere inferred by \citet{Hammar2014} and, therefore, are not reliable. 

In the rest of the sunspot, the well-known penumbral fine structure of the photospheric magnetic field  is clearly discernible in the bottom maps, whilst a more extended magnetic canopy can be seen in the upper maps, as expected for the chromospheric layers.
The general appearance of the penumbra is reasonable, with the magnetic field decreasing with height and the field inclination becoming more vertical in the chromosphere at a fixed radial distance, as expected. The weakening of the field with height is not dramatic since the compared layers are not so distant from each other.

Also, we must bear in mind that the spectral line  was observed in a relatively small wavelength range in order to get a sufficiently fast temporal cadence, which means that we have sampled only a small part of the wings and therefore, the photospheric information that can be retrieved from such a wavelength range is essentially limited to the upper photosphere.

\subsection{Inversions} \label{Sec:inv}

We compare the results obtained from the WFA in the sunspot (same FOV as displayed in Fig. \ref{Fig:5}) with inversions of the Stokes profiles carried out with the DeSIRe (Departure coefficient Stokes Inversion based on Response functions) code\footnote{Code description by Orozco et al. in prep.}. DeSIRe allows for the inversion of spectral lines that form under non-LTE conditions by combining the RH forward synthesis code  \citep{Uitenbroek2001} and the SIR inversion code \citep{Ruiz1992}.

We invert the four Stokes parameters of the Ca II 8542 $\r{A}$ line to infer the atmospheric properties within the formation region of the line. 
The inversion  setup considers a simple one-component model atmosphere with typical values of the magnetic field strength $B$ and  inclination $\gamma$, the LOS velocity $v_{LOS}$ and the temperature stratification between the photosphere and upper chromosphere \citep[FALF model, ][]{Fontela1993}. We use 4 cycles with 5, 6, 7, and 9 equidistant nodes in temperature for each cycle; 5 nodes in  micro-turbulence and  LOS velocity; 3 nodes in the magnetic field strength, inclination and azimuth angles; and 1 node in macro-turbulent velocity.

Similarly to what \citet{Bellot2006} did for bisector velocities, we have determined  the  optical depth at which the inversion results are best comparable with the results from the WFA applied to the line core wavelengths, which have a maximum sensitivity in the range $\log(\tau)=[-4.5,-5.5]$. The wings of the line are sensitive to a range of optical depths $\log(\tau)=[0,-4]$ \citep[][]{Quintero2016}. We restrict the comparison to those pixels where the inversions perform relatively well (with $\chi^2<3$, which represent almost $65\%$ of the pixels in the penumbra).
Figure \ref{Fig:5*} displays such a comparison. 
In the line core (upper panels), the WFA has a maximum correlation with the inversions at $\log(\tau)=-4.9$. In the wings (lower panels), the maximum correlation is obtained at $\log(\tau)=-3.3$.

 In spite of the considerably noisier maps obtained from the inversions, the solutions are highly consistent with the WFA results in both layers. The general appearance of the penumbra and most of the prominent penumbral structures can be reproduced with both methods with a large correlation between them. This is reflected in the  density plots on the right column of the figure, which show a strong correlation between the WFA in the line core and the inversion results at $\log(\tau)=-4.9$ (low chromosphere), with a slope of 0.96 and a correlation coefficient of 0.95.
  Likewise, there is a good correlation between the WFA in the wings and the inversion output at $\log(\tau)=-3.3$ (upper photosphere), with a correlation coefficient of 0.94.

 Therefore, the inversions provide a solution that is consistent with the WFA results. However,  given that we have scanned the line at only a few wavelength points  across a  small wavelength range in order to maximize the temporal resolution of the data, we consider that the spectral information is not sufficient to perform  inversions reliably. The WFA is a more robust method in this case, so from now on we use the WFA  to investigate the magnetic properties of our PMJs.

\section{Results}

In this section, we describe the temporal evolution of the magnetic field inside and around PMJs, as inferred with the WFA in the upper photosphere and low chromosphere.
We first concentrate on three different cases, and afterwards we classify the evolution of all 36 PMJs based on how the field behaves during the maximum brightness stage with respect to its configuration during the frame `before'. Some statistics are also discussed.

\begin{figure}
   \centering

\includegraphics[width=\hsize]{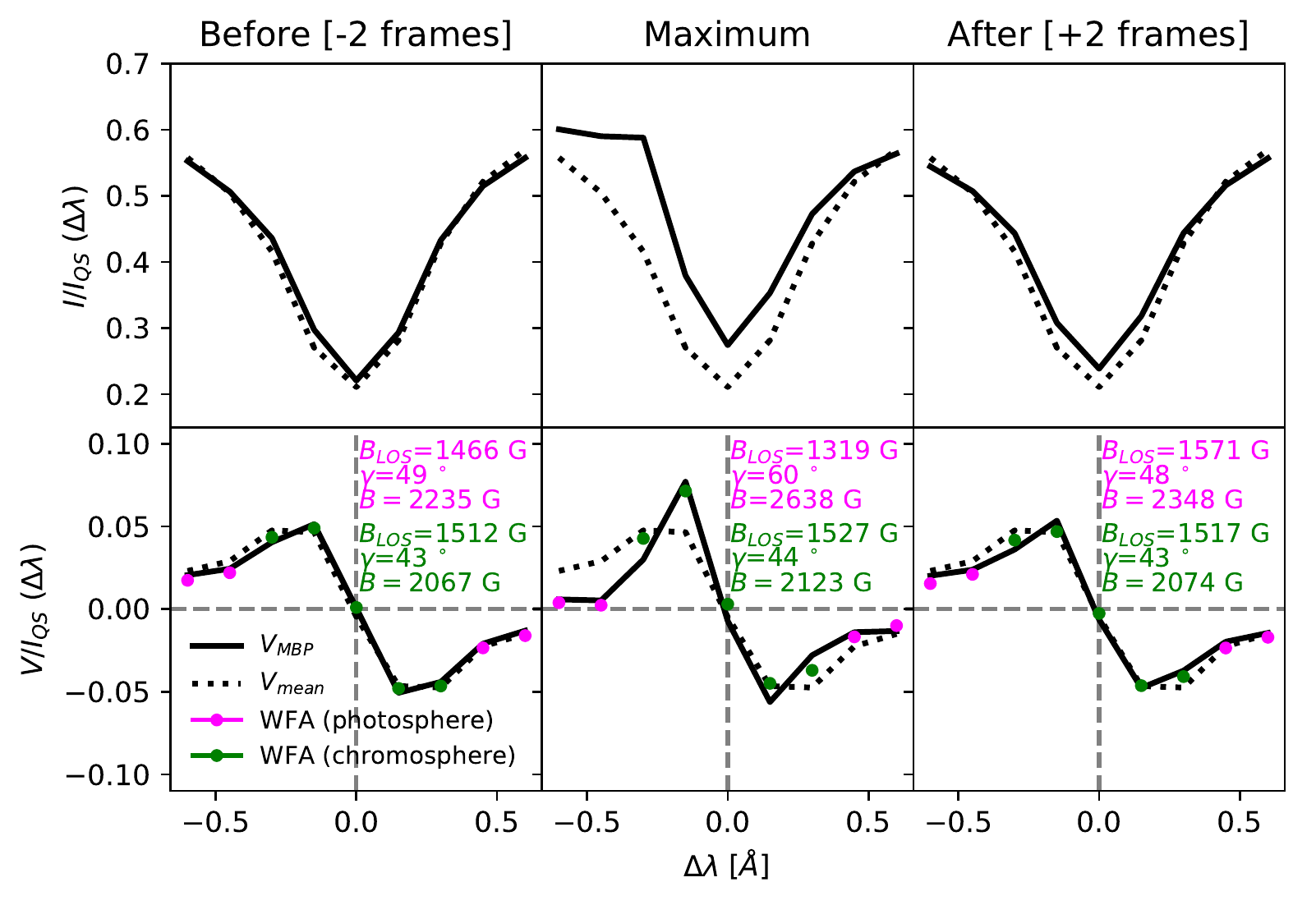}

      \caption{Observed Stokes $I$ and $V$ profiles (solid lines) and temporally averaged profiles (dotted lines) at the MBP in PMJ 1 during three different temporal stages. 
The labels on each panel indicate the $B_{LOS}$,  $\gamma$ and $B$ values inferred from Eqs. \ref{eq:2} and  \ref{eq:3}, evaluated in the line wings (magenta) and the line core (green).  The profiles resulting from the WFA fits are shown with magenta and green markers for the wings and the line core, respectively. 
}
         \label{Fig:6}
   \end{figure}
   
    \begin{figure}
   \centering
   
   \includegraphics[width=\hsize]{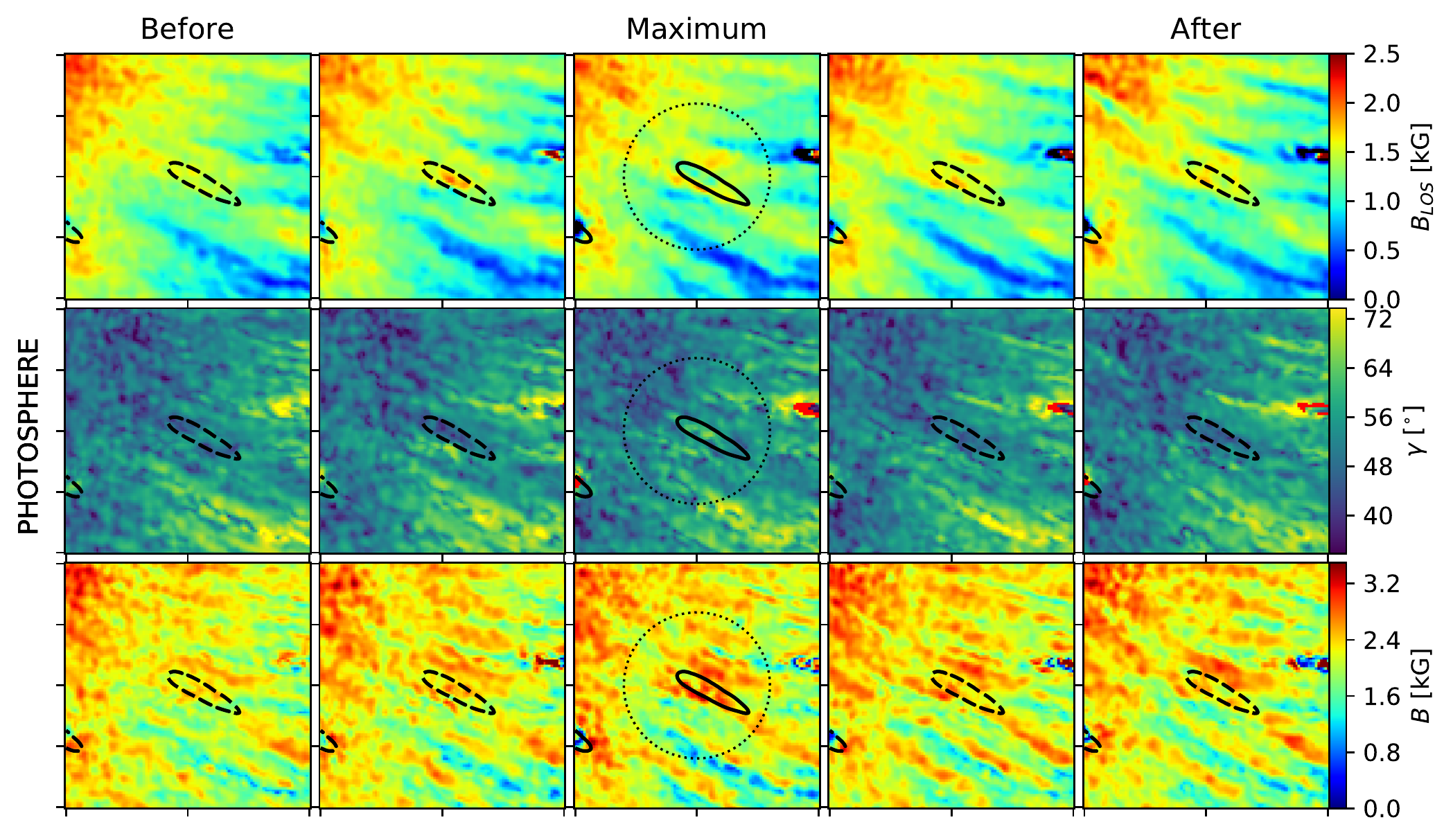}
   ~
   \includegraphics[width=\hsize]{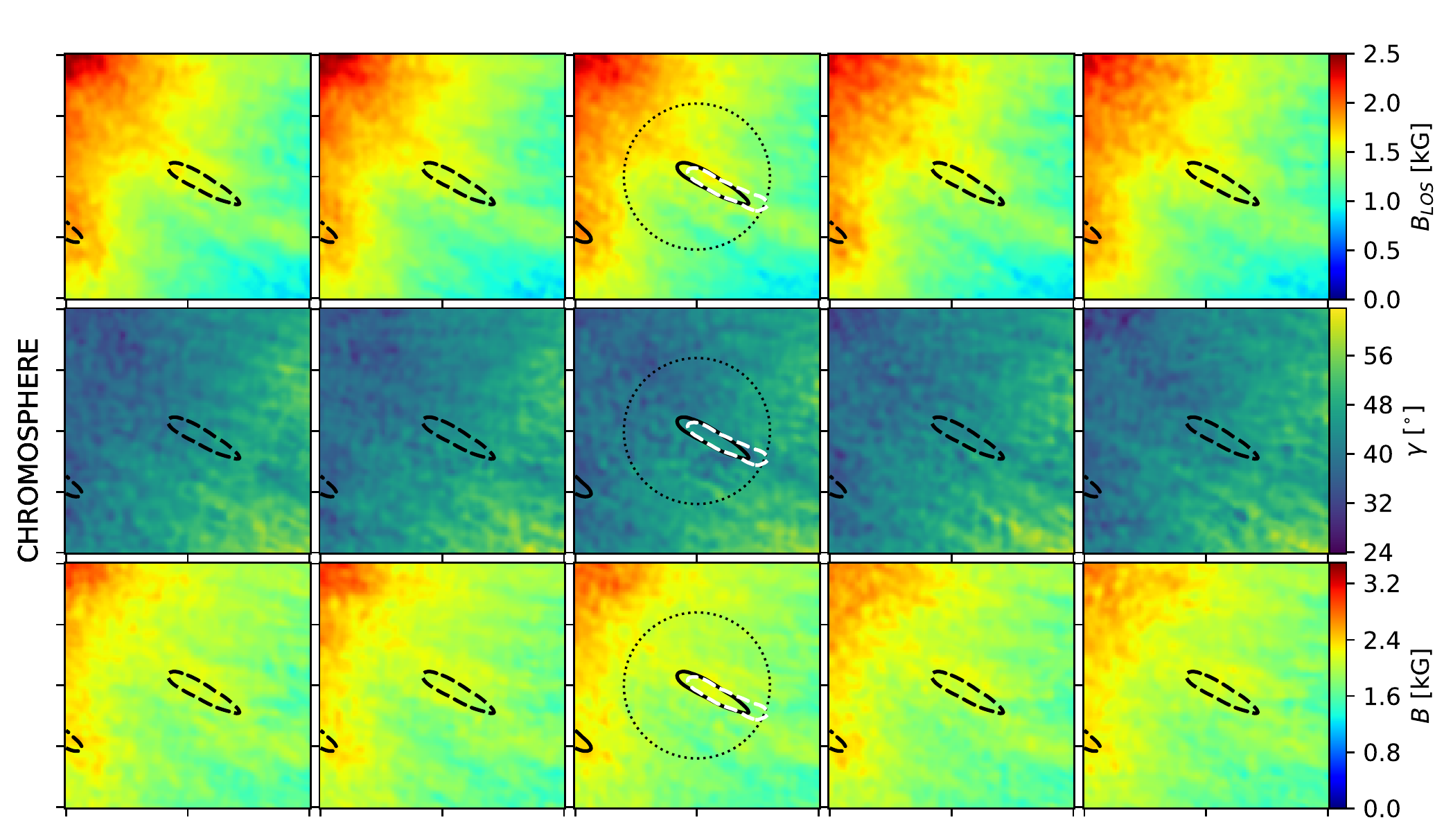}
   
      \caption{Maps of  $B_{LOS}$, $\gamma$, and $B$ during  the evolution of PMJ 1, computed by applying the WFA to the wings (rows 1-3) and to the line core (rows 4-6). The red patches in the inclination maps show regions where $\gamma>90^{\circ}$. 
      The black contours enclose the PMJ area. 
      Dotted circles  on the maps at `maximum'  define an arbitrarily chosen surrounding of the PMJ,  centered on the MBP with a radius of 1.5$\arcsec$. 
       The columns show consecutive frames separated by 17 s. The white contours enclose the CC region. 
       The boxes cover a subfield of $5\arcsec \times 5\arcsec$.}
         \label{Fig:7}
   \end{figure}

  \begin{figure}
   \centering
   
   \includegraphics[width=\hsize]{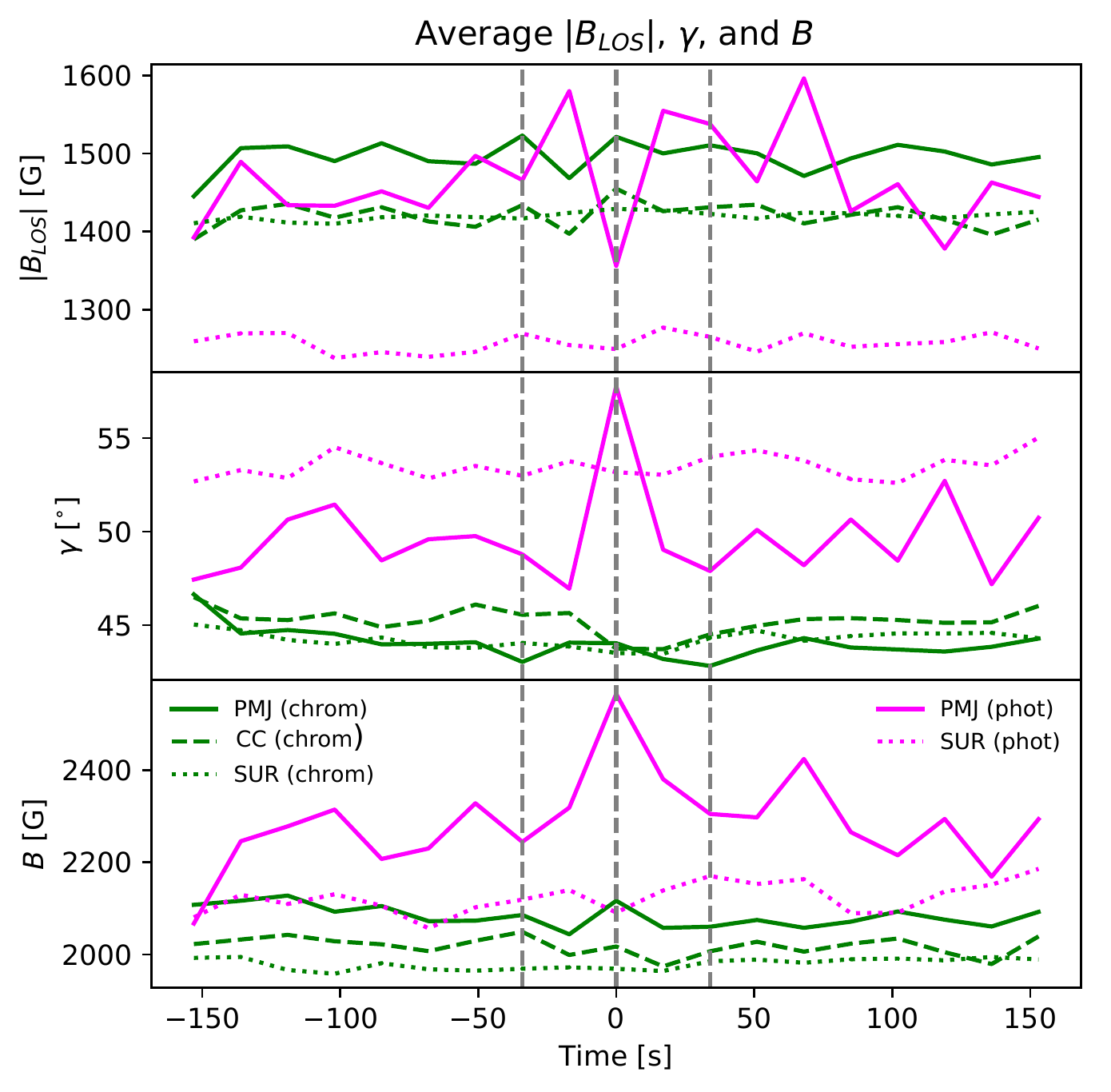}

      \caption{Temporal evolution of $B_{LOS}$, $\gamma$, and $B$ averaged within PMJ 1 (solid lines), the CC region (dashed lines), and in the surroundings along the dotted circles drawn in Fig. \ref{Fig:7} (dotted lines) in the lower chromosphere (green) and upper photosphere (magenta). 
       The vertical dashed lines mark the stages `before', `maximum', and `after'. }
         \label{Fig:7*}
   \end{figure}

   \subsection{Case 1}
   
   Figure \ref{Fig:6} shows the Stokes $I$ and $V$ profiles emerging from the MBP  of PMJ 1  at the times `before', `maximum', and `after', as well as the result of  applying the WFA to  the line wings (magenta markers) and  to the line core (green markers).
The shapes of the observed Stokes $V$ profiles can be nicely reproduced under the WFA, with  $\chi^2<2$  in both wavelength ranges.

As indicated in the figure (magenta labels),  there is a clear response of the photospheric field to 
the PMJ: the longitudinal field undergoes a reduction of nearly 150 G, which is accompanied by an increase of  the inclination of around 10$^{\circ}$, and by an increase of the total field strength of about 400 G during the `maximum’. The configuration of the photospheric field almost recovers its initial state after the PMJ disappears.

 The changes of the magnetic field in the lower chromosphere are substantially weaker than in the photosphere for the MBP (green labels),  with $B_{LOS}$, $\gamma$, and $B$ displaying changes of 15 G, $1^{\circ}$, and 50 G respectively, between the stages `before' and `maximum'. 
 The magnetic field configuration at the stage `after' is remarkably similar to the stage `before'.
The small magnetic field variations inferred for the low chromosphere are comparable to the uncertainties associated with the WFA, so they cannot be unambiguously related to the PMJ.
What is clear is that the  enhanced circular polarization signal observed at $\Delta\lambda=-0.15\r{A}$  is mainly a consequence of the emission peak in the blue wing of the intensity profile, given that the shape of Stokes $V$ follows the shape of the derivative of Stokes $I$ scaled to a $B_{LOS}$ of the order of 1.5 kG in the line core wavelengths.

 In Figure \ref{Fig:7}, we display maps of the magnetic field evolution in the upper photosphere (rows 1-3) and the low chromosphere (rows 4-6) during PMJ 1. 
 The entire PMJ region (delimited by elongated black contours) displays magnetic field changes  that are qualitatively similar to those observed at the MBP during the three stages of interest, i. e., the photospheric field becomes stronger and more inclined during the maximum brightness stage, when it displays the largest changes. 
This can be clearly observed in Figure \ref{Fig:7*}, where the temporal evolution of the average photospheric magnetic field inside the PMJ region is shown with magenta solid lines.

With an average magnetic field inclination of $\sim 50^{\circ}$ in the photosphere and a height difference of $\sim$500 km between the photosphere and the low chromosphere, the projection of the CCs would be about 15 pixels, in agreement with the observed shifts. 
The plots of Fig. \ref{Fig:7*} reveal that there are not clear changes occurring in the chromospheric field either along the LOS or along the magnetic field lines in this case. The field inside the PMJ and the CC regions 
 show a rather steady evolution (solid and dashed green lines, respectively).

To investigate how the magnetic field behaves in the surroundings during PMJ 1, we averaged the magnetic field parameters along the circles displayed in Fig. \ref{Fig:7}. The dotted lines in Fig. \ref{Fig:7*} show the temporal evolution of the field which, 
as expected, 
does not undergo any evident change that can be clearly related to the PMJ occurrence, neither in the photosphere nor in the chromosphere.  There are very small fluctuations  during the 300 s temporal range of the plots, which might be due to seeing or instrumental effects, and even  minor changes occurring in the penumbral field. 
In the photosphere, the fluctuations of the field inside the PMJ region are slightly larger than those in the surroundings. Nonetheless, the changes observed at `maximum' in the PMJ region clearly stand out, which means that such changes are not systematic variations.

\subsection {Case 2}

\begin{figure}
   \centering

   \includegraphics[width=\hsize]{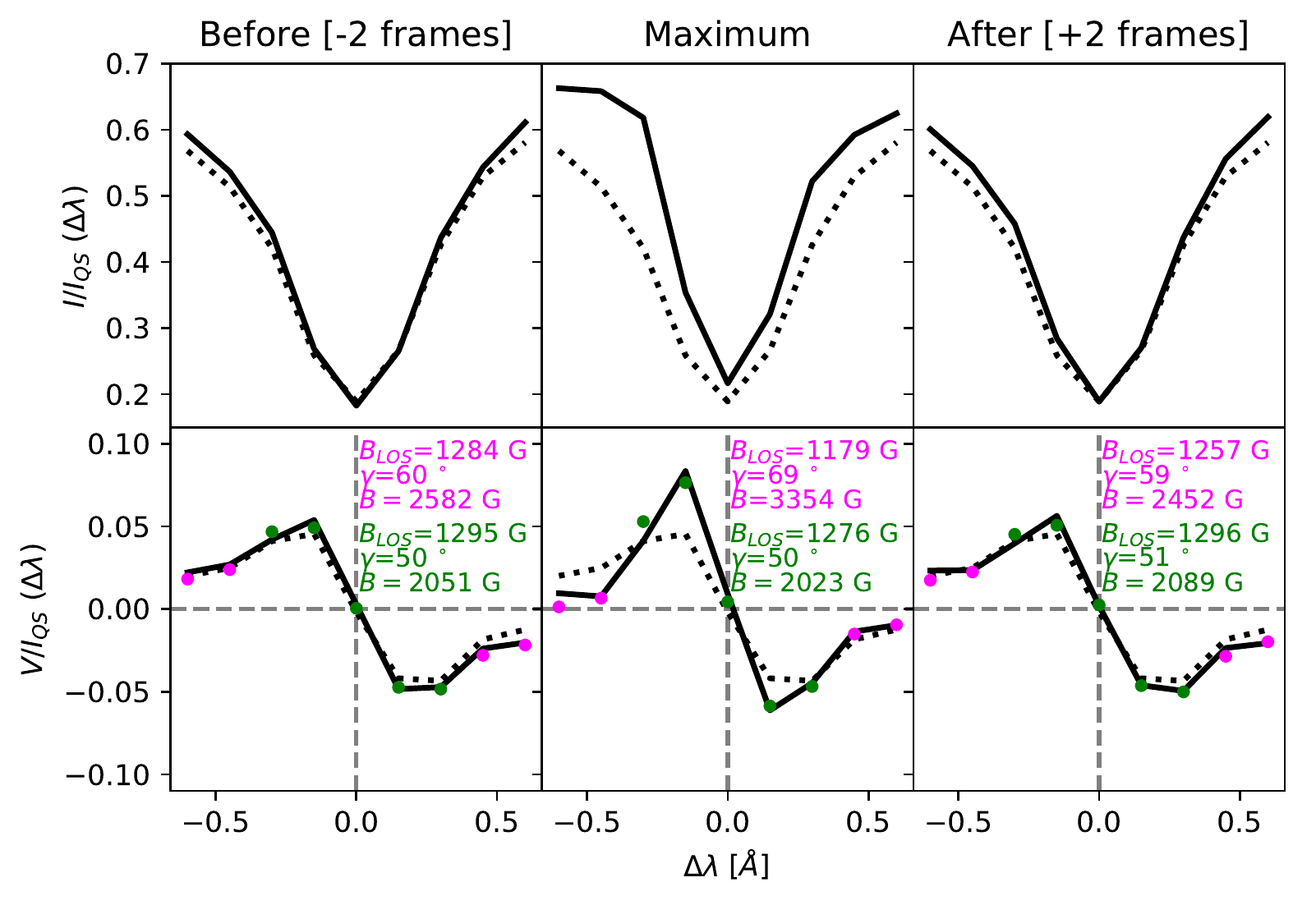}
      \caption{Observed Stokes $I$ and $V$ profiles  and results of the WFA at the MBP in PMJ 2. 
      Same format as  Fig. \ref{Fig:6}.
}
         \label{Fig:11}
   \end{figure}

\begin{figure}
   \centering
   
   \includegraphics[width=\hsize]{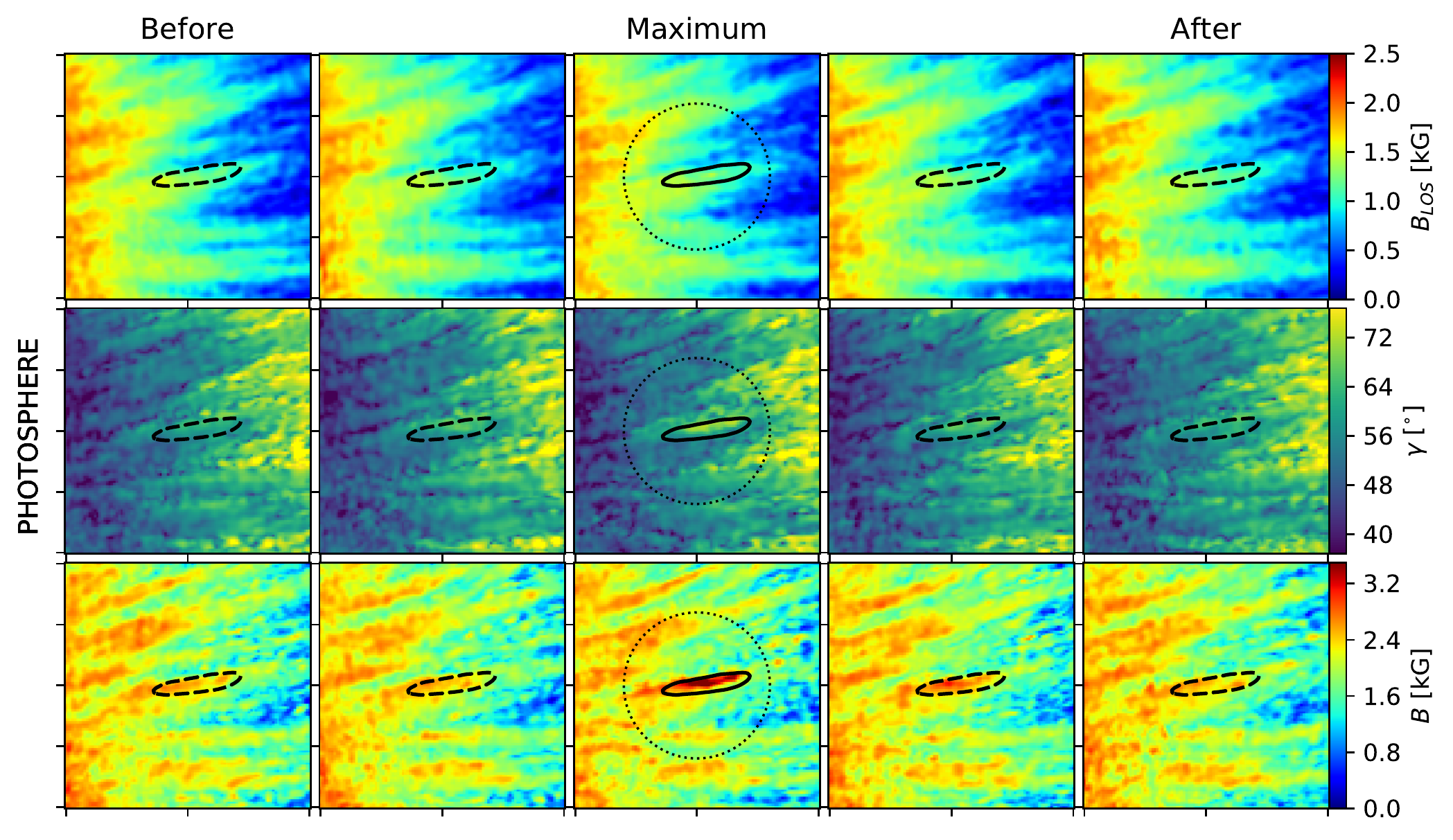}
   \includegraphics[width=\hsize]{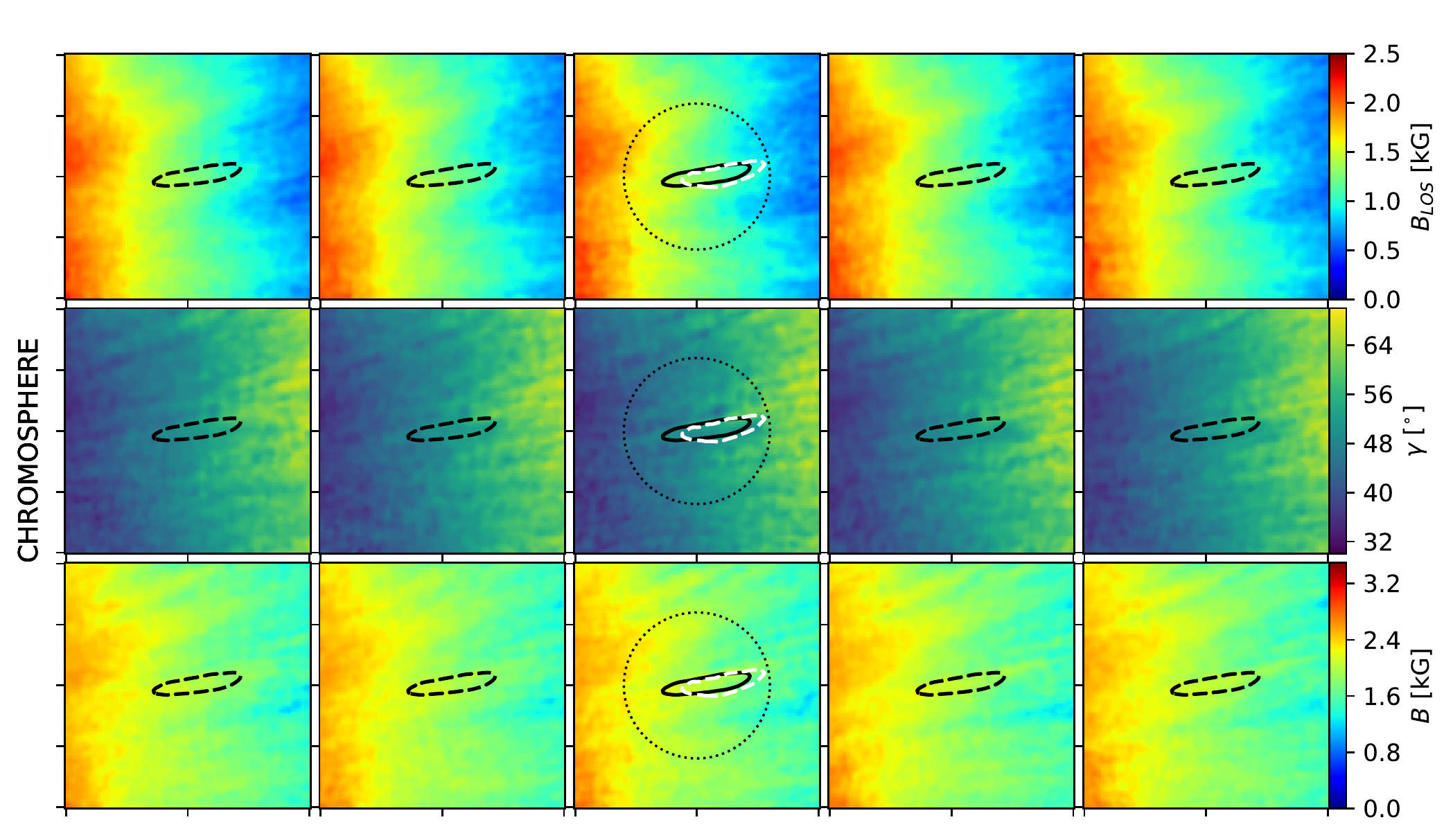}

      \caption{Maps of  $B_{LOS}$, $\gamma$, and $B$ during the different stages of evolution of PMJ 2. Same format as  Fig. \ref{Fig:7}. }
         \label{Fig:12}
   \end{figure}

Figures \ref{Fig:11} and \ref{Fig:12} show the temporal evolution of PMJ 2, whose brightening is observed during three consecutive frames in the limb-side penumbra (green marker in Fig. \ref{Fig:1}). It occurs near the end of an inner penumbral filament.
 As shown in Fig. \ref{Fig:11} for the MBP, the emergent Stokes $I$ profiles are remarkably similar during the stages `before' and `after'. In contrast, the entire profile displays an increase of the intensity during the `maximum' stage, and such brightening is notably larger in the blue wing wavelengths, similar to the MBP of PMJ 1.

  \begin{figure}
   \centering
   
   \includegraphics[width=\hsize]{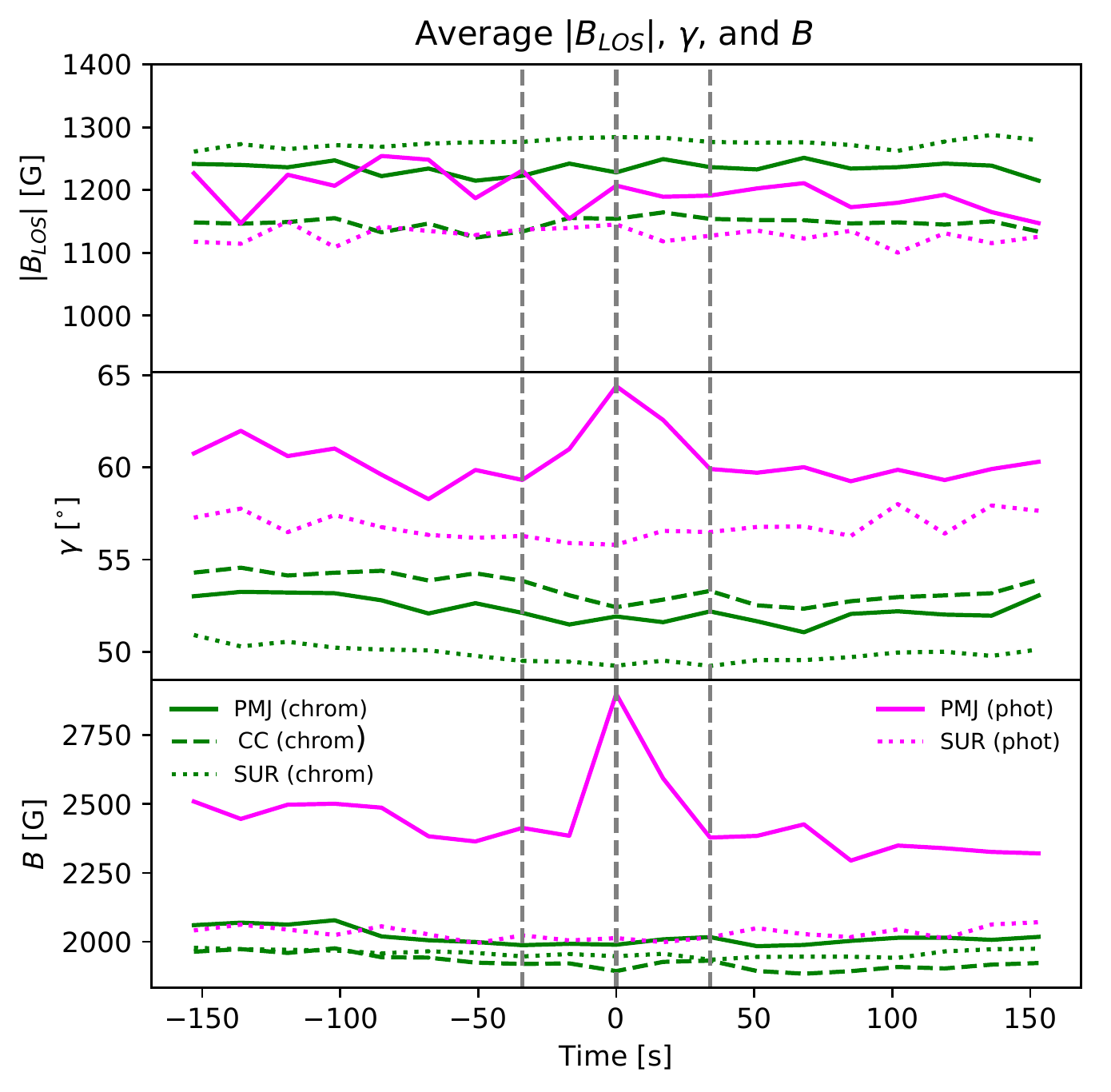}
  
      \caption{Temporal evolution of $B_{LOS}$, $\gamma$, and $B$ inside and in the surroundings of PMJ 2. Same format as Fig. \ref{Fig:7*}.}
         \label{Fig:12*}
   \end{figure}

 The results of the WFA  indicate that the photospheric longitudinal field decreases almost 100 G between the frames `before' and `maximum', whilst the field inclination increases by nearly 10$^{\circ}$ and the total field strength increases more than 700 G between the two stages.
Afterwards, the photospheric magnetic  configuration in the stage `after' returns to almost the same state as in the stage `before'.
Also, similarly to PMJ 1, there are not substantial changes of the chromospheric  field.

However, Figures \ref{Fig:12} and \ref{Fig:12*} show that, unlike PMJ 1, the average photospheric $B_{LOS}$  inside PMJ 2 does not display considerable changes with time. 
In this case, there is only a clear increase in the average field inclination and strength in the photosphere, while there are not clear  signatures in the evolution of the chromospheric field that can be associated with PMJ 2, either along the LOS or 
along the same field lines. 

Fig. \ref{Fig:12} also shows a  more inclined field inside the PMJ region than in the immediate surroundings in all the frames, a result that is consistent with the observation of an enhanced TLP signal inside PMJ 2 since the stage `before' (Fig. \ref{Fig:4tlp}).

Despite the differences in $B_{LOS}$, the evolution of the photospheric  field is similar for PMJs 1 and 2: there is a clear increase of $\gamma$ and $B$ in both cases. Slight differences in the behavior of the magnetic field are expected for PMJs that occur on different penumbral structures, which is the case of PMJs 1 and 2 as indicated by their different photospheric magnetic  configuration.

\subsection {Case 3}

\begin{figure}
   \centering

\includegraphics[width=\hsize]{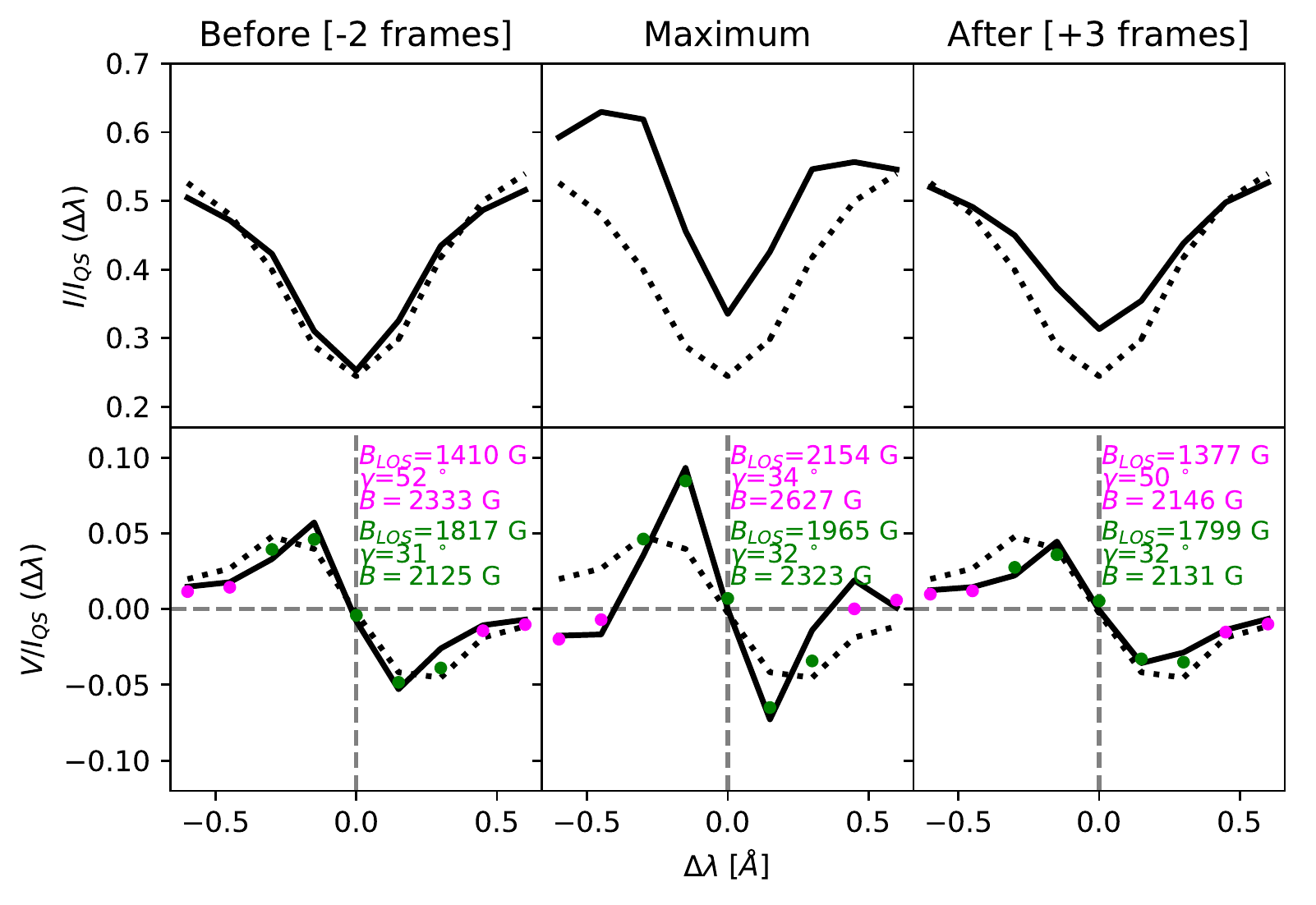}
      \caption{Observed Stokes $I$ and $V$ profiles at the MBP in PMJ 3 and results of the WFA. Same format as Fig. \ref{Fig:6}.
}
         \label{Fig:13}
   \end{figure}

\begin{figure}
   \centering

\includegraphics[width=\hsize]{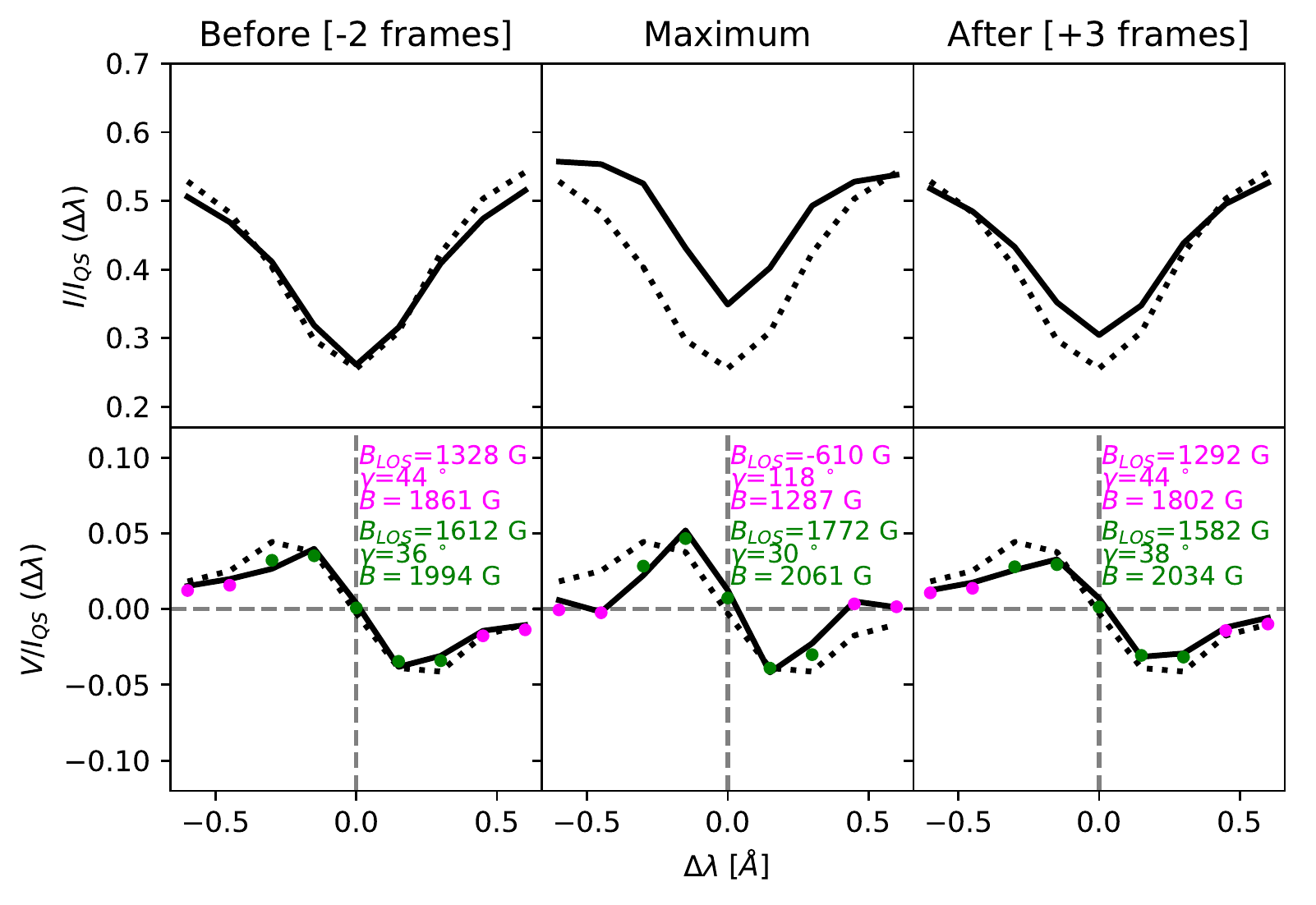}

      \caption{Observed Stokes $I$ and $V$ profiles in a pixel near the border of PMJ 3 and results of the WFA. Same format as Fig. \ref{Fig:6}.
}
         \label{Fig:13*}
   \end{figure}

PMJ 3 was observed during four consecutive frames nearly perpendicular to the symmetry line of the sunspot  (cyan marker in Fig \ref{Fig:1}), above the interface between spines and intra-spines. 
Figures \ref{Fig:13} and \ref{Fig:13*} show  the evolution of the polarization signals in  the MBP, located in the central part of the brightening region, and in a pixel located near the border of the PMJ, respectively. 
The MBP shows intensity enhancements in the entire profile at `maximum', with both wings displaying emission, but a larger increase is observed in the blue wing. As a consequence of the strong emissions in the wings, the Stokes $V$ profile displays two extra-lobes at `maximum’. At the stage `after', the intensity profile displays a more regular shape with no wing emission, similar to the profile observed in the frame `before'. However, the line core  remains notably  brighter  than  `before'.

For this pixel, the WFA infers an increase of the photospheric magnetic field strength of nearly 300 G and a strong decrease of the inclination of 18$^{\circ}$  that results in an enhancement of $B_{LOS}$ by about 700 G  between the stages `before' and `maximum'.

  \begin{figure}
   \centering
  
  \includegraphics[width=\hsize]{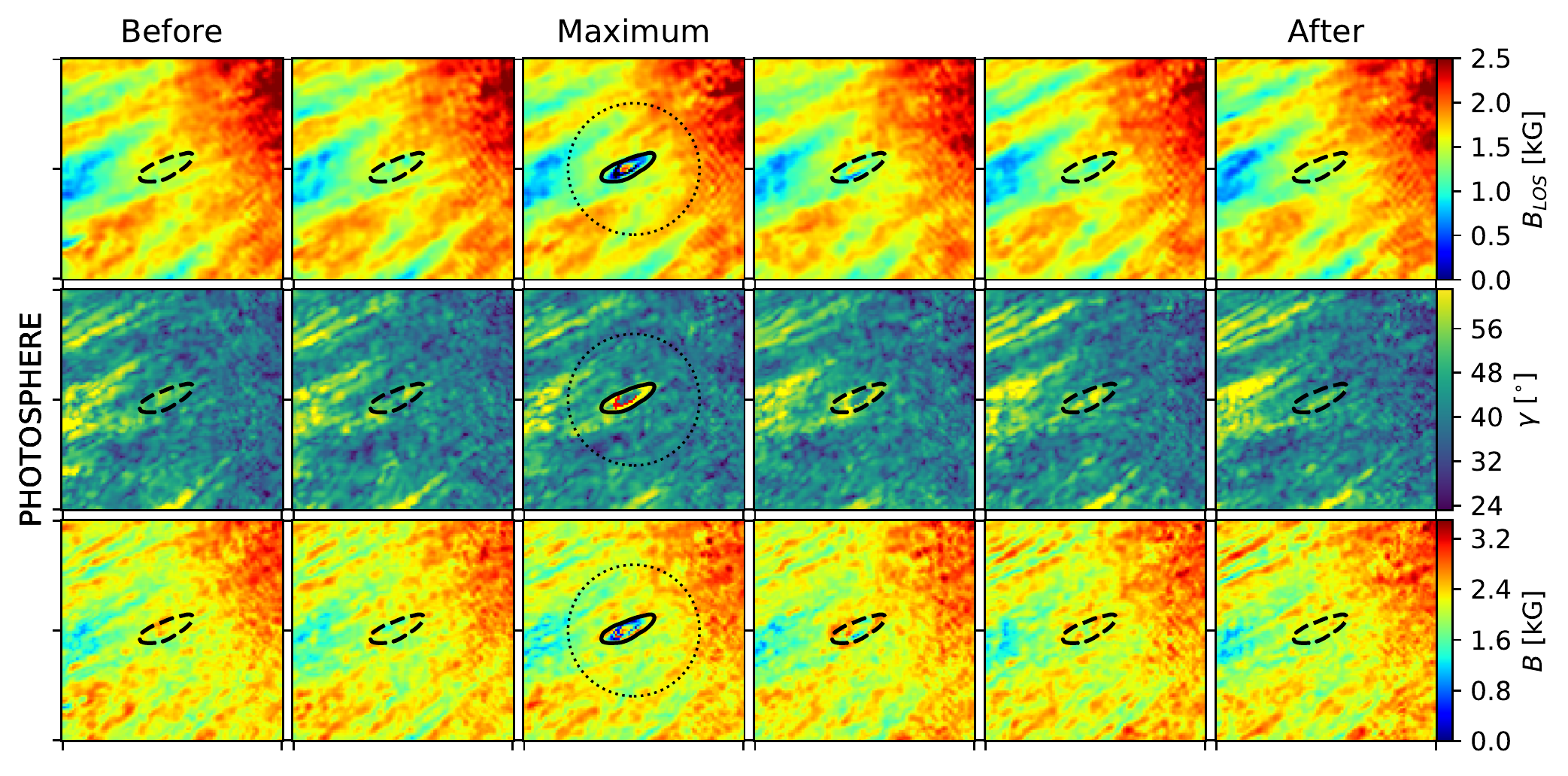}
   \includegraphics[width=\hsize]{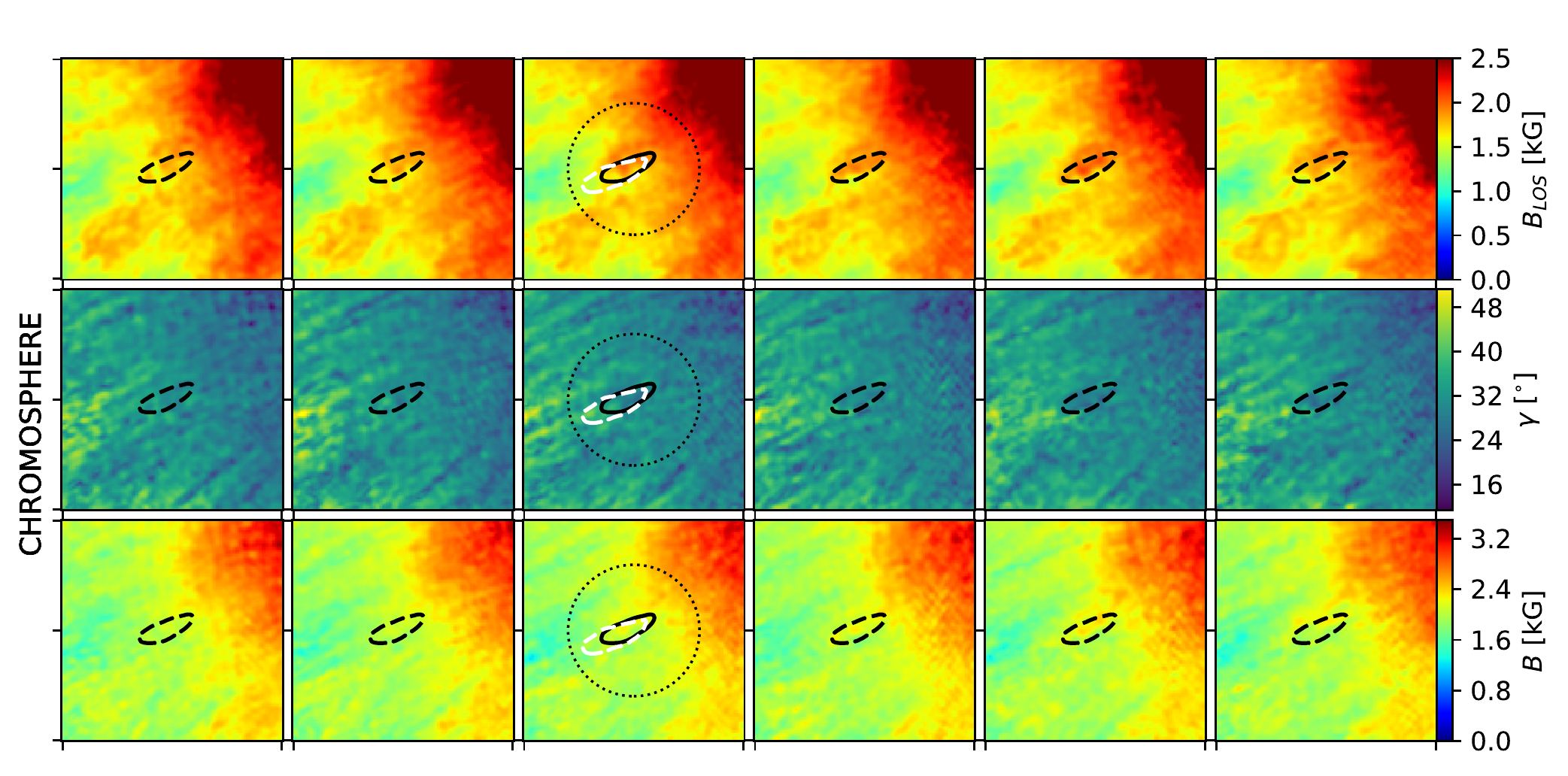}

      \caption{Maps of  $B_{LOS}$, $\gamma$, and $B$ during the different stages of the evolution of PMJ 3. Same format as  Fig. \ref{Fig:7}. }
         \label{Fig:14}
   \end{figure}

  \begin{figure}
   \centering
   
   \includegraphics[width=\hsize]{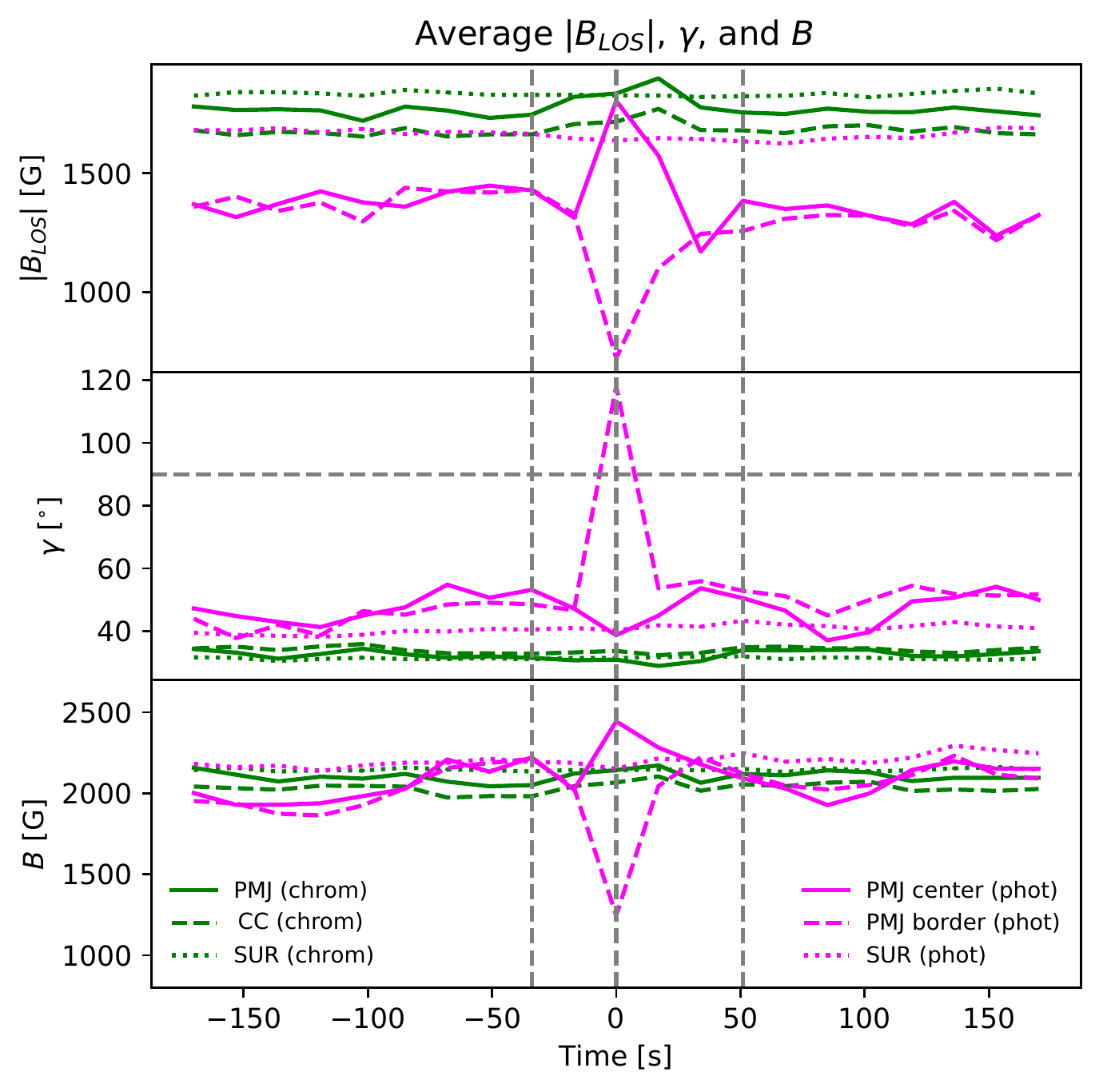}
  
      \caption{Temporal evolution of $B_{LOS}$, $\gamma$, and $B$  in different regions during PMJ 3: for the photosphere (magenta lines), in the central part of the PMJ (solid), edges of the PMJ region (dashed), and surroundings (dotted); for the chromosphere (green), in the PMJ (solid), the  CC region (dashed), and the surroundings (dotted). 
      }
         \label{Fig:14*}
   \end{figure}

The profiles at the edge of the PMJs  (Fig. \ref{Fig:13*}) also show intensity enhancements in the entire wavelength range at `maximum', but unlike the MBP, in this case the wings are not in emission, i. e. there is not a change of sign in the derivative of $I$. Nonetheless, the Stokes $V$ profile displays reduced signals in the wings with opposite signs to the profiles from the stages `before’ and `after’. This is interpreted as a change of polarity in the photospheric magnetic field by the WFA, which also infers a decrease of the magnetic field strength of more than 500 G and an increase of the inclination in the photosphere.

The different behaviors observed in  PMJ 3 at `maximum’ can also be noticed in the maps of Fig. \ref{Fig:14}: in the upper photosphere, the magnetic field becomes stronger and more vertical in the central part of the PMJ region (similar to the MBP), while it turns weaker and more inclined in the inner borders. Furthermore, the photospheric field seems to change polarity in some pixels at the edges of the PMJ region. It is difficult to judge if such polarity changes are real given that the Stokes $V$ wing signals  are very small in those pixels and even fall within the level of the noise, similarly to the profiles in Fig. \ref{Fig:13*}.

The magenta solid lines in Figure \ref{Fig:14*} show the temporal evolution of the average photospheric $B_{LOS}$, $\gamma$, and $B$ in the central part of the PMJ region, which contains the MBP. The evolution of the average field in the borders of the PMJ region is shown with dashed lines. The differences in the behavior of the photospheric field become more evident in these plots.
However, the two regions do not display big differences before and after the PMJ occurrence.
In the low chromosphere (green), the whole PMJ region displays an increase of $B_{LOS}$ and $B$ (solid lines). 
A similar response is observed in the CC region  (dashed lines in Fig. \ref{Fig:14*}).
Although these changes are still weaker than in the upper photosphere, they appear to be clearly associated with the PMJ brightening.

As in PMJs 1 and 2, the surroundings of PMJ 3 
do not display changes of magnetic configuration during the PMJ occurrence (see the dotted lines in Fig. \ref{Fig:14*}).

\begin{figure}
   \centering

    \includegraphics[width=0.85\hsize]{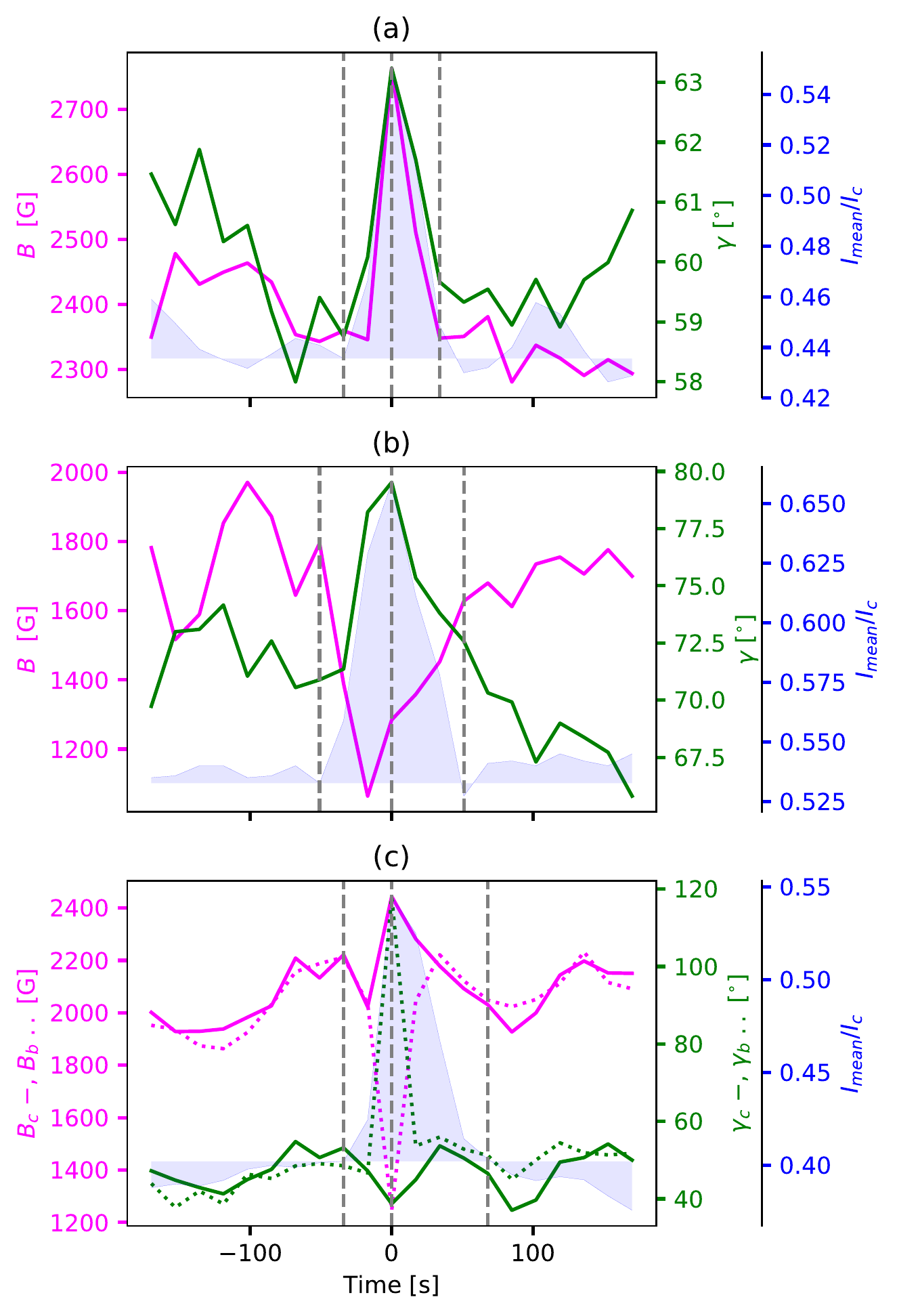}
  \caption{Three different types of temporal evolution of  $\gamma$ (green) and $B$ (magenta)  observed among the 36 PMJs  in the upper photosphere. Values are averaged inside the PMJ area for the cases shown in (a) and (b), but the averaging is done separately in the central region (solid lines) and in the borders of the brightening region (dotted lines) for case (c). Blue shades show the light curves in the PMJ calculated by averaging the intensity in the blue wing of the line. The vertical dashed lines in each panel indicate the three stages of interest, with the maximum brightness stage placed at the origin.}
         \label{Fig:diffev}
   \end{figure}

   \begin{figure*}
   \centering

   \includegraphics[width=0.33\hsize]{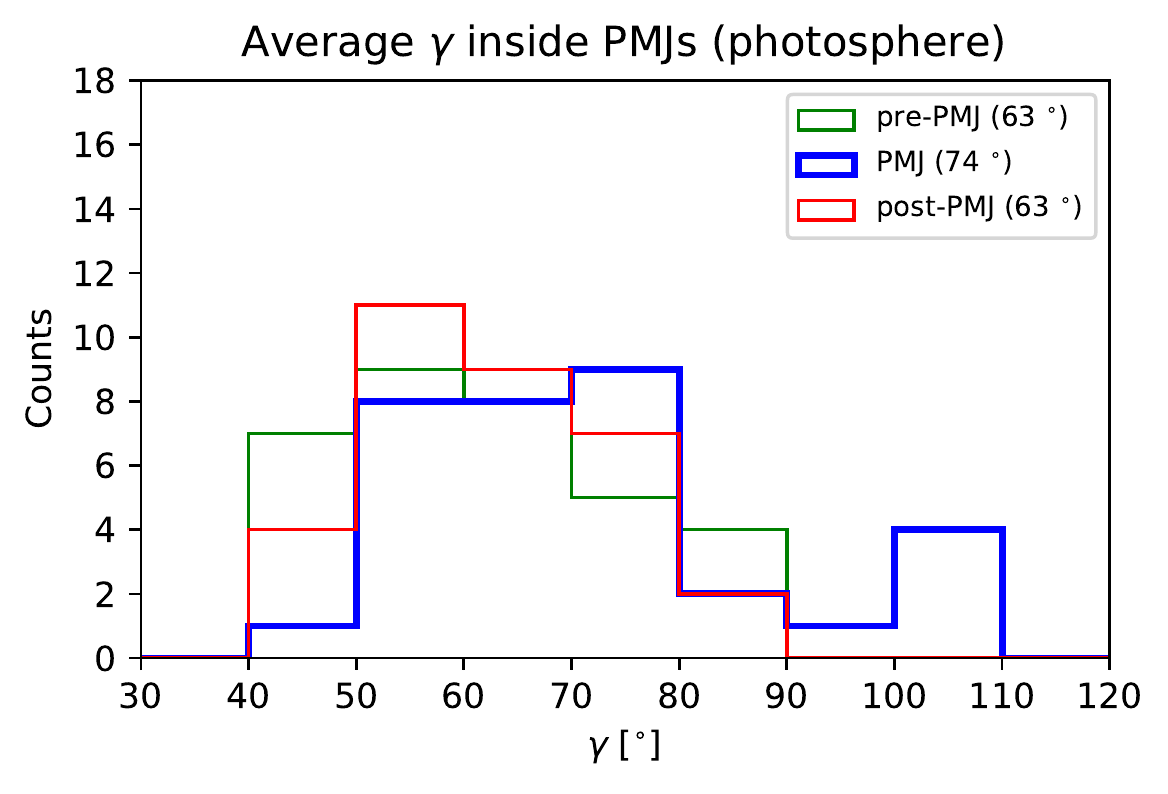}
   \includegraphics[width=0.33\hsize]{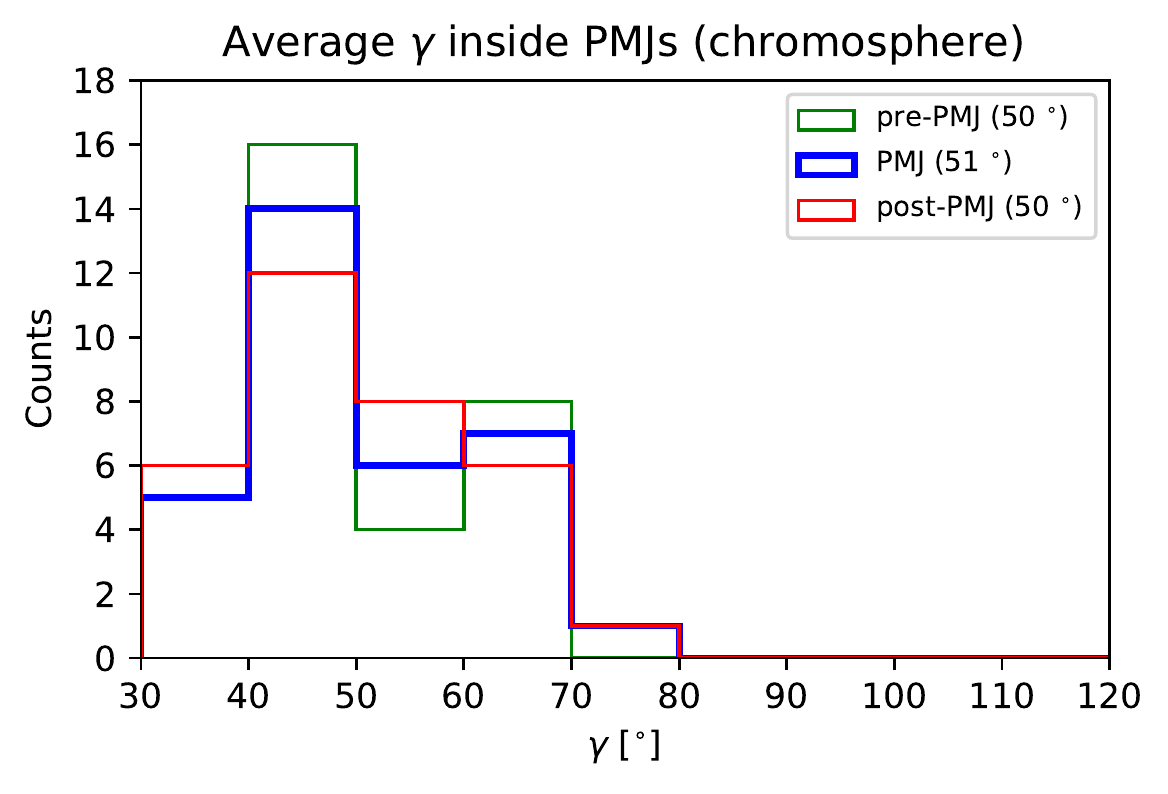}
   \includegraphics[width=0.32\hsize]{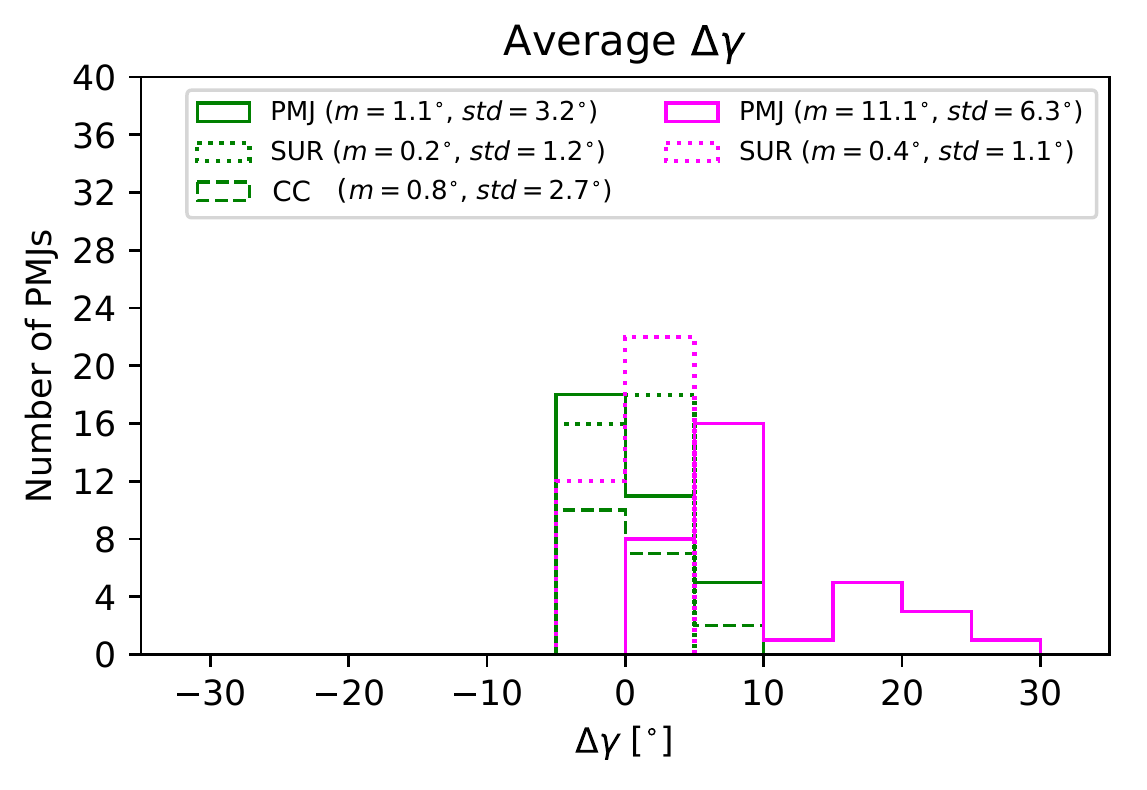}
   
   ~
     
   \includegraphics[width=0.33\hsize]{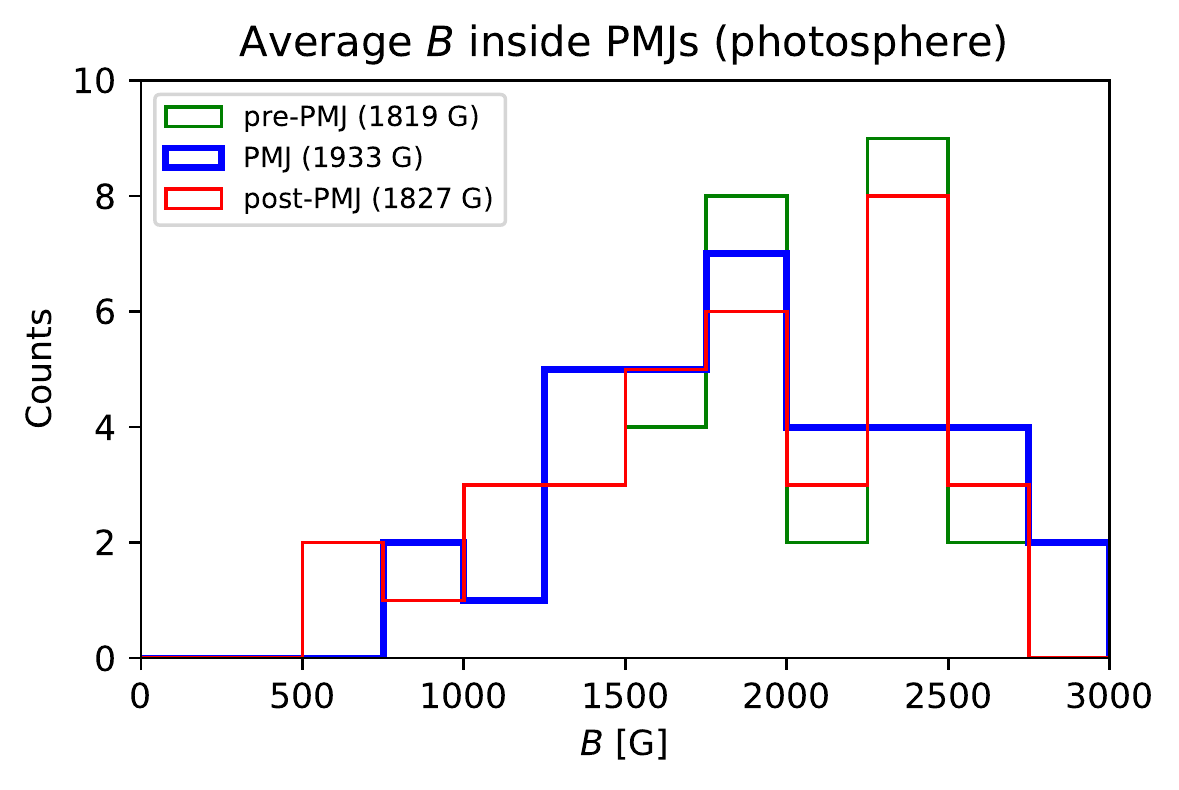}
   \includegraphics[width=0.33\hsize]{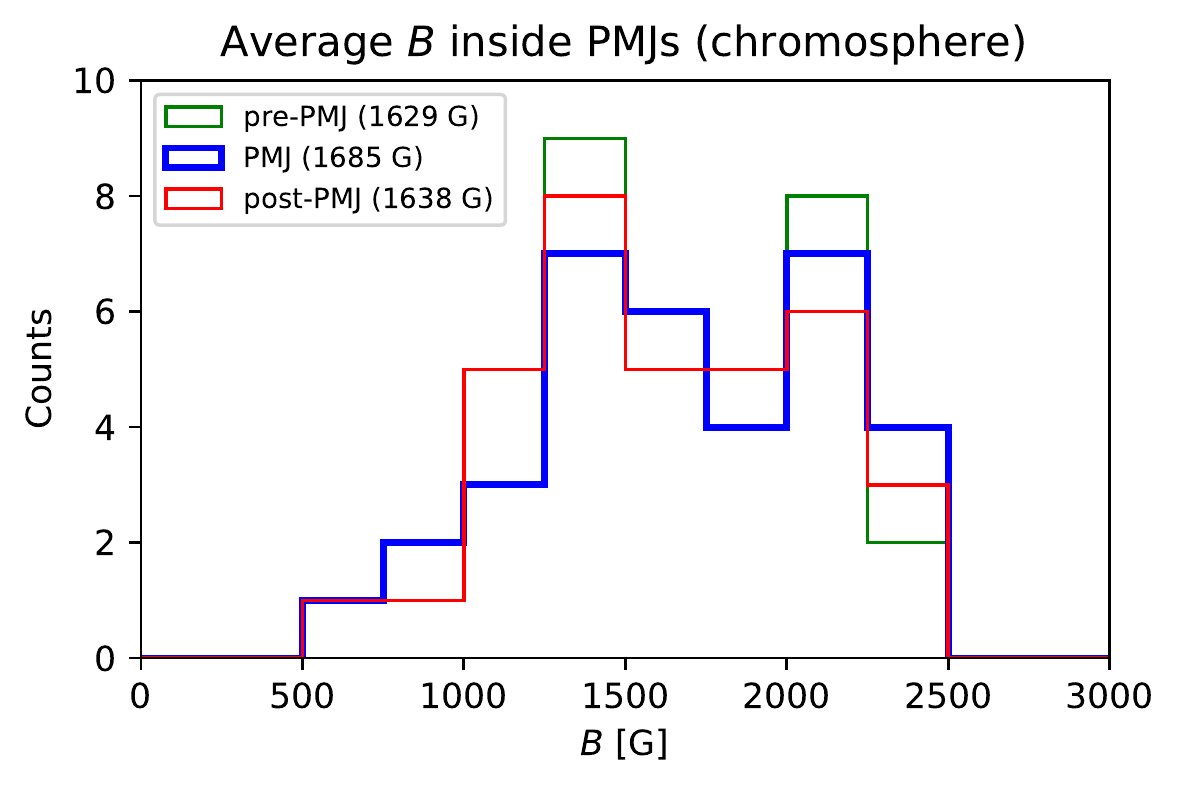}
   \includegraphics[width=0.32\hsize]{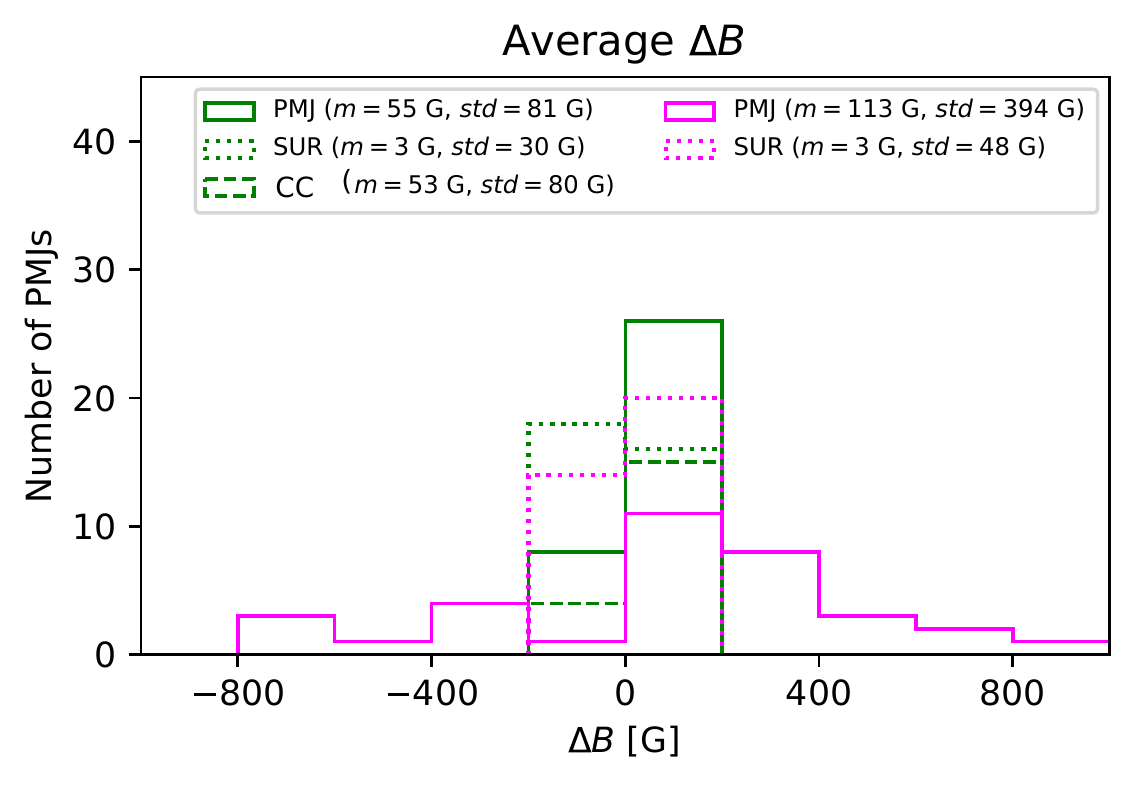}

  \caption{ Magnetic field configuration of 34 PMJs in the upper photosphere (left) and low chromosphere (center). 
  The histograms show the average values of  $\gamma$ (top), and $B$ (bottom) inside the PMJ regions during three different stages:  the pre-PMJ phase (green), which corresponds to the average field during 2 min before the PMJ, the maximum brightness (blue), and the post- PMJ phase (red), which is the average field during 2 min after the PMJ. The legends give the mean values of each distribution. 
   Right: Distribution of magnetic field changes between the pre-PMJ phase and the maximum brightness  in the photosphere (magenta) and in the chromosphere (green)  inside the PMJ regions (solid) and in the surroundings (dotted). The green dashed distributions show the chromospheric field changes occurring inside the CC regions for the 19 PMJs marked with circles in Fig. \ref{Fig:2}.
   The  legends give the mean $m$ and the standard deviation $std$ of each distribution. 
   }
         \label{Fig:st_ph}
   \end{figure*}

\subsection{Magnetic field evolution in the photosphere}

All 36 PMJs displayed considerable changes of the photospheric
magnetic field configuration. Taking into account the evolution of the
average $\gamma$ and $B$ inside the brightening regions, we identify three
types of PMJs as shown in Figure \ref{Fig:diffev}.

Most of the PMJs (25 out of 36 PMJs, or 69$\%$) show a magnetic field
evolution of type (a) similar to those of PMJs 1 and 2, i.e., they
undergo a transient increase of the field inclination and the field
strength during the maximum brightness phase. PMJs of type (a) are
indicated with red markers in Fig. \ref{Fig:2}. These PMJs tend to appear
slightly clustered in the limb-side penumbra, but they can be seen all
over the penumbra and in two cases even right at the outer penumbral
boundary.

Small differences are detected among the PMJs of type (a), mainly due
to the behavior of $B_{LOS}$. Specifically, in 16 of them the longitudinal
field decreases as a consequence of the increase of the inclination,
similar to PMJ 1. In 4 cases, $B_{LOS}$ does not display clear changes in
spite of the increase of $\gamma$ and $B$, as happens in PMJ 2. In 5 PMJs of type (a),
observed near the outer limb-side penumbral boundary where the
magnetic field is weak and largely horizontal, the WFA infers a change
of polarity during the maximum brightness stage. However, in all cases
the  Stokes $V$ profiles show reduced signals in the wings,
down to the noise level, which makes these polarity changes uncertain.

Another type of evolution is shown in Fig. \ref{Fig:diffev}b and was observed in 9
PMJs (25$\%$ of the cases; blue markers in Fig. \ref{Fig:2}). In this type of
evolution, the photospheric magnetic field vector also becomes more
inclined, but the field strength shows an overall decrease
during the brightening phase.

Finally, there are two PMJs whose brigtening regions show mixed types
of magnetic field evolution (5$\%$ of the sample; green markers in
Fig. \ref{Fig:2}). One of them is PMJ 3. In these cases, the photospheric field
becomes stronger and slightly more vertical in the central part of the
brightening, but weaker and more inclined near the borders, as shown
in Figure \ref{Fig:diffev}c. Interestingly, the WFA infers changes of magnetic
polarity in some pixels near the border of the brightening region in
both PMJs, but again this result requires further confirmation due to the
very small Stokes $V$ signals observed in the line wings.

\begin{figure}
   \centering
   
  \includegraphics[width=0.85\hsize]{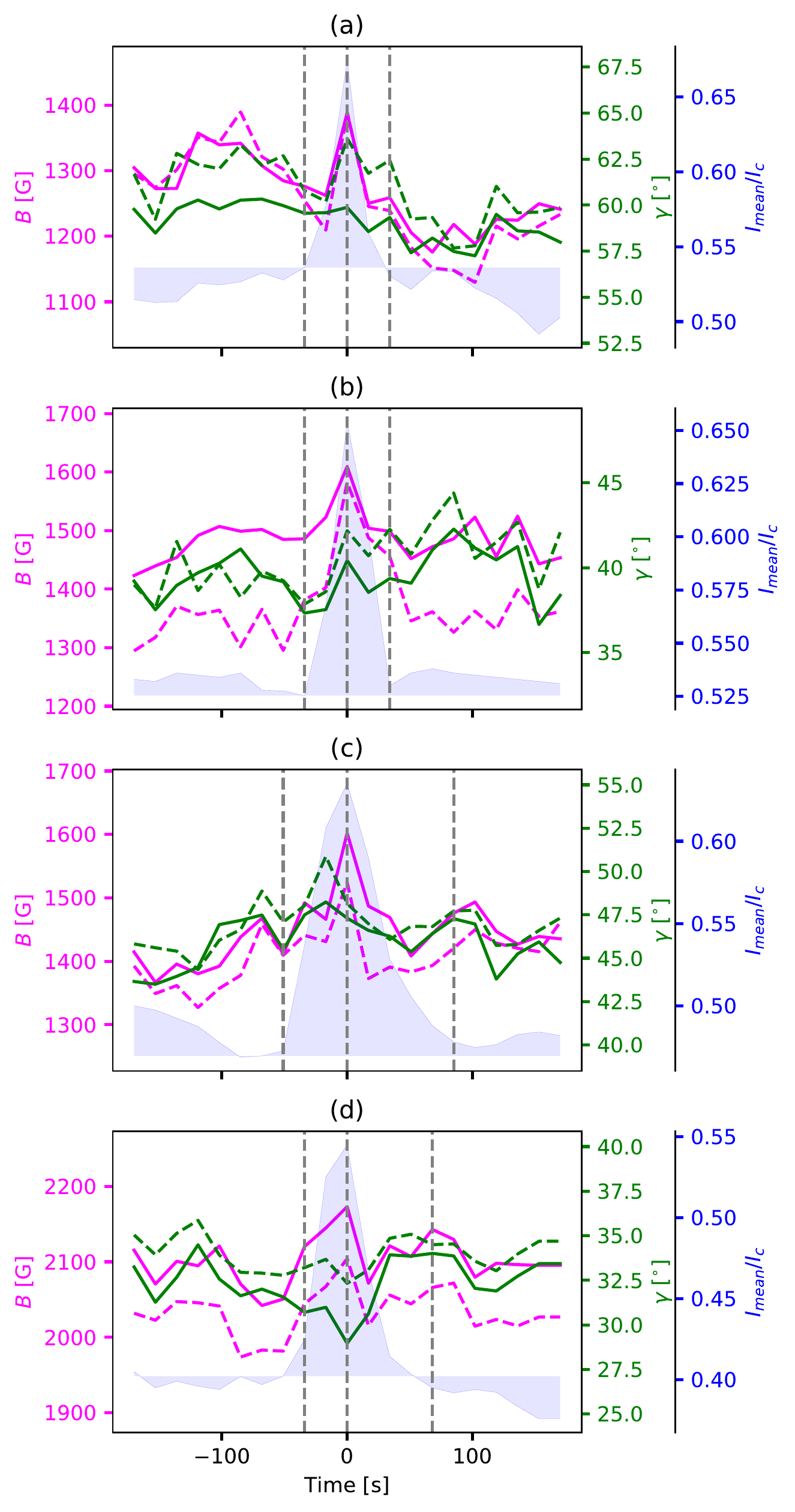}
  \caption{Temporal evolution of $\gamma$ (green) and $B$ (magenta) in the chromosphere as inferred from the WFA for the  individual PMJs marked with crossed symbols in Fig. \ref{Fig:2}. The solid lines show the average values inside the PMJ regions and the dashed lines show averages inside the CC regions. 
  Same format as  Fig. \ref{Fig:diffev}.}
         \label{Fig:diffev_cr}
   \end{figure}

The left panels of Figure \ref{Fig:st_ph} show distributions of the average
photospheric magnetic field in 34 PMJs (we exclude the two PMJs of
type (c) to avoid mixing different magnetic configurations) during
three stages of interest: the pre-PMJ phase, the maximum brightness,
and the post-PMJ phase. The pre- and post-PMJ phases correspond to the
average magnetic field over nearly 2 minutes (7 consecutive frames)
before the start and after the end of the PMJs, respectively. The
two-minute averaging is made to smear out the largest fluctuations.

In all three phases, the distributions cover a wide range of values
of $\gamma$ and $B$, due to the fact that the PMJs occur at different
radial distances within the penumbra. However, the distributions for
the maximum brightness stage (blue) stand out clearly from the pre-
and post-PMJ distributions (green and red, respectively), since they
contain more inclined and stronger fields. Specifically, the mean
values of $\gamma$ and $B$ at the time of maximum brightness are larger
that those observed during the pre- and post-PMJ phases by $\sim10^{\circ}$ 
and $\sim$100 G, respectively.

In contrast, the distributions for the pre- and post-PMJ phases are
quite alike and have very similar mean values, which means that no strong permanent changes in the local photospheric field were produced
by the occurrence of the PMJs.

The right panels of Fig. \ref{Fig:st_ph} show distributions of the photospheric
magnetic field changes occurring in the surroundings of the PMJs
(magenta dotted lines), defined by those pixels that lie along
circumferences of 1.5$\arcsec$ radius centered at the MBPs. The changes in the
surroundings are always much smaller than the ones observed in the PMJ
regions, by a factor of at least 10.
 This shows that PMJs are
highly localized events that do not leave imprints on the adjacent
medium.

\subsection{Magnetic field evolution in the chromosphere}

The magnetic properties of the short-lived PMJs exhibit weaker changes in the chromosphere than in the photosphere (see green solid distributions in the right panels of Fig. \ref{Fig:st_ph}).  In most of the cases, it is difficult to judge if those changes are caused by the PMJs because either they are of the same order of magnitude as the errors associated with the WFA or they cannot be distinguished from other small-scale fluctuations that might be intrinsic of the penumbral field.

However, Figure \ref{Fig:diffev_cr} shows four  PMJs (crossed symbols in Fig. \ref{Fig:2}) displaying some degree of chromospheric response  to the observed brightenings inside both the PMJ (solid lines) and the CC regions (dashed lined). 
These PMJs occur with a slight but clean increase of the total field strength in the chromosphere at `maximum', while there is not a clear trend in the evolution of the magnetic field inclination. 
The observed changes have similar amplitudes inside the PMJ and within the CC regions, except in the PMJ of Fig. \ref{Fig:diffev_cr}b, where the increase of the magnetic field strength is about twice as larger inside the 
CC region as that inside the PMJ.

   The central panels of Fig.  \ref{Fig:st_ph} show that the distributions of the average  $\gamma$ and $B$ in the chromosphere inside the PMJ regions are remarkably similar during the pre-PMJ, `maximum', and post-PMJ phases. They have  mean values of $\sim50^{\circ}$ and $\sim1600$ G respectively. However, the average value of $B$ for the distributions at `maximum' is  slightly larger than for the pre- and post-PMJ phases by about 50 G.

The right panels of Fig. \ref{Fig:st_ph} show distributions of the chromospheric magnetic field changes occurring inside the PMJ regions (solid green lines). 
There is about the same number of PMJs becoming 
 slightly more vertical and more inclined than  the pre-PMJ phase, 
 with a mean $\Delta\gamma\sim 1^{\circ}$. This value is well below the mean $\Delta\gamma$ of the photospheric field (magenta solid distribution). 
In contrast, 
 the chromospheric field tends to be stronger  during the maximum brightness stage 
 in most PMJ regions. The distribution has a mean $\Delta B$ and a standard deviation of  55 and 81 G, respectively. However,
these  values are about two and five times smaller than those in the upper photosphere.

Unlike the four PMJs of Fig. \ref{Fig:diffev_cr}, the majority of the identified CC regions 
did not show clear changes of the chromospheric field configuration, similar to the PMJ regions.
The distribution of their changes are shown with green dashed lines in the right panels of Fig. \ref{Fig:st_ph} and display remarkably similar trends and mean values as those of the green solid distributions.

The chromospheric field changes occurring in the surroundings of the PMJs (dotted green distributions) are yet smaller than the chromospheric changes inside the PMJ and CC regions.
This means that even if very small, the  changes of the chromospheric field observed during the PMJs along the LOS and along the field lines are likely not a result of systematic errors.

\section{Summary}
We have presented a detailed analysis of the magnetic field evolution during the lifetime of 36 short-lived PMJs. Our investigation focused on the configuration of the magnetic field in the upper photosphere and low chromosphere based on the WFA applied separately to the wings and core of the Ca II 8542 $\r{A}$ line. The main results of this analysis can be summarized as follows:
\begin{enumerate}
\item The PMJs showed brightness enhancements  in at least one of the wings and frequently also in the line core, but the enhancement was always larger in the blue wing than in the red wing (above 10$\%$ and up to 63$\%$ of the temporally averaged intensity over the full time series).

\item 19 PMJs displayed line-core intensity enhancements exceeding 10$\%$ of their temporally averaged values in regions that are shifted outward by $10-25$ pixels in the radial direction from the PMJ regions, which correspond to distances of $\sim400-1000$ km. They roughly match the expected projections on the POS of the CCs
of field lines with inclinations of $40-65^{\circ}$ in the upper photosphere. 
In many other cases, the line core remained practically unchanged in the PMJ region and  surroundings so that we could not detect the CCs. It could be that the field lines are largely horizontal and do not reach the chromosphere within our FOV.
\item In the CC regions,
 the intensity profiles commonly showed slightly brighter cores even after the PMJ disappeared, lasting  at least 85 s longer (5 frames). 

\item All 36 PMJs displayed enhanced polarization signals in  Stokes $Q$, $U$, and $V$ 
at `maximum'. In particular, the Stokes $V$ profiles presented enhanced signals within the line core region ($|\Delta\lambda| \le0.30 \r{A}$),   
but in most cases the  Stokes $V$ signals in the wings ($0.45 \r{A}\le|\Delta\lambda| \le0.60 \r{A}$) became weaker at `maximum'. In some cases, the PMJs display extra lobes in the wings due to emission or possible changes of polarity of the photospheric field. This resulted in different types of changes of the TCP during the `maximum' stage, with most of the PMJ regions showing a slight decrease or no clear changes in TCP and others displaying an enhanced TCP with respect to the stages `before' and `after'. In contrast, all PMJs exhibited enhanced TLP  at `maximum'.

\item Positive NCP values were commonly observed inside the PMJs. 
They can be understood as a result of the asymmetries of the intensity profiles, which display larger emission peaks in the blue wing than in the red wing. Such asymmetries affect  the shapes of the Stokes $V$ profiles because they follow the shapes of the partial derivatives of  Stokes $I$ with wavelength. Hence, since the sunspot has positive polarity and the positive blue lobe amplitudes and areas were predominantly larger than those of the negative red lobes, the NCP values turned out to be positive during the PMJs. They could also be partly the result of $B$ and $v_{LOS}$ gradients, as it is the case of photospheric lines. Inversions are needed to confirm or disprove this mechanism.
\item The WFA applied to the wing wavelengths reveals clear changes of the photospheric magnetic field configuration inside the  PMJ regions. These changes were generally strongest at the maximum brightness stage. 
\item We identified 3 different types of evolution of the photospheric magnetic field, but in most cases the field inclination and the field strength increase at `maximum'. 
\item The WFA applied to the line core wavelengths suggests very small changes of the chromospheric field. 
However, there is a trend of  having slightly stronger chromospheric fields inside the PMJ  and CC regions at `maximum' compared to their pre- and post-PMJ phases. 
Therefore, one could say that the enhanced polarization signals of the line core wavelengths are mostly induced by the observed intensity enhancements  rather than by strong changes in the chromospheric magnetic field.
\item The chromospheric magnetic field changes are remarkably similar when we look along the LOS or along the same field lines. This is because the projection effects were relatively small in most of the  identified  CC regions.
\item Only 4 PMJs displayed clear changes of the chromospheric field, which had a neat correlation with the brightness enhancement of the PMJs. In these cases, the field strength experienced an increase at `maximum'. However, such changes were much weaker than in the photosphere.
\item The changes of the photospheric and chromospheric magnetic field in the surroundings of the PMJs were negligible compared to the changes inside the PMJ regions, which rules out the possibility that the latter  are caused by noise or intrinsic fluctuations in the penumbra. 
\end{enumerate}

These results show that the PMJ brightenings are indeed accompanied by a perturbation in the magnetic field. 
 Such perturbation must originate closer to the height where the wings of the Ca  II 8542 $\r{A}$ line are formed, i. e. the upper photosphere.
These magnetic perturbations likely propagate but rapidly decrease toward the immediate surroundings of the PMJs, which show negligible changes beyond the brightening region.
 They  likely also propagate upwards to chromospheric heights where they leave weaker imprints. However,  since the only chromospheric change detected seems to be an increase of the magnetic field strength, it is possible that only the  compressive phase of the perturbation reaches the chromosphere.

\section{Discussion and Conclusions}

This work provides the first analysis of the temporal evolution of the magnetic field vector in short-lived PMJs using spectropolarimetric observations with very high temporal resolution.
In this section, we discuss our results and revisit some important aspects of PMJs reported in previous works.

\subsection{Location}

Most of the false detections  in our dataset were caused mainly by distorted profiles due to the strong inverse Evershed flow or strongly red-shifted profiles 
repeatedly observed along superpenumbral filaments and appearing as elongated bright structures on the blue-wing image sequences as well as on the running difference images. Therefore, after a careful examination of nearly two hundred pre-selected events, we finally identified only 36 events as short-lived PMJs.

Similar to the findings of \citet{Esteban2019} for PMJs of longer duration, our short-lived PMJs occurred above regions that are expected to harbor strong horizontal gradients of the magnetic field inclination in the photosphere. This supports the scenario of magnetic reconnection taking place in the lower photosphere as a possible driver of PMJs \citep[e. g.,][]{Tiwari2016}.

\subsection{Lifetimes}

\citet{Reardon2013} found the presence of a `precursor' phase observed up to 1 minute before the rapid impulsive brightening of some PMJs. This could be an indicator of disturbances  in density or temperature occurring prior to a reconnection event, such as those due to bow-shocks. In our data, we have also observed a large fraction of PMJs displaying a discernible `precursor' phase. However, we have discarded such events  given that their lifetimes were longer than the  upper limit of two minutes chosen for this study.

The lifetime distribution of the 36 PMJs peaks at 68 s and has a mean value of 71 s, which is lower than the 90 s mean lifetime found by  \citet{Drews2017}. However, the number of events in our sample  and the upper cutoff limit of 2 minutes for the durations are both much smaller than those considered by those authors. Hence, our samples are not comparable and  differences in the lifetime distributions can be expected.

\subsection{Magnetic field evolution}

The shortest-lived PMJs show enhanced polarization signals in the Ca II 8542 $\r{A}$ profiles with respect to their surroundings like long duration PMJs \citep{Esteban2019}. 
 However, we found that the polarization signals do change with time inside the PMJ regions, displaying the largest amplitudes during the maximum brightness stage.  
According to the WFA, the short-lived PMJs produce different types of changes in the photospheric magnetic field, which are generally strongest at the maximum brightness stage. 
The different types of evolution displayed by the photospheric magnetic field do not seem to be related to the position of the PMJs within the penumbra. Thus, it is possible that not all the short-lived PMJs have the same nature.
However, due to our temporal resolution of 17~s, we may have captured different phases of the same type of perturbation, which we misinterpreted as different types of evolution of the magnetic field.

The changes identified in the chromospheric magnetic field, 
of the order of 100 G, are in agreement with the findings of \citet{Buehler2019}, who observed an increase of nearly 100 G in the LOS magnetic field associated with two PMJs when compared to the average  field during the stages `before' and `after'. These authors also applied the WFA to Ca II 8542 $\r{A}$, but using a wider spectral range ($\pm 765$ m$\r{A}$). 
They also noticed that the magnetic  perturbation seemed to propagate on the POS with similar projected velocities as the PMJ brightenings move on such a plane.

\subsection{Origin}
According to our results, it is possible that the observed changes of the magnetic field are caused by perturbation fronts that have a photospheric origin. This is supported by the fact that they are larger in the upper photosphere than in the chromosphere. 
Furthermore, the changes of the orientation and strength of the field suggest that the perturbations have a compressive nature, and therefore they likely propagate at oblique directions with respect to the magnetic field.
Such upwardly propagating fronts would likely involve a compressive leading phase followed by a rarefaction phase which spreads out with time, similarly to a shock wave. 
Since we only observed a strengthening of the magnetic field in the chromosphere, it is possible that the rarefaction phase fades out as the front propagates upwards in the atmosphere.

 The magnetic field changes we have detected in PMJs cannot be caused by transverse waves propagating along the magnetic field since in that case we would  observe only
a change of orientation but not of the strength of the field. 
Therefore, we interpret the observed magnetic field variations  as due to perturbations propagating across or obliquely to the field, such as compressional Alfv\'en waves or MHD fast-mode wavefronts, 
 which likely steepen quickly as they travel upwards in the photosphere, thus leading to the formation of a  shock front similar to what occurs during umbral flashes \citep{Grant2018}. \citet{Buehler2019} also suggested that the magnetic perturbations observed in the chromosphere are likely produced by Alfv\'en waves.
 
 \subsection{Temperature}

A temperature increase along the magnetic field lines that connect the upper photosphere with the low chromosphere would explain the intensity changes observed in different wavelengths and in different positions on the POS, and this is in agreement with the magnetic reconnection scenario.
In particular, it is well known that the Ca II 8542 $\r{A}$ line core can be a reasonable proxy for the temperature in the chromosphere \citep{Cauzzi2009}. Therefore,  the slightly brighter line cores observed during the post-PMJ phases 
may indicate that the chromospheric plasma is generally hotter than during the pre-PMJ stages. 
This is surprising because, due to the large chromospheric radiative losses, the dissipation timescales are expected to be considerably shorter than in the photosphere unless 
there is a continuous energy input. This aspect of PMJs should be investigated 
more thoroughly in the future.

Based on inversion results, previous works have reported an increase in the temperature associated with the PMJ brightening. \citet{Esteban2019} found that the PMJ region had a temperature $200-500$ K larger than the surrounding environment above $\log(\tau)=-2$. This could be also true for the short-lived PMJs which at some places displayed intensity enhancements in the entire analyzed wavelength range. However, one would have to confirm that by performing non-LTE inversions of the  Ca II 8542 $\r{A}$ line.

Inversions are necessary to understand how the temperature, velocity, and magnetic field stratifications in the atmosphere evolve with time during the PMJs, but they require full Stokes observations in multiple lines while keeping a high temporal cadence. 
This is a particularly suitable task to be approached by the new generation of 4-m telescopes such as DKIST \citep{Rimmele2019} and EST \citep{Collados2013}, which are expected to provide unprecedented spectropolarimetric capabilities to overcome many of the current observational limitations.




\begin{acknowledgements}
We thank the referee for useful suggestions that greatly improved the paper. We also thank Sara Esteban Pozuelo for discussions and valuable comments.
This work has been funded by the Spanish Ministry of Science and Innovation through project RTI2018-096886-B-C51, including a percentage from FEDER funds, and through the Centro de Excelencia Severo Ochoa grant SEV-2017-0709 awarded to the Instituto de Astrof\'isica de Andaluc\'ia-CSIC in the period 2018-2022.
The Swedish 1-m Solar Telescope is operated on the island of La Palma by the Institute for Solar Physics of Stockholm University in the Spanish Observatory del Roque de los Muchachos of the Instituto de Astrof\'isica de Canarias.
\end{acknowledgements}

\bibliography{aabib}

\begin{thebibliography}{40}
\expandafter\ifx\csname natexlab\endcsname\relax\def\natexlab#1{#1}\fi

\bibitem[{{Bellot Rubio} {et~al.}(2006){Bellot Rubio}, Schlichenmaier, \&
  Tritschler}]{Bellot2006}
{Bellot Rubio}, L.~R., Schlichenmaier, R., \& Tritschler, A. 2006, A\&A, 453,
  1117

\bibitem[{Borrero \& Ichimoto(2011)}]{Borrero2011}
Borrero, J.~M. \& Ichimoto, K. 2011, Living Rev. Solar Phys., 8, 98

\bibitem[{Buehler {et~al.}(2019)Buehler, {Esteban Pozuelo}, {de la Cruz
  rodr\'iguez}, \& Scharmer}]{Buehler2019}
Buehler, D., {Esteban Pozuelo}, S., {de la Cruz rodr\'iguez}, J., \& Scharmer,
  G.~B. 2019, ApJ, 876, 47

\bibitem[{Cauzzi {et~al.}(2009)Cauzzi, Reardon, Rutten, Tritschler, \&
  Uitenbroek}]{Cauzzi2009}
Cauzzi, G., Reardon, K., Rutten, R.~J., Tritschler, A., \& Uitenbroek, H. 2009,
  A\&A, 503, 577

\bibitem[{Cauzzi {et~al.}(2008)Cauzzi, Reardon, Uitenbroek, Cavallini, Falchi,
  Falciani, Janssen, Rimmele, Vecchio, \& W\"{o}ger}]{Cauzzi2008}
Cauzzi, G., Reardon, K.~P., Uitenbroek, H., {et~al.} 2008, A\&A, 480, 515

\bibitem[{Centeno(2018)}]{Centeno2018}
Centeno, R. 2018, ApJ, 866, 89

\bibitem[{Collados {et~al.}(2013)Collados, Bettonvil, Cavaller, \& {EST
  Team}}]{Collados2013}
Collados, M., Bettonvil, F., Cavaller, L., \& {EST Team}. 2013, in {Highlights
  of Spanish Astrophysics VII, Proceedings of the X Scientific Meeting of the
  Spanish Astronomical Society (SEA), held in Valencia, July 9 - 13, 2012}, ed.
  J.~Guirado, L.~Lara, V.~Quilis, \& J.~Gorgas, 808--819

\bibitem[{de~la Cruz~Rodr\'iguez {et~al.}(2015)de~la Cruz~Rodr\'iguez,
  L\"ofdahl, S\"utterlin, Hillberg, \& {Rouppe van der Voort}}]{delaCruz2015}
de~la Cruz~Rodr\'iguez, J., L\"ofdahl, M.~G., S\"utterlin, P., Hillberg, T., \&
  {Rouppe van der Voort}, L. 2015, A\&A, 573, A40

\bibitem[{{de la Cruz Rodr\'iguez} {et~al.}(2013){de la Cruz Rodr\'iguez},
  {Rouppe van der Voort}, {Socas-Navarro}, \& {van Noort}}]{delaCruz2013}
{de la Cruz Rodr\'iguez}, J., {Rouppe van der Voort}, L., {Socas-Navarro}, H.,
  \& {van Noort}, M. 2013, A\&A, 556, 1

\bibitem[{Drews \& {Rouppe van der Voort}(2017)}]{Drews2017}
Drews, A. \& {Rouppe van der Voort}, L. 2017, A\&A, 602, A80

\bibitem[{{Esteban Pozuelo} {et~al.}(2016){Esteban Pozuelo}, {Bellot Rubio}, \&
  {de la Cruz Rodr\'iguez}}]{Esteban2016}
{Esteban Pozuelo}, S., {Bellot Rubio}, L.~R., \& {de la Cruz Rodr\'iguez}, J.
  2016, ApJ, 832, 170

\bibitem[{{Esteban Pozuelo} {et~al.}(2019){Esteban Pozuelo}, {de la Cruz
  Rodr\'iguez}, Drews, {Rouppe van der Voort}, Scharmer, \&
  Carlsson}]{Esteban2019}
{Esteban Pozuelo}, S., {de la Cruz Rodr\'iguez}, J., Drews, A., {et~al.} 2019,
  ApJ, 870, 88

\bibitem[{Fontela {et~al.}(1993)Fontela, Avrett, \& Loeser}]{Fontela1993}
Fontela, J.~M., Avrett, E.~H., \& Loeser, R. 1993, ApJ, 406, 319

\bibitem[{Grant {et~al.}(2018)Grant, Jess, Zaqarashvili, Beck, {Socas-Navarro},
  Aschwanden, Keys, Christian, Houston, \& Hewitt}]{Grant2018}
Grant, S. D.~T., Jess, D.~B., Zaqarashvili, T.~V., {et~al.} 2018, Nature
  Physics, 14, 480

\bibitem[{Hammar(2014)}]{Hammar2014}
Hammar, J. 2014, PhD thesis, Uppsala Universitet, Sweden

\bibitem[{J\v{u}rc\'ak \& Katsukawa(2008)}]{Jurcak2008}
J\v{u}rc\'ak, J. \& Katsukawa, Y. 2008, A\&A, 488, L33

\bibitem[{J\v{u}rc\'ak \& Katsukawa(2010)}]{Jurcak2010}
J\v{u}rc\'ak, J. \& Katsukawa, Y. 2010, A\&A, 524, A21

\bibitem[{Katsukawa {et~al.}(2007)Katsukawa, Berger, Ichimoto, Lites, Nagata,
  Shimizu, Shine, Suematsu, Tarbell, Title, \& Tsuneta}]{Katsukawa2007}
Katsukawa, Y., Berger, T.~E., Ichimoto, K., {et~al.} 2007, Science, 318, 1594

\bibitem[{Katsukawa \& Jurc\v{a}k(2010)}]{Katsukawa2010}
Katsukawa, Y. \& Jurc\v{a}k, J. 2010, A\&A, 524, A20

\bibitem[{Kosugi {et~al.}(2007)Kosugi, Matsuzaki, Sakao, Shimizu, Sone,
  Tachikawa, Hashimoto, Minesugi, Ohnishi, Yamada, Tsuneta, Hara, Ichimoto,
  Suematsu, Shimojo, Watanabe, Shimada, Davis, Hill, Owens, Title, Culhane,
  Harra, Doschek, \& Golub}]{Kosugi2007}
Kosugi, T., Matsuzaki, K., Sakao, T., {et~al.} 2007, Sol. Phys., 243, 3

\bibitem[{{Landi Degl'innocenti} \& Landolfi(2004)}]{Deglinnocenti2004}
{Landi Degl'innocenti}, E. \& Landolfi, M. 2004, Polarization in spectral lines
  (United States of America: Kluwer Academic Publishers)

\bibitem[{L\"{o}fdahl \& Sharmer(1994)}]{Lofdhal1994}
L\"{o}fdahl, M.~G. \& Sharmer, G.~B. 1994, A\&AS, 107, 243

\bibitem[{Magara(2010)}]{Magara2010}
Magara, T. 2010, ApJL, 715, L40

\bibitem[{{Mart\'inez Gonz\'alez} \& {Bellot Rubio}(2009)}]{Martinez2009}
{Mart\'inez Gonz\'alez}, M.~J. \& {Bellot Rubio}, L.~R. 2009, ApJ, 700, 1391

\bibitem[{{Quintero Noda} {et~al.}(2016){Quintero Noda}, Shimizu, {de la Cruz
  Rodr\'iguez}, Katsukawa, Ichimoto, Anan, \& Suematsu}]{Quintero2016}
{Quintero Noda}, C., Shimizu, T., {de la Cruz Rodr\'iguez}, J., {et~al.} 2016,
  MNRAS, 459, 3363

\bibitem[{Reardon {et~al.}(2013)Reardon, Tritschler, \&
  Katsukawa}]{Reardon2013}
Reardon, K., Tritschler, A., \& Katsukawa, Y. 2013, ApJ, 779, 143

\bibitem[{Rempel \& Schlichenmaier(2011)}]{Rempel2011b}
Rempel, M. \& Schlichenmaier, R. 2011, Living Rev. Solar Phys., 8, 3

\bibitem[{Rimmele(2019)}]{Rimmele2019}
Rimmele, T. 2019, BAAS, 51

\bibitem[{{Ruiz Cobo} \& {del Toro Iniesta}(1992)}]{Ruiz1992}
{Ruiz Cobo}, B. \& {del Toro Iniesta}, J.~C. 1992, ApJ, 398, 375

\bibitem[{Ryutova {et~al.}(2008)Ryutova, Berger, Frank, \& Title}]{Ryutova2008}
Ryutova, M., Berger, T., Frank, Z., \& Title, A. 2008, ApJ, 686, 1404

\bibitem[{Samanta {et~al.}(2017)Samanta, Tian, Banerjee, \&
  Schanche}]{Samanta2017}
Samanta, T., Tian, H., Banerjee, D., \& Schanche, N. 2017, ApJL, 835, L19

\bibitem[{Scharmer {et~al.}(2003)Scharmer, Bjelksjo, Korhonen, Lindberg, \&
  Petterson}]{Scharmer2003}
Scharmer, G.~B., Bjelksjo, K., Korhonen, T.~K., Lindberg, B., \& Petterson, B.
  2003, SPIE Proceedings, 4853, 341

\bibitem[{{Scharmer} {et~al.}(2008){Scharmer}, Narayan, Hillberg, de~la
  Cruz~Rodriguez, L\''ofdahl, Kiselman, S\''utterlin, van Noort, \&
  Lagg}]{Scharmer2008b}
{Scharmer}, G.~B., Narayan, G., Hillberg, T., {et~al.} 2008, ApJL, 689, L69

\bibitem[{Tiwari {et~al.}(2018)Tiwari, Moore, {De Pontieu}, Tarbell, Panesar,
  Winebarger, \& Sterling}]{Tiwari2018}
Tiwari, S.~K., Moore, R.~L., {De Pontieu}, B., {et~al.} 2018, ApJ, 869, 147

\bibitem[{Tiwari {et~al.}(2016)Tiwari, Moore, Winebarger, \&
  Alpert}]{Tiwari2016}
Tiwari, S.~K., Moore, R.~L., Winebarger, A.~R., \& Alpert, S.~E. 2016, ApJ,
  816, 92

\bibitem[{Tsuneta {et~al.}(2008)Tsuneta, Suematsu, Ichimoto, Shimizu, Otsubo,
  Shimizu, Suematsu, Nakagiri, Noguchi, Tarbell, Title, Rosenberg, Hoffmann,
  Jurcevich, Kushner, Levay, Lites, Elmore, Matsushita, Kawaguchi, Saito,
  Mikami, Hill, \& Owens}]{Tsuneta2008}
Tsuneta, S., Suematsu, Y., Ichimoto, K., {et~al.} 2008, Sol. Phys., 249, 167

\bibitem[{Uitenbroek(2001)}]{Uitenbroek2001}
Uitenbroek, H. 2001, ApJ, 557, 389

\bibitem[{{van Noort} {et~al.}(2005){van Noort}, {Rouppe van der Voort}, \&
  L\"ofdahl}]{Vannoort2005}
{van Noort}, M., {Rouppe van der Voort}, L., \& L\"ofdahl, M.~G. 2005, Sol
  Phys., 228, 191

\bibitem[{Vissers {et~al.}(2015)Vissers, {Rouppe van der Voort}, \&
  Carlsson}]{Vissers2015}
Vissers, G. J.~M., {Rouppe van der Voort}, L. H.~M., \& Carlsson, M. 2015,
  ApJL, 811, L33

\bibitem[{Vissers {et~al.}(2013)Vissers, {Rouppe van der Voort}, \&
  Rutten}]{Vissers2013}
Vissers, G. J.~M., {Rouppe van der Voort}, L. H.~M., \& Rutten, R.~J. 2013,
  ApJ, 774, 32

\end{thebibliography}


\begin{thebibliography}{51}
\expandafter\ifx\csname natexlab\endcsname\relax\def\natexlab#1{#1}\fi

\bibitem[{{Asensio Ramos} {et~al.}(2012){Asensio Ramos}, {Manso Sainz},
  {Mart\'inez Gonz\'alez}, Viticchi\'e, {Orozco Su\'arez}, \&
  Socas-Navarro}]{Asensio2012}
{Asensio Ramos}, A., {Manso Sainz}, R., {Mart\'inez Gonz\'alez}, M.~J.,
  {et~al.} 2012, ApJ, 748, 83

\bibitem[{{Bellot Rubio} {et~al.}(2004){Bellot Rubio}, Balthasar, \&
  Collados}]{Bellot2004}
{Bellot Rubio}, L.~R., Balthasar, H., \& Collados, M. 2004, A\&A, 427, 319

\bibitem[{Borrero \& {Bellot Rubio}(2002)}]{Borrero2002}
Borrero, J.~M. \& {Bellot Rubio}, L.~R. 2002, A\&A, 385, 1056

\bibitem[{Borrero {et~al.}(2011)Borrero, Tomczyk, Kubo, {Socas-Navarro}, Schou,
  Couvidat, \& Bogart}]{Borrero2011b}
Borrero, J.~M., Tomczyk, S., Kubo, M., {et~al.} 2011, Sol. Phys., 273, 267

\bibitem[{{Cabrera Solana} {et~al.}(2005){Cabrera Solana}, {Bellot Rubio}, \&
  {del Toro Iniesta}}]{Cabrera2005}
{Cabrera Solana}, D., {Bellot Rubio}, L.~R., \& {del Toro Iniesta}, J.~C. 2005,
  A\&A, 439, 687

\bibitem[{Cauzzi {et~al.}(1993)Cauzzi, Smaldone, Balasubramaniam, \&
  Keil}]{Cauzzi1993}
Cauzzi, G., Smaldone, L.~A., Balasubramaniam, K.~S., \& Keil, S.~L. 1993, Sol.
  Phys., 146, 207

\bibitem[{Danilovic {et~al.}(2008)Danilovic, Gandorfer, Lagg, Sch\"ussler,
  Solanki, V\"ogler, Katsukawa, \& Tsuneta}]{Danilovic2008}
Danilovic, S., Gandorfer, A., Lagg, A., {et~al.} 2008, A\&A, 484, L17

\bibitem[{Evershed(1909)}]{Evershed1909}
Evershed, J. 1909, MNRAS, 69, 454

\bibitem[{Frisch(1963)}]{Frisch1963}
Frisch, I.~E. 1963, {Optical Spectra of Atoms} (Moscow, Leningrad: Fizmatgiz)

\bibitem[{Frutiger(2000)}]{Frutiger2000}
Frutiger, C. 2000, PhD thesis, Institute of Astronomy, ETH Z\"urich,
  Switzerland, no. 13896

\bibitem[{Frutiger {et~al.}(2000)Frutiger, Solanki, Fligge, \&
  Bruls}]{Frutiger2000b}
Frutiger, C., Solanki, S.~K., Fligge, M., \& Bruls, J. H. M.~J. 2000, A\&A,
  358, 1109

\bibitem[{Georgoulis(2005)}]{Georgoulis2005}
Georgoulis, M.~K. 2005, ApJ, 629, L69

\bibitem[{Keppens \& {Mart\'inez Pillet}(1996)}]{Keppens1996}
Keppens, R. \& {Mart\'inez Pillet}, V. 1996, A\&A, 316, 229

\bibitem[{Lagg {et~al.}(2009)Lagg, Ishikawa, Merenda, Wiegelmann, Tsuneta, \&
  Solanki}]{Lagg2009}
Lagg, A., Ishikawa, R., Merenda, L., {et~al.} 2009, Astronomical Society of the
  Pacific Conference Series, 415, 327

\bibitem[{Lagg {et~al.}(2004)Lagg, Woch, Kripp, \& Solanki}]{Lagg2004}
Lagg, A., Woch, J., Kripp, N., \& Solanki, S.~K. 2004, A\&A, 414, 1109

\bibitem[{Lites {et~al.}(2013)Lites, Akin, \& et~al.}]{Lites2013}
Lites, B.~W., Akin, D.~L., \& et~al., G.~C. 2013, Sol. Phys., 283, 579

\bibitem[{Lites {et~al.}(2001)Lites, Elmore, \& Streander}]{Lites2001}
Lites, B.~W., Elmore, D.~F., \& Streander, K.~V. 2001, in Advanced Solar
  Polarimetry - Theory, Observation, and Instrumentation - ASP Conference
  Proceedings, ed. M.~Sigwarth, Vol. 236, 33

\bibitem[{Lites \& Ichimoto(2013)}]{Lites2013b}
Lites, B.~W. \& Ichimoto, K. 2013, Sol. Phys., 283, 601

\bibitem[{Lites {et~al.}(1990)Lites, Skumanich, \& Scharmer}]{Lites1990}
Lites, B.~W., Skumanich, A., \& Scharmer, G.~B. 1990, ApJ, 355, 329

\bibitem[{Livingston(2002)}]{Livingston2002}
Livingston, W. 2002, Sol. Phys., 207, 41

\bibitem[{Livingston \& Harvey(2006)}]{Livingston2006}
Livingston, W. \& Harvey, J.~W. 2006, Sol. Phys., 239, 41

\bibitem[{Lucy(1974)}]{Lucy1974}
Lucy, L.~B. 1974, AJ, 79, 745

\bibitem[{{Mart\'inez Pillet}(2000)}]{Martinez2000}
{Mart\'inez Pillet}, V. 2000, A\&A, 361, 734

\bibitem[{Mathew {et~al.}(2003)Mathew, Lagg, Solanki, Collados, Borrero,
  Berdyugina, Krupp, Woch, \& Frutiger}]{Mathew2003}
Mathew, S.~K., Lagg, A., Solanki, S.~K., {et~al.} 2003, A\&A, 410, 695

\bibitem[{Mathew {et~al.}(2004)Mathew, Solanki, Lagg, Collados, Borrero, \&
  Berdyugina}]{Mathew2004}
Mathew, S.~K., Solanki, S.~K., Lagg, A., {et~al.} 2004, A\&A, 422, 693

\bibitem[{Moore(1945)}]{Moore1945}
Moore, C.~E. 1945, {A Multiplet Table of Astrophysical Interest} (Princeton,
  New Jersey: Princeton University Observatory)

\bibitem[{Okamoto \& Sakurai(2018)}]{Okamoto2018}
Okamoto, T.~J. \& Sakurai, T. 2018, ApJL, 852, L16

\bibitem[{Rees \& Semel(1979)}]{Rees1979}
Rees, D.~E. \& Semel, M.~D. 1979, A\&A, 74, 1

\bibitem[{Rempel(2012)}]{Rempel2012b}
Rempel, M. 2012, ApJ, 62, 21

\bibitem[{Rempel(2015)}]{Rempel2015}
Rempel, M. 2015, ApJ, 814, 125

\bibitem[{Richardson(1972)}]{Richardson1972}
Richardson, W.~H. 1972, JOSA, 62, 55

\bibitem[{{Ruiz Cobo}(2007)}]{Ruiz2007}
{Ruiz Cobo}, B. 2007, in Modern Solar Facilities-Advanced Solar Science, ed.
  F.~Kneer, K.~G. Puschmann, \& A.~D. Wittmann, 287--296

\bibitem[{Sabatier(2000)}]{Sabatier2000}
Sabatier, P.~C. 2000, Journal of Mathem. Phys., 41, 4082

\bibitem[{Schlichenmaier \& Collados(2002)}]{Schlichenmaier2002b}
Schlichenmaier, R. \& Collados, M. 2002, A\&A, 381, 668

\bibitem[{Schwarz(1978)}]{Schwarz1978}
Schwarz, G.~E. 1978, Ann. Stat., 6, 461

\bibitem[{Semel(1967)}]{Semel1967}
Semel, M. 1967, AnAp, 30, 513

\bibitem[{Semel(1970)}]{Semel1970}
Semel, M. 1970, A\&A, 5, 330

\bibitem[{{Siu-Tapia} {et~al.}(2017){Siu-Tapia}, Lagg, Solanki, {van Noort}, \&
  J\v{u}rc\'ak}]{siu2017}
{Siu-Tapia}, A.~L., Lagg, A., Solanki, S.~K., {van Noort}, M., \& J\v{u}rc\'ak,
  J. 2017, A\&A, 607, A36

\bibitem[{{Siu-Tapia} {et~al.}(2018){Siu-Tapia}, Rempel, Lagg, \&
  Solanki}]{Siu2018}
{Siu-Tapia}, A.~L., Rempel, M., Lagg, A., \& Solanki, S.~K. 2018, ApJ, 852, 66

\bibitem[{Skumanich \& Lites(1987)}]{Skumanich1987}
Skumanich, A. \& Lites, B.~W. 1987, ApJ, 322, 473

\bibitem[{Skumanich {et~al.}(1994)Skumanich, Lites, \& {Mart\'inez
  Pillet}}]{Skumanich1994}
Skumanich, A., Lites, B.~W., \& {Mart\'inez Pillet}, V. 1994, Solar Surface
  Magnetism, 99

\bibitem[{Solanki(1987)}]{Solanki1987b}
Solanki, S.~K. 1987, PhD thesis, Institute of Astronomy, ETH Z\"urich,
  Switzerland, no. 8309

\bibitem[{Solanki(1993)}]{Solanki1993}
Solanki, S.~K. 1993, Space Sci. Rev., 63, 1

\bibitem[{Solanki \& Montavon(1993)}]{Solanki1993b}
Solanki, S.~K. \& Montavon, C. A.~P. 1993, A\&A, 275, 283

\bibitem[{Stenflo(1993)}]{Stenflo1993}
Stenflo, J.~O. 1993, in Proc. Int. Conf. (Feidburg, Germany: Cambridge
  University Press), 301

\bibitem[{Suematsu {et~al.}(2008)Suematsu, Tsuneta, Ichimoto, Shimizu, Otsubo,
  Katsukawa, Nakagiri, Noguchi, Tamura, Kato, Hara, Kubo, Mikami, Saito,
  Matsushita, Kawaguchi, Nakaoji, Nagae, Shimada, Takeyama, \&
  Yamamuro}]{Suematsu2008}
Suematsu, Y., Tsuneta, S., Ichimoto, K., {et~al.} 2008, Sol. Phys., 249, 197

\bibitem[{Tiwari {et~al.}(2013)Tiwari, {van Noort}, Lagg, \&
  Solanki}]{Tiwari2013}
Tiwari, S.~K., {van Noort}, M., Lagg, A., \& Solanki, S.~K. 2013, A\&A, 557,
  A25

\bibitem[{Tiwari {et~al.}(2015)Tiwari, {van Noort}, Solanki, \&
  Lagg}]{Tiwari2015}
Tiwari, S.~K., {van Noort}, M., Solanki, S.~K., \& Lagg, A. 2015, A\&A, 583,
  A119

\bibitem[{{van Noort}(2012)}]{vannoort2012}
{van Noort}, M. 2012, A\&A, 548, A5

\bibitem[{{van Noort} {et~al.}(2013){van Noort}, Lagg, Tiwari, \&
  Solanki}]{vannoort2013}
{van Noort}, M., Lagg, A., Tiwari, S.~K., \& Solanki, S.~K. 2013, A\&A, 557,
  A24

\bibitem[{{Westendorp Plaza} {et~al.}(2001){Westendorp Plaza}, {del Toro
  Iniesta}, {Ruiz Cobo}, \& {Mart\'inez Pillet}}]{Westendorp2001b}
{Westendorp Plaza}, C., {del Toro Iniesta}, J.~C., {Ruiz Cobo}, B., \&
  {Mart\'inez Pillet}, V. 2001, ApJ, 547, 1148

\end{thebibliography}


\begin{thebibliography}{3}
\expandafter\ifx\csname natexlab\endcsname\relax\def\natexlab#1{#1}\fi

\bibitem[{Cauzzi {et~al.}(2008)Cauzzi, Reardon, Uitenbroek, Cavallini, Falchi,
  Falciani, Janssen, Rimmele, Vecchio, \& W\"{o}ger}]{Cauzzi2008}
Cauzzi, G., Reardon, K.~P., Uitenbroek, H., {et~al.} 2008, A\&A, 480, 515

\bibitem[{Scharmer {et~al.}(2003)Scharmer, Bjelksjo, Korhonen, Lindberg, \&
  Petterson}]{Scharmer2003}
Scharmer, G.~B., Bjelksjo, K., Korhonen, T.~K., Lindberg, B., \& Petterson, B.
  2003, SPIE Proceedings, 4853, 341

\bibitem[{{Scharmer} {et~al.}(2008){Scharmer}, Narayan, Hillberg, de~la
  Cruz~Rodriguez, L\''ofdahl, Kiselman, S\''utterlin, van Noort, \&
  Lagg}]{Scharmer2008b}
{Scharmer}, G.~B., Narayan, G., Hillberg, T., {et~al.} 2008, ApJL, 689, L69

\end{thebibliography}
\bibliographystyle{aa}
%
%
%
%
%
%
%

\end{document}